%% file: main.tex
\documentclass[USenglish]{book}

\usepackage[utf8]{inputenc}
\usepackage{graphicx}
\usepackage[]{natbib}
\usepackage{chapterbib}
\usepackage{wileyvch}
\usepackage[breaklinks]{hyperref}
\author{Thomas Schultz and Marc Vrakking}\title{Attosecond and Free electron Laser Science}

\begin{document}
\frontmatter
\mainmatter
\include{06_Smirnova}
\backmatter
\end{document}

%% file: 06_Smirnova.tex
\newcommand{\be}{\begin{equation}}
\newcommand{\ee}{\end{equation}}
\newcommand{\bea}{\begin{eqnarray}}
\newcommand{\eea}{\end{eqnarray}}
\newcommand{\ra}{\rangle}
\newcommand{\la}{\langle}
\newcommand{\bef}{\begin{figure}}
\newcommand{\enf}{\end{figure}}
\newcommand{\p}{\mathbf{p}}
\newcommand{\rmbf}{\mathbf{r}}
\newcommand{\pc}{\mathbf{p}_{\rm c}}
\newcommand{\x}{\mathbf{x}}
\newcommand{\y}{\mathbf{y}}
\newcommand{\q}{\mathbf{q}}
\newcommand{\R}{\mathbf{R}}
\newcommand{\w}{\omega}
\let\k\undefined
\newcommand{\k}{\mathbf{k}}
\newcommand{\F}{\mathbf{F}}
\newcommand{\A}{\mathbf{A}}
\newcommand{\D}{\mathbf{D}}
\let\d\undefined
\newcommand{\d}{\mathbf{d}}

\chapterauthor{Olga Smirnova and Misha Ivanov}

\chapter{Multielectron High Harmonic Generation: simple man on a complex plane\label{Chapter_7_Smirnova}}\normalsize

\section{Introduction}

Attosecond science has emerged with the discovery of coherent
electron-ion collisions induced by a strong laser field, usually
referred to as "re-collisions" (\cite{Corkum93}). This discovery
was initiated by the numerical experiments of K. Schafer, J.
Krause and  K. Kulander (see \cite{Kulander92}). The work by
\cite{Corkum93} drew on the concepts developed in the earlier work
of \cite{Brunel87,Brunel90} and \cite{Brunel89}. It has also been
predated by the concept of the 'Atomic Antenna' (\cite{Kuchiev87}).
With the benefit of hindsight, we now see the work by
\cite{Kuchiev87} as the earliest quantum counterpart of the
classical picture developed by \cite{Corkum93} and
\cite{Kulander93} \footnote{While the quantum vision of
\cite{Kuchiev87} has predated the classical picture, at that time
it lacked the striking clarity and transparency of the
classical model (\cite{Corkum93}),
which linked several key -- and seemingly disparate strong-field
phenomena -- high harmonic generation, production of very high
energy electrons, and extreme efficiency of double ionization. The
history of this discovery is rich and interesting in its own
right. Some of it is recounted, from a more historical perspective,
in Chapter 4.  Our purpose
here is different -- we simply urge our reader to read the papers
by \cite{Brunel87,Brunel90}, \cite{Brunel89}, \cite{Kuchiev87},
\cite{Schafer93}, as well as a seemingly unrelated paper of \cite{Gallagher88}.}.

The classical picture of strong-field-induced ionization dynamics
is summarized as follows.  Once ionization removes an electron
from an atom or a molecule, this electron finds itself in the
strong oscillating laser field. Newton's equations of motion show that,
within one or few cycles after ionization, the oscillating
electron can be driven back by the laser field to re-encounter the
parent ion. During this re-encounter, referred to as re-collision,
the electron can do many things: scatter elastically (diffract),
scatter inelastically (excitation or ionization of the parent
ion), or radiatively recombine into one of the ion's empty states.
It is this latter process that we will focus on here. The
classical picture is usually referred to as the three-step model,
or the simple man model \footnote{As far as one of us (M.I.) can
remember, the latter term has been used by K. Kulander, K. Schafer
and H.-G. Muller, who have contributed a lot to the development of
this classical model.}.

If the recombination occurs to the exact same state that the
electron has left from, then the phase of the emitted radiation is
the same from one atom to another, leading to the generation of
coherent radiation in the medium. This process is known as high
harmonic generation (HHG). It produces tens of eV-broad coherent
spectra and has two crucial applications. First, high harmonic
emission is used to generate attosecond pulses of light (see e.g.
\cite{Krausz09}), which can then be used in time-resolved
pump-probe experiments. Second, the ultra-broad coherent harmonic
spectrum carries attosecond information about the underlying
nonlinear response, which can be extracted. The second direction
is the subject of high harmonic spectroscopy (see e.g.
\cite{Lein05,Baker06,Smirnova09,Stefan}) -- a new imaging technique with
a combination of sub-Angstrom spatial and attosecond temporal
resolution.

In the language of nonlinear optics, high harmonic generation is
a frequency up-conversion process that results from the
macroscopic response of the medium. The nonlinear polarization is
induced in the medium by (i) the response of the atoms and the molecules,
(ii) the response of the free electrons, (iii) the response of the guiding
medium. Here we focus on the theory of single atom or single
molecule response. The description of macroscopic propagation effects,
which determine how coherent radiation from different atoms or
molecules add together, can be found in \cite{Gaarde2008}.

From the famous simple man model to the recent multichannel model,
we will try to guide you through the several landmarks in our
understanding of high harmonic generation. We hope to provide
recipes and insight for modelling the harmonic response in complex
systems. The chapter includes the following sections:
\begin{itemize}
  \item 1.2 The \textbf{simple man model} of high harmonic generation
  (HHG);
  \item 1.3 Formal approach for one-electron systems;
  \item 1.4 \textbf{The Lewenstein model}: stationary phase equations for
  HHG;
   \item 1.5 Analysis of complex trajectories;
   \item 1.6 Factorization of the HHG dipole: simple man on a complex
   plane;
  \item 1.7 \textbf{The photoelectron model} of HHG:  the improved 'simple
  man';
  \item 1.8 \textbf{The multichannel model} of HHG: Tackling multi-electron systems;
  \item 1.9  Outlook;
  \item 1.10  Acknowledgements;
  \item 1.11 Appendix A: Supplementary derivations;
  \item 1.12 Appendix B: The saddle point method;
  \item 1.13 Appendix C: Treating the cut-off region: regularization of the divergent stationary phase
  solutions;
 \item 1.14  Appendix D: Finding saddle points for the Lewenstein
 model.
\end{itemize}
Atomic units $\hbar=m=e=1$ are used everywhere, unless specified
otherwise.

\section{The simple man model of high harmonic generation (HHG)}

Experiments in the eighties and the early nineties of the last century
yielded an astounding result: shaken with sufficiently intense
infrared laser radiation, the atomic medium was found to up-convert
the frequency of the driving infrared laser light by up to two
orders of magnitude
(see e.g. \cite{Huillier93,Macklin93}).
The
observed harmonic spectrum formed a long plateau, with many
harmonic orders, followed by a sharp cut-off. This observation has
to be placed in the context of what has been routinely seen in the
traditional nonlinear optics: in the absence of resonances, the
nonlinear response would decrease dramatically with increasing
harmonic order, and the harmonic numbers would hardly ever reach
double digits, let alone form a plateau extending beyond N=101.

To generate very high harmonics of the driving frequency, the atom
has to absorb lots of photons. Generation of harmonics with
numbers like N=21,...,31,..., etc. means that at least that many
photons (21, ..., 31, ...) had to be absorbed by the atom.

The minimal amount of photons required for ionization is
$N_0=I_p/\omega$, where $I_p$ is the ionization potential and $\omega$
is the infrared laser frequency. For $I_p\sim 12-15$~eV and an 800
nm driving IR laser field (the standard workhorse in many HHG
experiments), $N_0\sim 10$. One would have thought that once ten
or so photons are absorbed, the electron should be free. And since
it is well-known that a free electron should not absorb any more
photons, the emission should stop around $N=11$ or so,
in stark contrast with experimental
observations.

Why and how many additional photons are absorbed? What is the underlying
mechanism? The liberated electron oscillates in the laser field,
and its instantaneous energy can be very high. Can this
instantaneous electron energy be converted into the harmonic
photons? Where is the source of non-linearity, if the free
electron oscillates with the frequency of the laser field?

The  physical picture that clearly answered these questions is the
classical three-step model. It is simple, remarkably accurate, and
is also intrinsically sub-cycle: within one optical period, an
electron is (i) removed from an atom or molecule, (ii) accelerated
by the oscillating laser field, and (iii) driven back to
re-collide with the parent ion. This picture connects the key
strong-field phenomena: above-threshold ionization, non-sequential
double ionization, and high harmonic generation. It reveals the
source of non-linearity in HHG: the recombination of the
accelerated electron with the ion.

\begin{figure}
\includegraphics[width=0.498\textwidth]{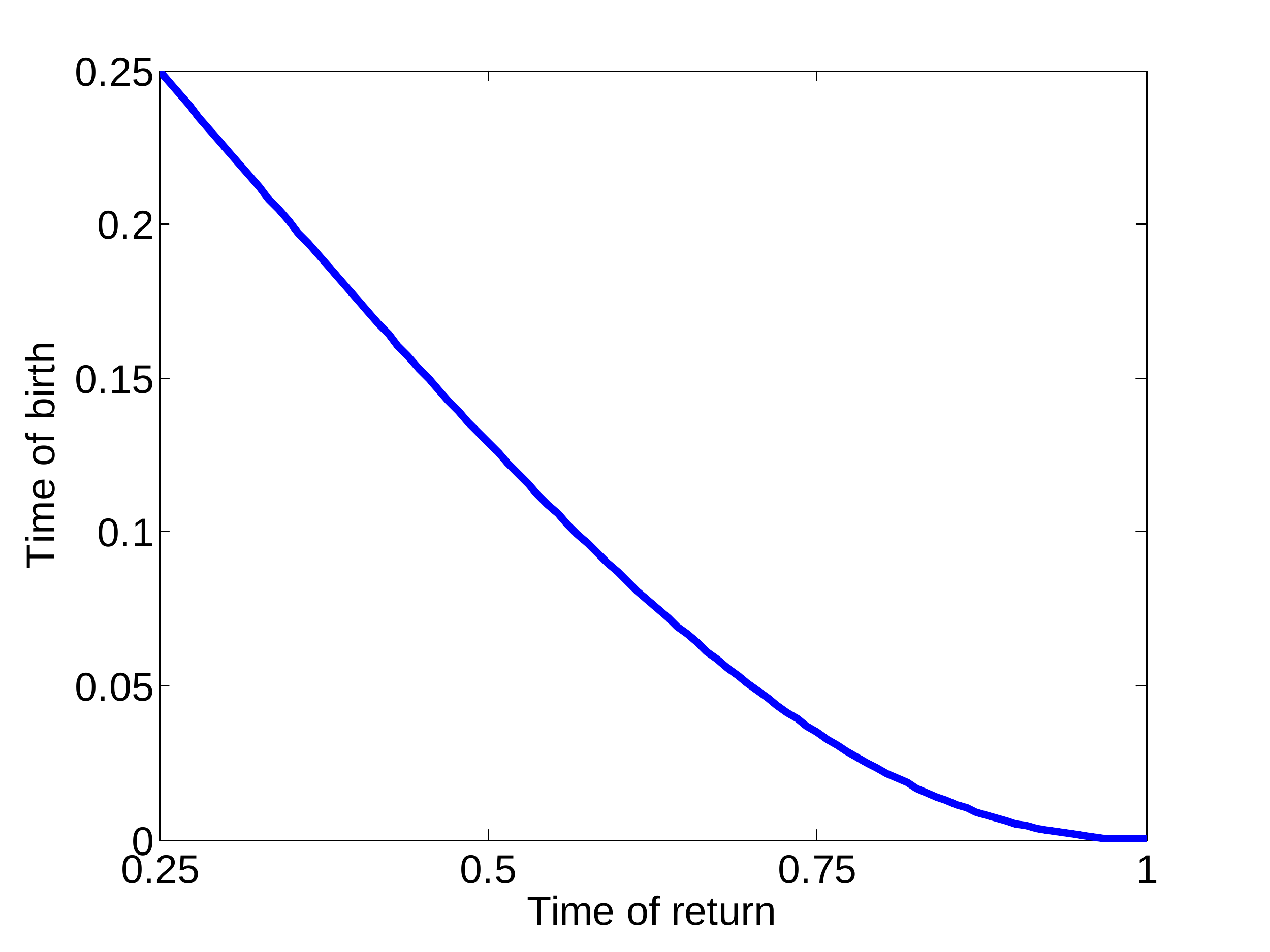}
\includegraphics[width=0.498\textwidth]{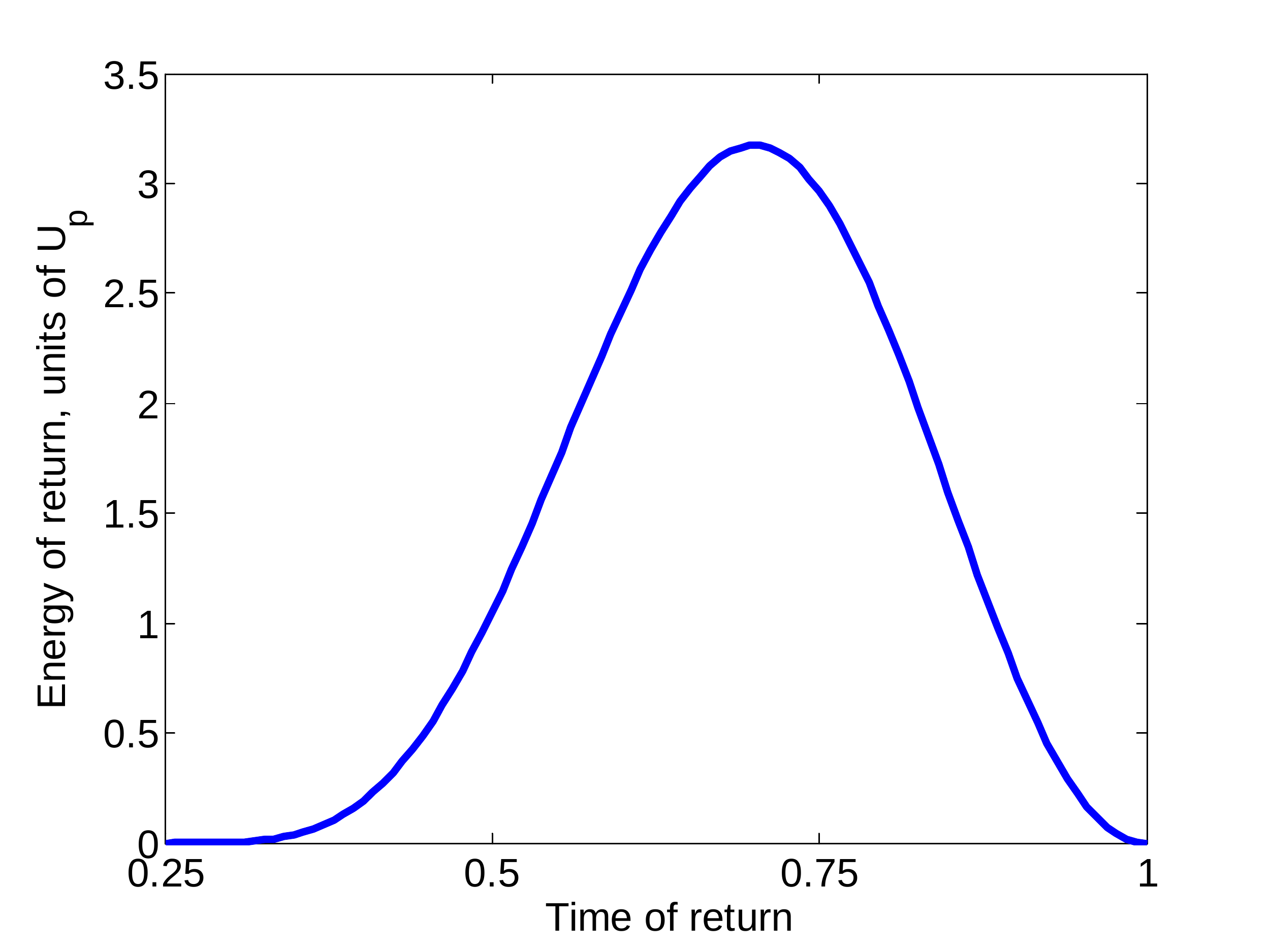}
\caption{Window of classical 'birth' times and the return energy. Left panel: Time of birth vs. time of return.
Right panel: Energy of the electron at the time of return.} \label{fig:3step}
\end{figure}

How can one check that this mechanism is indeed responsible for
HHG? The key thing test is whether or not this picture
explains the cut-off of the harmonic spectra, that is, the highest
harmonic order that can be efficiently produced. Numerically, the
empirical cut-off law was found to be $\Omega_{\rm max}=I_p+3U_p$
(\cite{Krause92}), where $U_p$ is the cycle-averaged energy of the
electron's oscillatory motion in the
laser field. To calculate the classical cut-off,  we should
calculate the maximal instantaneous energy of the returning
electron, but to do so we need to know the initial
conditions for the electron just after ionization. These
conditions are specified within the three-step (simple man) model
of HHG, which makes the following assumptions:
\begin{itemize}
  \item SM1: The electron is born in the continuum at any time within the laser cycle;
  \item SM2: The electron is born near the ionic core (i.e., near the origin of the reference frame) with zero
  velocity;
   \item SM3: If the electron returns to the ionic core (origin), its instantaneous energy at the moment
   of return is converted into the harmonic photon.
\end{itemize}
The pull of the ionic core on the liberated electron is neglected
in the model, which is not unreasonable considering the very large
excursions that the electron makes in the strong driving laser
field. The possibility of the electrons return to the core is
dictated by the phase of the laser field at which it is launched
on its classical orbit, and the time-window for the returning
trajectories -- the range of the 'birth' times $t_B$  -- shown in
Fig.~\ref{fig:3step}.

The calculation is done as follows: for
each $t_B$, we find the time of return $t_R$ to the electron's
original position (Fig.~\ref{fig:3step}, left panel) and the energy at the
moment of return (Fig.~\ref{fig:3step}, right panel).
The assumption that the strong laser field dominates the electron's
motion after ionization simplifies our calculations. Once the
ionic core potential is neglected, the kinetic momentum (velocity)
at the time of birth $t_B$ can be written as
$\mathbf{k}(t_B)=\mathbf{p}+\mathbf{A}(t_B)$, where $\p$ is the canonical
momentum of the electron and $\mathbf{A}(t)$ is the vector potential of the laser
field, which is related to the electric field $\F(t)$ as
$\F(t)=-\partial \mathbf{A}/\partial t$. The condition $\k(t_B)=\mathbf0$ (SM2)
specifies $\p=-\mathbf{A}(t_B)$. Therefore, the electron kinetic momentum at all later times $t$ is
$\k(t)=-\mathbf{A}(t_B)+\mathbf{A}(t)$ and the electron energy at
the time of return is
    $$
    E_{\rm ret}(t_R)=
    \k^2(t_R)/2=(\mathbf{A}(t_B)-\mathbf{A}(t_R))^2/2 \ .
    $$
The zero displacement of the electron from the time of birth, $t_B$, to the time of return, $t_R$,
(SM3) defines the return time
$t_R$:
    \bea
    \label{eq:3step}
    && \int_{t_B}^{t_R}dt~(\A(t)-\A(t_B))=0.
    \eea
According to this model, the maximal return energy is about $3.17~
U_p$, where $U_p=F^2/4\w ^2$ and $F$ is the electric field amplitude (see Fig. 1.1).
Then the maximum energy of the emitted harmonic photon is  $3.17~
U_p+I_p$, where $I_p$ is the binding energy of the ground state to
which the electron recombines, is in excellent agreement with the
empirical cut-off law found numerically by \cite{Kulander92}.

The formal quantum approach considered in the next section will
first take us away from the simple classical model. However, just
like the re-colliding electron revisits the ion, we will revisit
the simple man model several times in this chapter, refining it at
each step.

\section{Formal approach for one-electron systems}

The response of an individual atom or a molecule ${\bf P}(\rmbf,t)= n{\bf D}(t)$
is proportional to the induced dipole ${\bf D}(t)$:
\bea
 \label{eq:D}
 && \D(t)=\langle \Psi(t)|\hat {\bf d}|\Psi(t)\rangle ,
 \eea
where $n$ is the number density, ${\bf \hat d}$ is the dipole operator, and
$\Psi(t)$ is the wavefunction of the system  obtained by solving
the time-dependent Schr\"odinger equation (TDSE) with the
Hamiltonian $\hat H(t)$:
    \bea
 && i\partial\Psi(t)/\partial t=\hat{H}(t)\Psi(t).
      \label{eq:TDSE}
    \eea

We will first focus on the single-active-electron approximation (see
section 1.8 for the multielectron case). This approximation
assumes that only one electron feels the laser field -- the one
that is liberated via strong-field ionization and subsequently re-collides with
the parent ion. All other electrons are frozen in the ion,
unaffected by the laser field. The Hamiltonian of our system in
the single-active-electron approximation is \bea
 \label{eq:H}
 && \hat{H}(t)=\hat{\p}^2/2+U(\hat \rmbf)+\hat{V}_L(t),
 \eea
where $\hat{\p}=-i\nabla_\rmbf$ is the momentum operator, $U(\hat \rmbf)$
describes the interaction of the electron with the ionic core, and
$\hat{V}_L(t)$ describes the interaction between the electron and the laser field.
In the dipole approximation and in the length gauge, $
\hat{V}_L(t)=-\hat {\bf d} \cdot \F(t)=\hat\rmbf \cdot \F(t)$ (see Chapter 8 to
learn about different gauges or read Section 2.2.4 in
the excellent book by \cite{Grynberg} for a more detailed discussion).


Formally, the solution of the Schr\"odinger equation
(\ref{eq:TDSE}) can be written in the integral form (see e.g.
\cite{SmirnovaJMO07} for a simple derivation): 
\bea
&&|\Psi(t)\rangle= -i \int_{t_0}^{t} dt' \;\hat{U}(t,t')\hat V_{L}(t')\hat{U}_0(t',t_0)|g\rangle+\hat{U}_0(t,t_0)|g\rangle,
\label{eq:integral}
\eea
where the ket-vector $|g\rangle$ represents  the
wavefunction of the electron in the ground state
at initial time $t=t_0$,
 $\hat{U}(t,t')$ is the full propagator, while  $\hat{U}_0(t',t_0)$ is the field-free propagator.
The propagators are the operators that describe the time evolution of the
wavefunction. The propagator $\hat{U}_0(t',t_0)$ governs the
electron dynamics from time $t_0$ to time $t'$ without the laser
field, and is determined by the following equations:
  \bea
  \label{eq:u0}
 && i\partial\hat{U}_0(t,t_0)/\partial t=\hat H_0 \hat{U}_0(t,t_0),\\
 \label{eq:h0}
 &&\hat{U}_0(t_0,t_0)=1,\\
 && \hat H_0=\hat{\p}^2/2+U(\hat \rmbf).
  \eea
Symbolically, the solution of Eq.~(\ref{eq:u0}) can be written in the
compact form
    \begin{equation}
        \label{eq:propU0}
    \hat{U}_0(t',t_0)= e^{-i\int_{t_0}^{t'}\hat{H}_0(\xi)\,d\xi},
    \end{equation}
where the integral is \emph{time-ordered}, that is, the
contribution of later times to the evolution follows the
contribution of the earlier times.

The full propagator $\hat{U}(t,t')$ governs the electron dynamics
from time $t'$ to the observation time $t$, driven by the combined
action of the laser field and of the ionic core potential $U(\hat \rmbf)$.
It is given by
 \bea
 && i\partial\hat{U}(t,t')/\partial t=\hat H \hat{U}(t,t'),\\
 &&\hat{U}(t,t')= e^{-i\int_{t'}^t \hat{H}(\xi)\,d\xi},\\
 &&\hat{U}(t',t')=1.
  \eea
The propagation without the laser field is straightforward.
Denoting the  ground state energy $E_g=-I_p$ (ionization
potential) and the stationary ground state wavefunction
$\Psi_g(\rmbf)=\langle \rmbf|g\rangle$, we have:
 \bea
  &&\Psi_g(\rmbf,t')=\emph{U}_0(t',t_0)\Psi_g(\rmbf)=e^{iI_p(t'-t_0)}\Psi_g(\rmbf).
\label{eq:laser_free_prop}
 \eea

The full propagator  $\hat{U}(t,t')$, on the other hand,  is just
as hard to find as the solution of the original equation
(\ref{eq:TDSE}). The advantage of the integral
expression Eq.~(\ref{eq:integral}) is that making meaningful
approximations is technically easier and physically more transparent.

Remembering that the laser field is strong, we can try to neglect
the ionic potential in the full propagator. In this case the
electron is free from time $t'$ to time $t$. Its motion is only
affected by the laser field and is described by the Hamiltonian
$\hat H_V(t)=\hat{\p}^2/2+ \hat V_L(t)$. The corresponding approximation is
called the Strong Field Approximation (SFA), and the propagator
corresponding to $\hat H_V(t)$ is often called the Volkov
propagator. The main advantage of the SFA is that the Volkov
propagator can  be found analytically.
In the length gauge used here,
the result of acting with the Volkov propagator
$\hat U_V(t,t')$ on the plane wave with kinetic momentum
$\k(t')=\p+\mathbf{A}(t')$ is
  \bea
  \label{eq:volkovprop}
  &&\hat U_V(t,t')|\p+\mathbf{A}(t')\rangle=e^{-iS_V(\p,t,t')}|\p+\mathbf{A}(t)\rangle,
  \nonumber \\
  &&\langle\rmbf|\p+\mathbf{A}(t)\rangle=\frac{1}{(2\pi)^{3/2}} e^{i[\p+\mathbf{A}(t)]\cdot \rmbf},
  \nonumber \\
  &&S_{V}(\p,t,t')=\frac{1}{2}\int _{t'}^t d\xi \,[\p+\mathbf{A}(\xi)]^2.
  \eea
That is, the plane wave with the kinetic momentum
$\k(t')=\p+\mathbf{A}(t')$   turns into a plane wave with the
kinetic momentum $\k(t)=\p+\mathbf{A}(t)$ and accumulates the
phase $S_{V}(\p,t,t')$ on the way.

The Eqs.~(\ref{eq:volkovprop}) define the Volkov function
$$\Psi^V_{\p}(\rmbf,t;t')=\frac{1}{(2\pi)^{3/2}} e^{-iS_V(\p,t,t')}e^{i[\p+\mathbf{A}(t)] \cdot \rmbf} \ \ .$$
Formally, the Volkov function is an eigenstate of the
time-periodic Hamiltonian. It provides the quantum-mechanical
description of the behavior of the free electron in the laser
field. The coordinate part of the Volkov function is a plane wave,
and these plane waves form a complete basis at each moment of time:
   \bea
   &&\hat 1=\int d\p \,|\p+\mathbf{A}(t)\rangle\langle \p+\mathbf{A}(t)|.
   \label{eq:basis}
  \eea
Within the SFA, Eq.(\ref{eq:integral}) takes the form
  \bea
 |\Psi(t)\rangle= -i \int_{t_0}^{t} dt' \,\hat{U}_V(t,t')\hat V_{L}(t')\hat{U}_0(t',t_0)|g\rangle+\hat{U}_0(t,t_0)|g\rangle,
      \label{eq:integralV}
\eea
 and can be solved analytically. The first term describes
ionization, the second term describes the evolution of the non-ionized
part of the electron wavefunction.

Thus, it is natural to associate $t'$ with the time when
ionization is initiated: before $t'$ the electron is bound, after
$t'$ the electron is becoming free. Substituting Eq.~(\ref{eq:integralV})
into Eq.~(\ref{eq:D}) yields:
    \bea
    \D(t)\simeq -i \langle \hat{U}_0(t,t_0)g|\hat \d| \int_{t_0}^{t} dt'\,
    \hat{U}_V(t,t')\hat V_{L}(t')\hat{U}_0(t',t_0)|g\rangle +c.c.
    \label{eq:dipV1}
    \eea
Here we have assumed that there is no permanent dipole in the
ground state and that the contribution of the continuum-continuum transitions
to the dipole is negligible. The latter
assumption is fine as long as ionization is weak. Thus, the dipole
in Eq.~(\ref{eq:dipV1}) is evaluated between the bound
and the continuum components of the same wavefunction.

The propagator $\hat{U}_V(t,t')$ is known when it acts on the
Volkov states. Thus, we introduce the identity operator resolved
on the Volkov states, Eq.~(\ref{eq:basis}), into Eq.~(\ref{eq:dipV1}):
  \bea
  \label{eq:dipV2}
    \D(t)&=&-i \langle g|\hat\d| \int_{t_0}^{t}
    dt'\,e^{iI_p(t'-t)}
    \times
    \nonumber \\
    &&\times \int d\p\, \emph{U}_V(t,t')|\p+\mathbf{A}(t')\rangle\langle \p+\mathbf{A}(t')|\hat V_{L}(t')|g\rangle +c.c. \;\; .
    \eea
Finally, remembering that $\hat V_L(t)=-\hat \d \cdot {\bf F} (t)$, we
re-write Eq.~(\ref{eq:dipV2}) in the compact form:
    \bea
    \label{eq:dipV3}
    \hspace*{-1cm}
    && \D(t)= i \int_{t_0}^{t}dt'\int d\p\, \d ^{*}(\p+\A(t))\,e^{-iS(\p,t,t')}\,{\bf F}(t')\,\d(\p+\A(t')) + c.c.,
    \eea
where we have introduced the dipole matrix elements $\d(\p+\A(t))$
of the transitions between the ground state and the plane wave
continuum,
    \bea
    && \d(\p+\A(t))\equiv\langle \p+\A(t)|\hat \d|g\rangle .
    \label{eq:drec}
    \eea
The phase
  \bea
 && S(\p,t,t')\equiv\frac{1}{2}\int_{t'}^{t} [\p+\mathbf{A}(\tau)]^2 d\tau+I_p(t-t')
 \label{eq:Saction}
    \eea
is often referred to as action, and we will use this term below,
even though, strictly speaking, it is only the energy part of the
full classical action.

It is convenient to re-write the equation (\ref{eq:dipV3}) for the harmonic dipole $\D(t)$ by evaluating the integral over $t'$ by parts (see
e.g.  \cite{Gribakin97}, \cite{Wilhelm} and  Appendix A):
 \bea
    \label{eq:grib}
    \hspace*{-1cm}
    && \int_{t_0}^{t}dt' \,e^{-iS(\p,t,t')}\,{\bf F}(t')\,\d(\p+\A(t'))=\\\nonumber &&=\int_{t_0}^{t}dt' \,e^{-iS(\p,t,t')}\Upsilon(\p+\A(t')) ,\\
    &&\Upsilon(\p+\A(t'))=\left[ \frac{(\p+\A(t'))^2}{2}+I_p \right]\langle \p+\A(t')|g\rangle,
     \label{eq:Y}
 \eea
where $\langle \p+\A(t)|g\rangle$ is a Fourier transform of the ground state $|g\rangle$,
$\Upsilon(\p)$ reflects the  dependence of ionization rate on the angular structure of the ground state.
Equation (\ref{eq:dipV3}) takes the following form:
 \bea
    \label{eq:dipV3_new}
    \hspace*{-1cm}
    && \D(t)= i \int_{t_0}^{t}dt'\int d\p\, \d ^{*}(\p+\A(t))\,e^{-iS(\p,t,t')}\Upsilon(\p+\A(t')) + c.c.,
    \eea

The harmonic spectrum $I(N\omega)$ can be obtained from the Fourier
transform of $\D(t)$:
\bea
\label{eq:dipV4}
I(N\omega)\propto (N\w)^4|D(N\omega)|^2,
\nonumber \\
\D(N\omega)=  \int dt\, e^{iN\omega t}\, \D(t) .
\eea

Note that $S(\p,t,t')$ is large and the integrand is a highly
oscillating function, which is an advantage for the analytical
evaluation of this integral. The analytical approach
(\cite{Lewenstein94}) is based on the saddle point method (see
Appendix B),  which is the mathematical tool for evaluating
integrals from fast-oscillating functions. It provides the
physical picture of high harmonic generation as a three step
process involving ionization, propagation and recombination
(\cite{Ivanov96}). It also supplies the time-energy mapping
(\cite{Lein05,Baker06}) crucial for attosecond imaging, and it is
the basis for the extension of the above approach beyond the SFA
and beyond the single-active-electron approximation (see e.g.
\cite{Smirnova09}).

Let us now focus on the analytical saddle point approach to HHG.

\section{The Lewenstein model:
Saddle point equations for HHG}

The goal of this section is to evaluate the integral equations~(\ref{eq:dipV3_new},\ref{eq:dipV4})
using the saddle point method
(see Appendix B). We need to find saddle points for all three
integration variables $t'$, $t$ and $\p$, i.e. points where the
rapidly changing phase of the integrand has zero derivatives with
respect to all integration variables.

There are two ways to deal with the integrals Eqs.~(\ref{eq:dipV3_new},\ref{eq:dipV4}).
First, one can treat them as
multi-dimensional integral, i.e. one finds the saddle points for all the
integration variables 'in parallel', and then one follows the
multi-dimensional saddle point approach to deal with the whole
multi-dimensional integral 'at once'.

One can also take a different route and evaluate the
multiple integrals~(\ref{eq:dipV3_new},\ref{eq:dipV4}) step by
step, sequentially. First, we find the saddle points $t_i$ for the
integral over $t'$ from the saddle point equation:
 \bea &&
\frac{dS}{dt'}\equiv\frac{\partial S(t',\p,t)}{\partial t'}=0,
\label{eq:gion}
 \eea
where the phase $S$ is
given by Eq.~(\ref{eq:Saction}). We then evaluate the integral over
$t'$ treating it as a one-dimensional integral, with $\p$ and $t$
entering as fixed parameters.

Next, we move to the integral over $\p$. Dealing with its saddle
points, we should keep in mind that the saddle points of the
previous integral $t'=t_i\equiv t_i(\p,t)$ depend on $\p$:
$\frac{\partial t_i}{\partial p_\alpha}\neq 0$, $\alpha=x,y,z$.

Fortunately, thanks to Eq.~(\ref{eq:gion}), the explicit dependence
of $t_i(\p,t)$  on $\p$ does not affect the position of the saddle
points for the $\p$-integral:
 \bea
&& \frac{d S(t_i,\p,t)}{d p_\alpha} \equiv
\frac{\partial S(t_i,\p,t)}{\partial p_\alpha}+\frac{\partial S(t_i,\p,t)}{\partial t_i}
\frac{d t_i}{d p_\alpha}
=\frac{\partial S(t_i,\p,t)}{\partial p_\alpha}=0.
\label{eq:gprop}
 \eea
Note that the integral over $\p$ is multi-dimensional,
which leads to a slightly different form of the pre-exponential factor (prefactor) involving Hessian(see Appendix B).

Finally, we deal with the integral over $t$. Here, again,  the
saddle points $\p_s(t_i,t)$ depend on $t$: $ \frac{
\partial p_{s,\alpha}}{\partial t}\neq 0$. But once again the explicit dependence of
$\p_s(t_i,t)$  on $t$ does not affect the position of the saddle
points thanks to Eq.~(\ref{eq:gprop}):
 \bea \hspace*{-.5cm}
&& \frac{d S(t_i,\p_s,t)}{d t}
\equiv \frac{\partial S(t_i,\p_s,t)}{\partial t}+
\frac{\partial S(t_i,\p_s,t)}{\partial p_\alpha}
\frac{ d p_\alpha}{d t}
=\frac{\partial S(t_i,\p_s,t)}{\partial t}=0.
\label{eq:grec} \eea
 The fact that both routes yield the same
saddle point equations is, of course, not surprising -- one should
not get different answers depending on how the integral is
evaluated.

Using Eq.~(\ref{eq:Saction}), we obtain the explicit form of the
Eqs.~(\ref{eq:gion},\ref{eq:gprop},\ref{eq:grec}), which
define the saddle points $t_i$, $\p_s$, $t_r$:
\begin{eqnarray}
&&\frac{[\p_s+\textbf{A}(t_i)]^2}{2}+I_p=0,
 \label{eq:ion}
 \\
&&\int_{t_i}^{t_r}[\p_s+\textbf{A}(t')]\,dt'=0,
 \label{eq:ret}
 \\
&&\frac{[\p_s+\textbf{A}(t_r)]^2}{2}+I_p=N\omega  .
 \label{eq:rec}
\end{eqnarray}
Here, $\mathbf{p}_s$ is the electron drift (canonical) momentum,
$\k_s(t)=\textbf{p}_s+\textbf{A}(t)$ is the kinetic momentum (the
instantaneous electron velocity, up to the electron mass). The
trajectories that satisfy the Eqs.~(\ref{eq:ion},\ref{eq:ret},\ref{eq:rec}) are known
as quantum orbits, see \cite{Salieres01}, \cite{Kopold02}, \cite{Becker02}.

Equation~(\ref{eq:ret}) requires that the electron returns to the
parent ion -- the pre-requisite for recombination. Indeed, the
time integral of the electron velocity yields the electron displacement
from $t_i$ to $t_r$. Thus, Eq.~(\ref{eq:ret}) dictates that the
displacement is equal to zero.

Whereas Eq.~(\ref{eq:rec}) describes energy conservation during
recombination, Eq.~(\ref{eq:ion}) describes tunnelling. It shows
that the electron's kinetic energy at $t_i$ is negative, its
velocity $\k_s (t_i)=\textbf{p}_s+\textbf{A}(t_i)$ is complex, and
hence $t_i=t'_i+t''_i$ is also complex -- the hallmarks of the
tunnelling process.

\begin{figure}
\includegraphics[width=1\textwidth]{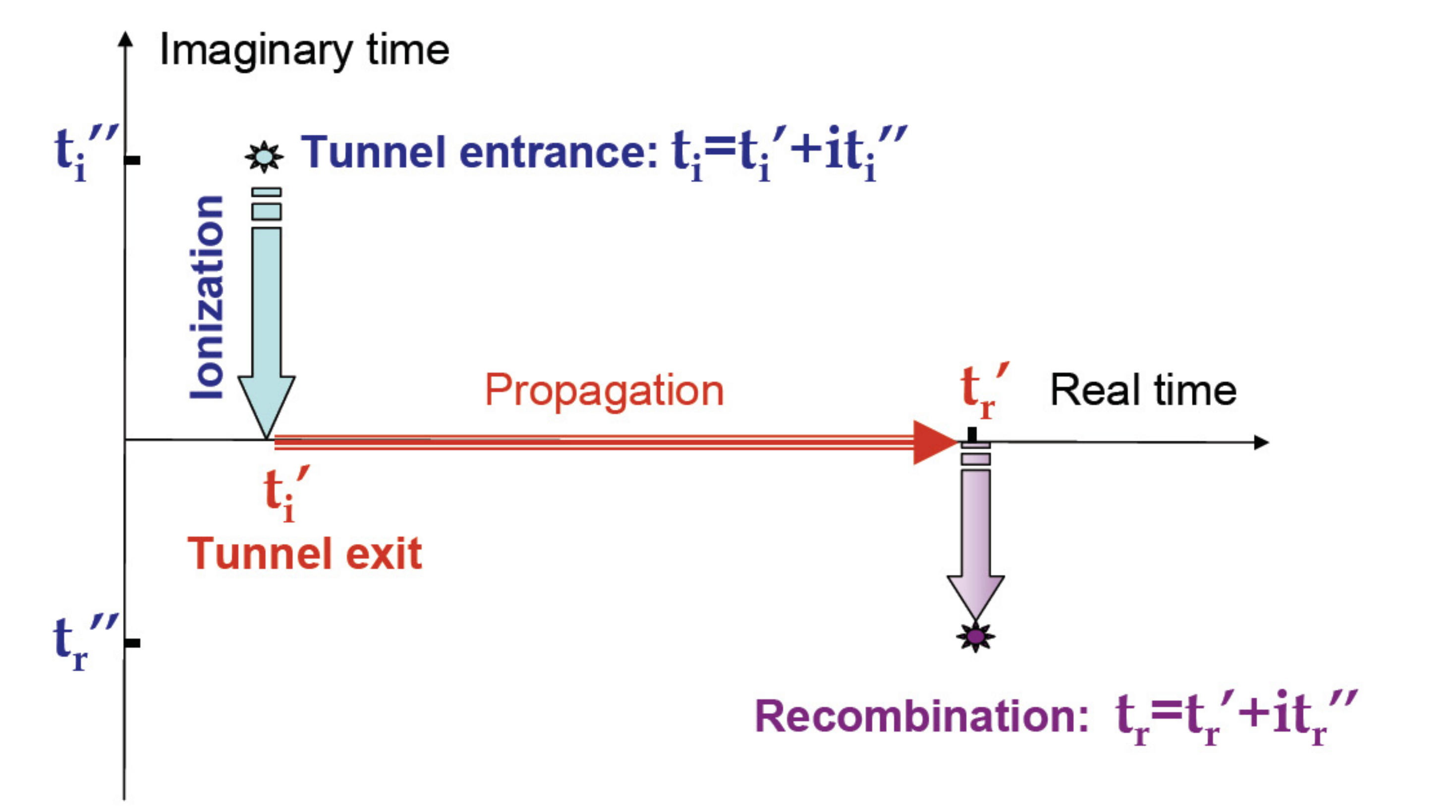}
\caption{Contour of the time integration in the action S. Ionization
occurs from the complex time $t_i$ to the real time $t'_i$. }
\label{fig:action}
\end{figure}

The time $t_i$ can be identified with the moment when the electron
\emph{enters the barrier}, see Fig.~\ref{fig:action}.
Its \emph{real part} will then
correspond to the time when the electron \emph{exits the barrier}.
The origin of this concept will be explained in the next section.

The electron displacement during this 'under-the-barrier' motion from
$t_i$ to ${\rm Re}(t_i)$ is, in general, complex.  Whether we like
it or not, it yields a complex coordinate of 'exit'
$\rmbf_{ex}=\rmbf'_{ex}+i\rmbf''_{ex}$ at ${\rm Re}(t_i)\equiv t'_i$ (see e.g. \cite{Lisa12}):
 \begin{eqnarray}
&&\int_{t_i}^{t'_i}[\textbf{p}+\textbf{A}(t')]dt'=\rmbf'_{ex}+i\rmbf''_{ex}.
 \label{eq:exit}
\end{eqnarray}
As a result, Eqs.~(\ref{eq:ret}) and (\ref{eq:rec}) cannot
be satisfied unless $\p$ or $t_r$ are complex. Indeed, $t_r$ must
be complex to compensate for the imaginary displacement accumulated
under the barrier. However, the energy conservation condition in Eq.~(\ref{eq:rec})
dictates that $\k_s
(t_r)=\textbf{p}_s+\textbf{A}(t_r)$ is real at the moment of
recombination. Therefore $\p_s$ must also be complex to
compensate for the imaginary part of $\textbf{A}(t_r)$.

Thus, we are forced to conclude that, in contrast to the classical
trajectories of the simple man model, the quantum orbits are the
trajectories with complex canonical momenta, complex velocities,
and complex displacements. These trajectories evolve in complex
time. The only quantity that is required to be real is the one we
measure -- the energy of the emitted photon, see Eq.~(\ref{eq:rec}).
Later in this chapter, we will see when and how one can replace
these trajectories with a different set of trajectories that do not involve
complex canonical momenta and therefore better correspond to the classical
picture. But for the moment, let us deal with the problem at hand.

For a linearly polarized field, it is convenient to rewrite the Eqs.~(\ref{eq:ion},\ref{eq:ret},\ref{eq:rec})
 in terms of electron momenta parallel, $p_{s,\parallel}$, and perpendicular, $p_{s,\bot}$,
to the polarization vector of the laser field:
\begin{eqnarray}
&&\frac{[p_{s,\parallel}+A(t_i)]^2}{2}+I_{p,\rm eff}=0,
 \label{eq:ion1} \\
&&\int_{t_i}^{t_r}[p_{s,\parallel}+A(t')]\,dt'=0, \hspace{0.5cm}
\int_{t_i}^{t_r}p_{s,\bot}\, dt'=0,
 \label{eq:ret1}\\
&&\frac{[p_{s,\parallel}+A(t_r)]^2}{2}+I_{p,\rm eff}=N\omega,
 \label{eq:rec1}
\end{eqnarray}
where we have introduced an "effective" ionization potential:
$I_{p,{\rm eff}}=I_p+p^2_{s,\bot}/2$. Equations~(\ref{eq:ret1}) dictate that
the stationary perpendicular canonical momentum is equal to zero
for the linearly polarized field, $p_{s,\bot}=0$ and hence $I_{p,{\rm eff}}=I_p$ . Then,
Eqs.~(\ref{eq:ion1},\ref{eq:ret1},\ref{eq:rec1}) reduce to:
\begin{eqnarray}
&&\frac{[p_{s,\parallel}+A(t_i)]^2}{2}+I_p=0,
 \label{eq:ion0}
 \\
&&\int_{t_i}^{t_r}[p_{s,\parallel}+A(t')]\,dt'=0,
 \label{eq:ret0}
 \\
&&\frac{[p_{s,\parallel}+A(t_r)]^2}{2}+I_p=N\omega.
 \label{eq:rec0}
\end{eqnarray}
Separating the real and the imaginary parts in
Eqs.~(\ref{eq:ion0}, \ref{eq:ret0}, \ref{eq:rec0}), we obtain six
equations for six unknowns: $t_i=t'_i+i t''_i$, $t_r=t'_r+it''_r$,
$p_{s,\parallel}=p'+i p''$. Our goal is to solve these equations
for each harmonic order $N$. Here is one way to do it, which we
find simple and visually appealing.

First, we use the Eqs.~(\ref{eq:ion0},\ref{eq:rec0}) to express all
variables via the real, $t'_r$, and the imaginary, $t''_r$, return times.
This can be done analytically. Second, we substitute the result
into the real part and the imaginary part of Eq.~(\ref{eq:ret0}):
 \bea
 \label{eq:F1}
 &&F_1(N,t'_r,t''_r)={\rm Re}\left[\int_{t_i}^{t_r}[p_{s,\parallel}+A(t')]dt'\right]=0,\\
 \label{eq:F2}
 &&F_2(N,t'_r,t''_r)={\rm Im}\left[\int_{t_i}^{t_r}[p_{s,\parallel}+A(t')]dt'\right]=0.
\eea
Third, we solve the Eqs.~(\ref{eq:F1},\ref{eq:F2}) to find
the only two remaining unknowns: the real, $t'_r$, and the imaginary, $t''_r$,
return times. While the Eqs.~(\ref{eq:F1},\ref{eq:F2}) cannot be
solved analytically, dealing with two equations is much easier
than dealing with the original six.

Solving the Eqs.~(\ref{eq:F1},\ref{eq:F2}) means that we need to find the
minima of the two-dimensional surface $F(N,t'_r,t''_r)$, defined in
the plane of the real, $t'_r$, and the imaginary, $t''_r$, return times: \bea
 \label{eq:surface}
&&F(N,t'_r,t''_r)\equiv
[F_1(N,t'_r,t''_r)]^2+[F_2(N,t'_r,t''_r)]^2=0.
 \eea
These minima can be easily found numerically using the gradient
method. The advantage of using $F(N,t'_r,t''_r)$ is the ability to
visualize the solutions: by simply plotting the surface given by
Eq.~(\ref{eq:surface}), see  Fig.~\ref{fig:surf}, one can examine the positions of the minima
versus the harmonic number $N$.

If we restrict our analysis to those solutions that lie within the
same cycle of the laser field as the moment of ionization, ${\rm
Re}(t_i)$, we will find two stationary solutions for each harmonic
number $N$. These solutions are discussed in detail in the next
section. They correspond to two families of quantum orbits,
called the 'short' and the 'long' trajectories. The trajectories merge for the
largest possible return energies, i.e. near the cut-off  of
the harmonic spectrum.

There are also solutions that lie outside the laser cycle during
which the electron was 'born' into the continuum. These
'super-long' trajectories describe second, third, and higher-order
returns of the electron to the origin. In typical experimental
conditions, their contribution to the high harmonic emission is
negligible thanks to the macroscopic effects -- very long
trajectories do not phase match well (see e.g. \cite{Salieres01}). Only very recently, the
beautiful experiments of \cite{Zair08} have been able to clearly
resolve the contribution of these trajectories, and even identify their
interference with the contribution from the long and the short
trajectories.

The stationary phase method for the integral over the return time
$t$ breaks down when these two stationary points merge and the
second derivative of the action with respect to the return time is
equal to zero, $\partial^2 S/\partial t^2=0$. At this point, one
needs to replace the standard saddle point method with the
regularization procedure, discussed in Appendix C.

Outside the cut-off region, and up to a global phase factor, the saddle
point method yields the following expression for the harmonic
dipole (\ref{eq:dipV3_new},\ref{eq:dipV4}):
 \bea
 \label{eq:anal_dip0}
  \D(N\omega)&=&\sum_{j=1}^{4M}
  \left[\frac{2\pi}{iS''_{t_i,t_i}}\right]^{\frac{1}{2}}
  \left[\frac{2\pi}{iS''_{t_r,t_r}}\right]^{\frac{1}{2}}
  \frac{(2\pi)^{3/2}}{\sqrt{{\rm det}(iS''_{\p_s,\p_s})}}
  \times
  \nonumber \\
&&\hspace*{-1cm}\times \d^*(\p_s+\A(t^{(j)}_r))\,e^{-iS(\p_s, t^{(j)}_r, t^{(j)}_i)}\,\Upsilon(\p_s+\A(t_i))\,e^{iN\omega t^{(j)}_r},
\eea
where the Hessian ${\rm det}(iS''_{\p_s,\p_s})$ appears due to the
 multi-dimensional nature of the integral over $\p$. The sum runs over all stationary
points $j$ for $M$ periods of the laser light, and the corresponding ionization and
recombination times are labelled with the superscript $j$. Since there are two
trajectories for each half-cycle of the laser field, i.e. for each
ionization 'burst', and since there are 2M ionization bursts for
$M$ laser cycles, the number of stationary points is $4M$.

The length gauge SFA presents a good approximation for short range potentials \cite{Frolov_prl}.
However, it misses polarization, the Stark shift and the depletion of the bound state,
 which can be introduced into Eq. ( \ref{eq:anal_dip0}) if necessary.

Note that the expression (\ref{eq:anal_dip0}) \textbf{can not} be directly ported to long range potentials.
Indeed, in the long range Coulomb  potential the ground state has different radial structure (compare Eqs.(\ref{eq:asymp_wf}) and (\ref{eq:asymp_wf_l}) below)
and $\Upsilon(\p_s+\A(t_i))$
is singular exactly at the saddle point
$[\p_s+\A(t_i)]^2/2+I_p=0$. As a consequence, the saddle point calculations should be
modified, (see e.g. \cite{Keldysh64,Gribakin97}) to accommodate for the presence of such singularity.
Once the singularity is treated correctly
 (\cite{Keldysh64,Gribakin97}), the result is still incomplete and unsatisfactory,
  because the long range potential also affects the structure of the continuum states, which can not
  be accurately represented by the Volkov states.
Consistent treatment of long-range effects including  modifications of  both bound and continuum
states can be found in \cite{Lisa12,jivesh13}. A practical recipe for incorporating the effects of the Coulomb potential
is discussed in the next section and in the Outlook section.


\section{Analysis of the  complex trajectories}

Let us now show how the above method of finding the saddle points
 works for a linearly polarized laser field
$F=F_0\cos(\omega t)$, which corresponds to the vector potential
$A=-A_0\sin(\omega t)$. We shall introduce the dimensionless variables
$p_1={\rm Re}(p_{s,\parallel})/A_0$, $p_2={\rm
Im}(p_{s,\parallel})/A_0$, $\phi_i=\omega t_i=\phi'_i+i\phi''_i$,
$\phi_r=\omega t_r=\phi'_r+i\phi''_r$,
  $\gamma^2=I_p/(2U_p)$, $\gamma_N^2=(N\omega-I_p)/(2Up)$.

In terms of these variables, Eqs. (\ref{eq:F1},\ref{eq:F2}) for
the linearly polarized field yield:
 \bea
 \label{eq:F11prim}
\nonumber
F_1&=&p_1(\phi'_r-\phi'_i)-p_2(\phi''_r-\phi''_i)-\cos(\phi'_i)\cosh(\phi''_i)\\
&& +\cosh(\phi''_r)\cos(\phi'_r)=0,
\\
 \label{eq:F21prim}
\nonumber
F_2&=&p_1(\phi''_r-\phi''_i)+p_2(\phi'_r-\phi'_i)+\sin(\phi'_i)\sinh(\phi''_i)\\
&& -\sinh(\phi''_r)\sin(\phi'_r)=0.
 \eea
The real and the imaginary parts of Eq.~(\ref{eq:rec0}) allow us to
express the real, $p_1$, and the imaginary, $p_2$, components of the canonical
momentum via the real and the imaginary parts of the return time (for above threshold harmonics):
 \bea
 \label{eq:p1}
&&p_1=\cosh(\phi''_r)\sin(\phi'_r)+\gamma_N,\\
 \label{eq:p2}
 &&p_2=\sinh(\phi''_r)\cos(\phi'_r).
\eea
The real and the imaginary parts of Eq.~(\ref{eq:ion0}),
 \bea
 \label{eq:eion_r}
&&p_1=\cosh(\phi''_i)\sin(\phi'_i),\\
 \label{eq:eion_i}
&&p_2+\gamma=\sinh(\phi''_i)\cos(\phi'_i),
\eea
 allow us to express the real, $\phi'_i$, and the imaginary, $\phi''_i$, ionization times via $p_1$ and $p_2$:
 \bea
 \label{eq:ion_r}
&& \phi'_i=\arcsin(\sqrt{(P-D)/2}),\\
 \label{eq:ion_i}
 &&\phi''_i=\mathrm{arcosh}(\sqrt{(P+D)/2}),
\eea
where
 \bea
 &&P=p_1^2+\tilde{\gamma}^2+1,
 \nonumber \\
 &&D=\sqrt{P^2-4p_1^2},
 \nonumber \\
 &&\tilde{\gamma}=\gamma+p_2.
    \label{eq:PandD}
 \eea

\begin{figure}
\includegraphics[width=0.498\textwidth]{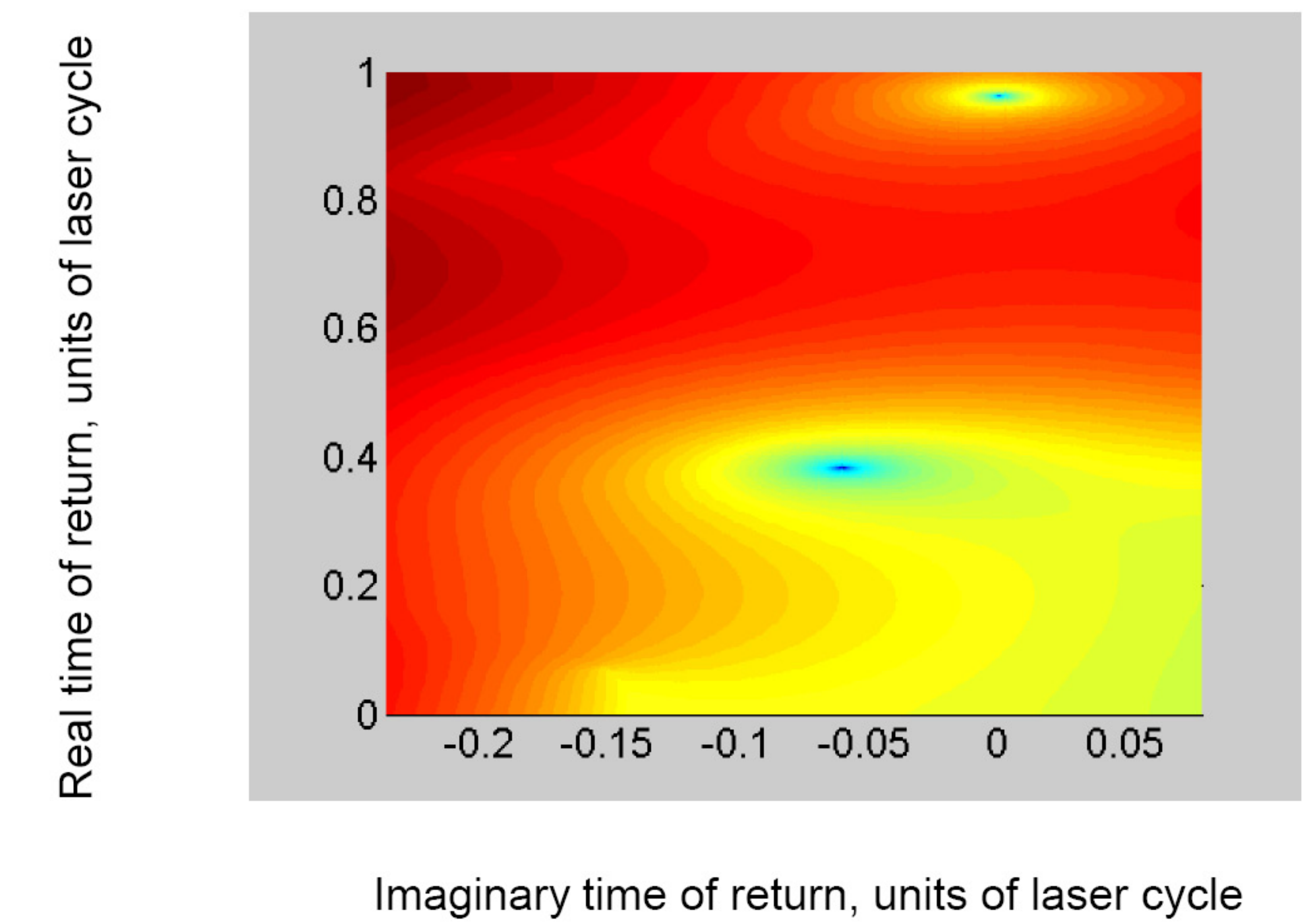}
\includegraphics[width=0.498\textwidth]{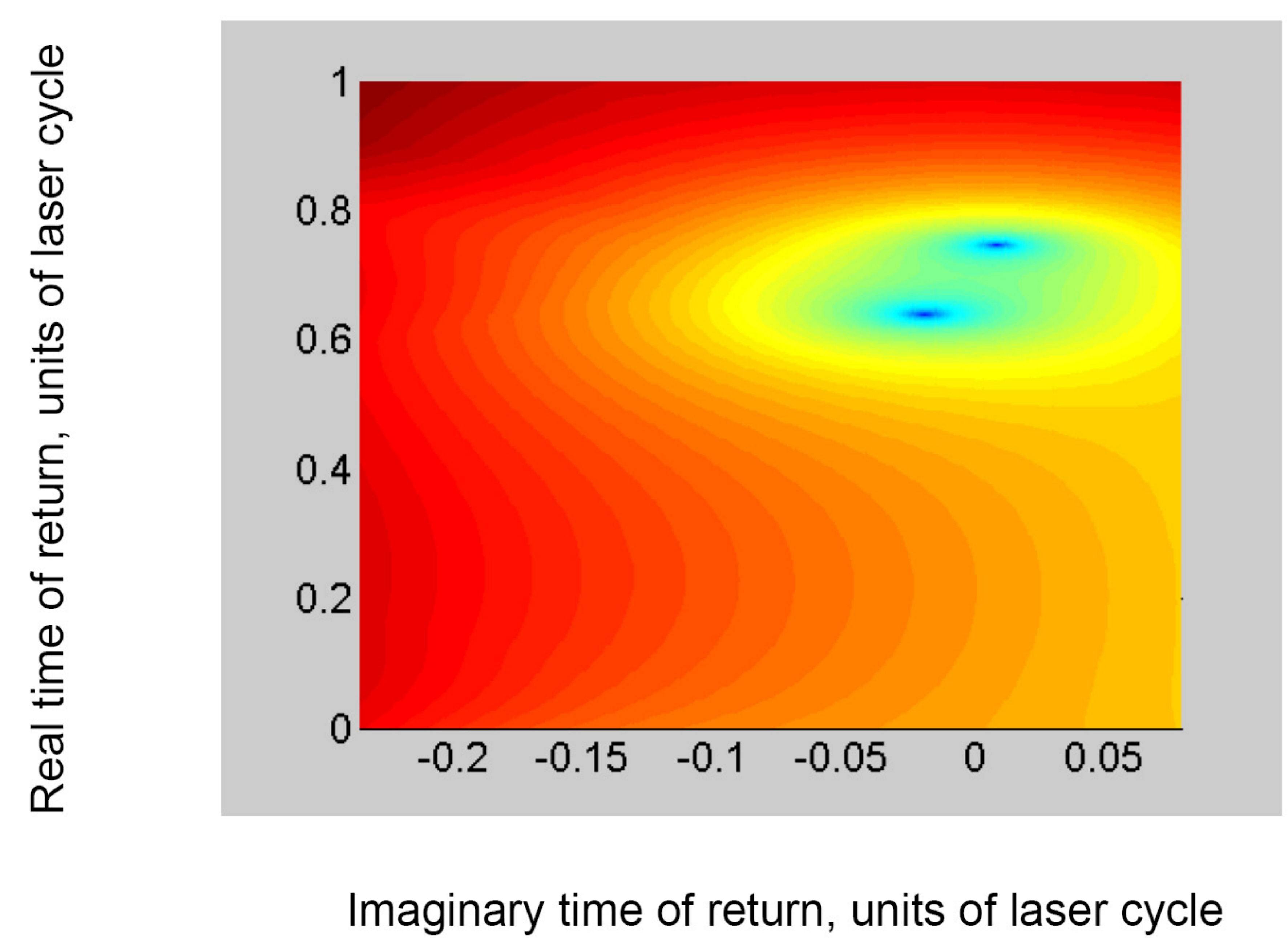}
\caption{Surface in Eq.~(\ref{eq:surface}) for $I_p=15.6$~eV, $I=1.3\cdot 10^{14}$ W/cm$^2$, $\hbar\omega=1.5$~eV.
Left panel: $N=11$; two minima corresponding to short ($\phi'_r\sim 0.4$) and to long ($\phi'_r\sim1$) trajectories, respectively.
Right panel: $N=27$; the two minima corresponding to short and to long trajectories are merging together.} \label{fig:surf}
\end{figure}

 Now we can use our recipe:
\begin{itemize}
    \item Pick a grid of values $\phi'_r,~\phi''_r$ in the complex
    plane of the return times $\phi_r$;
    \item Pick a point $\phi'_r,~\phi''_r$ and calculate
    $p_1$, $p_2$ using Eqs.~(\ref{eq:p1},\ref{eq:p2}) and $\phi'_i$, $\phi''_i$ using
    Eqs.~(\ref{eq:ion_r},\ref{eq:ion_i});
    \item Substitute $p_1$, $p_2$, $\phi'_i$, $\phi''_i$ into Eqs.
    (\ref{eq:F11prim},\ref{eq:F21prim});
    \item Plot the function $F \equiv
    F^2_1+F^2_2$ in the plane of the real and the imaginary return times;
    \item Look for the minima, see Fig.~\ref{fig:surf}.
\end{itemize}

Instead of reading out the solutions from the graph, one can find
the minima using the gradient method. An alternative algorithm using the same ideas is described in Appendix D.

\begin{figure}
\includegraphics[width=0.498\textwidth]{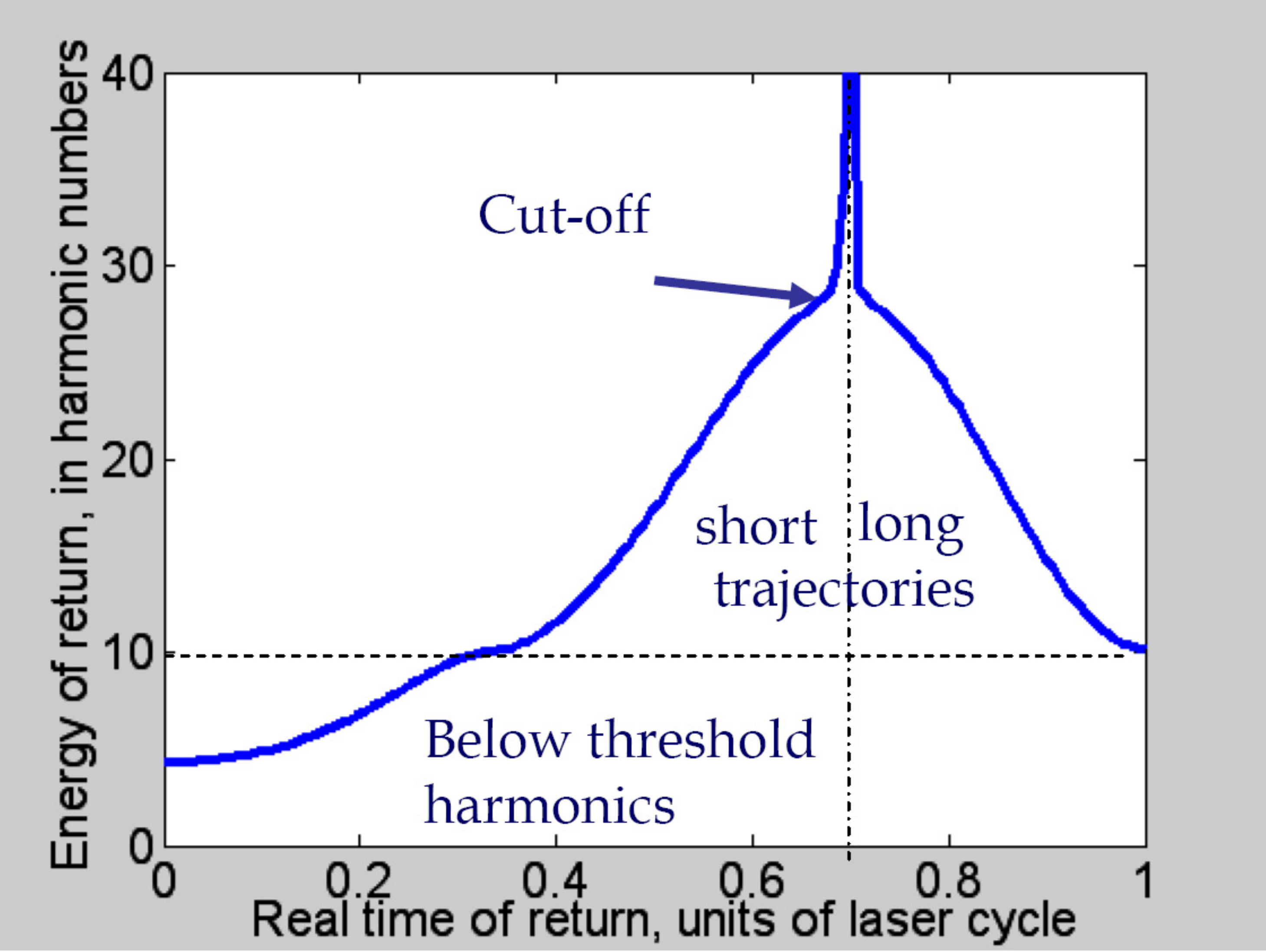}
\includegraphics[width=0.498\textwidth]{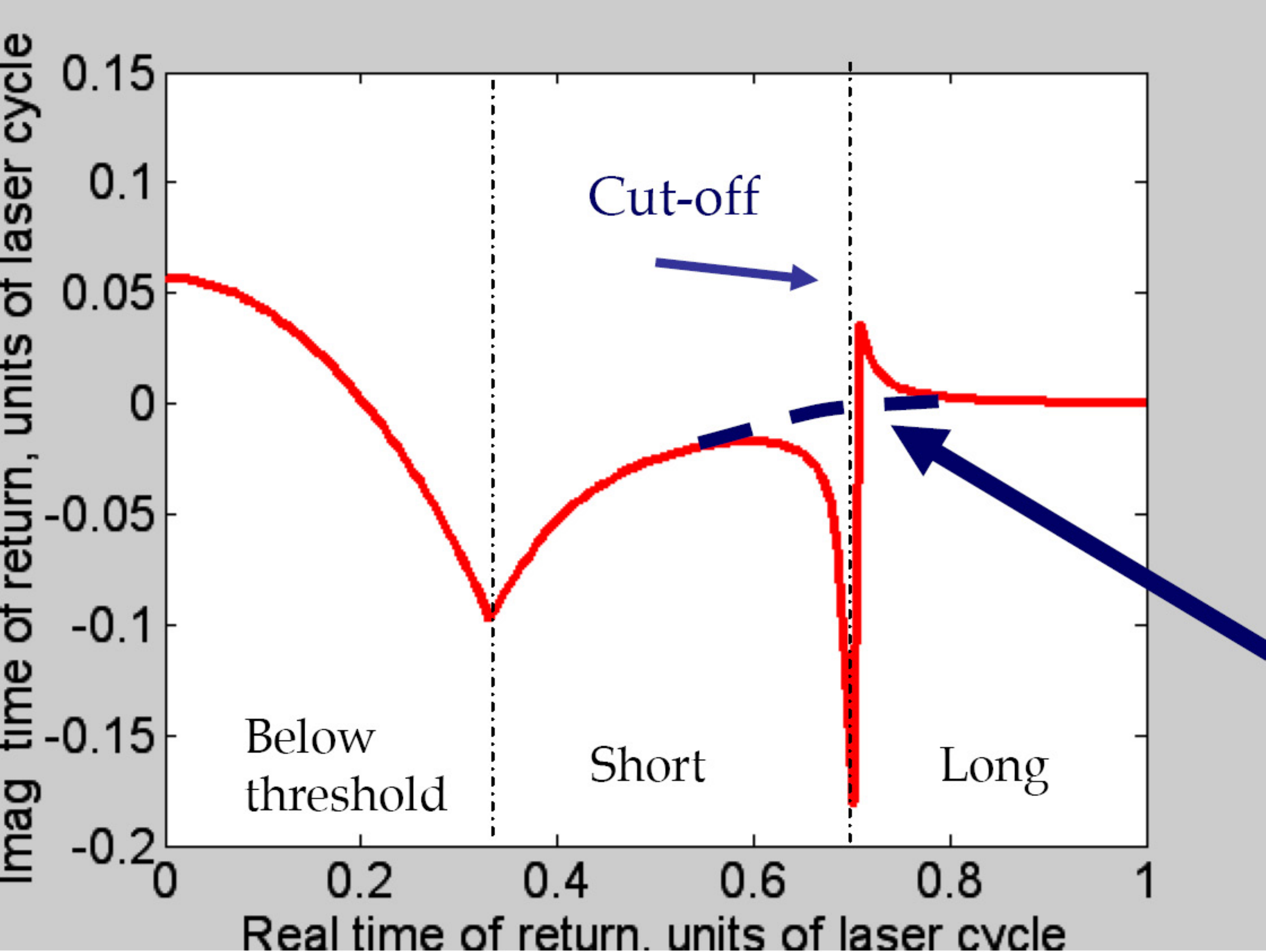}
\caption{Left panel: Emission energy, $E(t_r)+I_p=N\omega$, vs real time of return for $I_p=15.6$~eV, $I=1.3\cdot10^{14}$~W/cm$^2$.
 Right panel: Imaginary time of return vs real time of return. The solution diverges in the cut-off region.
 The thick blue line schematically shows the desired outcome of the regularization procedure.} \label{fig:rec_en}
\end{figure}

The imaginary and the real return times (Fig.~\ref{fig:rec_en}, right
panel) define the integration contour in the complex plane: only
along this contour the energy of return and therefore the energy
of the emitted photon are real.
This energy is shown in
Fig.~\ref{fig:rec_en} (left panel) vs the real component of the
return time for typical experimental conditions.

The cut-off (maximal energy) corresponds to about
$3.17~U_p+1.32~I_p$, see \cite{Lewenstein94}. There are two different
trajectories returning at different times that lead to the same
re-collision energy. Those returning earlier correspond to shorter
excursion and are called 'short trajectories',  those returning
later are called 'long trajectories' as they correspond to larger
excursions and longer travel times.

Fortunately for attosecond imaging, the contributions of the long
and the short trajectories to the harmonic emission separate in the
macroscopic response: the harmonic light diverges differently for
those trajectories, and thus the signals coming from short and long
trajectories can be collected separately. As a result, each
harmonic $N$ can be associated with a particular time delay between
ionization and recombination, $t'_r-t'_i$, and therefore each
harmonic takes a snapshot of the recombining system at a
particular moment of time. This time-energy mapping, see
(\cite{Lein05}, \cite{Baker06}, \cite{Shafir12}), is the basis for attosecond time resolution
in high harmonic spectroscopy.

As mentioned in the previous section, the stationary phase (saddle
point) method for the integral over return times $t$ breaks down
near the cut-off, where the two stationary points (short and long
trajectories) begin to coalesce and the second derivative of the
phase $S$ with respect to the return time is equal to zero,
$\partial^2 S/\partial t^2=0$. The regularization of the solutions
in the cut-off region is discussed in Appendix C. Here, we shall
proceed with the analysis of the stationary phase equations and
turn to the ionization times.

The concept of ionization time together with the semiclassical
(trajectory) perspective on ionization has been first introduced
by V. Popov and co-workers (see \cite{PPT1,PPT2,PPT3,PPT4}). Just
like in the Lewenstein model described above, the concept of
trajectories arises from the application of the saddle point method to
the integral describing ionization
 \footnote{ Eq. (\ref{eq:aion_int}) corresponds to the length-gauge SFA result for ionization.
Eq. (\ref{eq:aion_int}) also results from the PPT approach under the approximation, substituting the
 laser-dressed bound wave-function by the field-free ground state. Thus, the PPT approach
  allows one to identify the approximations in Eq. (\ref{eq:aion_int}) 
  for short-range potentials. The  SFA is inaccurate, because the Vokov states are not sufficiently accurate even for short-range potentials.}
  
 \bea
\label{eq:aion_int} a_{ion}(\p,t)= -i  \int_{0}^{T} dt'\,
e^{-iS(\p,T,t')}\,\Upsilon(\p+\A(t')). \eea
Here the upper limit of the integral in the action $S(\p,T,t')$ Eq.(\ref{eq:Saction})
is the real time $T$, at which the liberated photoelectron is
observed (detected), function $\Upsilon(\p+\A(t'))$ (see \ref{eq:Y}) contains the Fourier transform of the bound state wave-function.
The saddle point method applied to Eq.(\ref{eq:aion_int}) yields the
ionization amplitude:
 \bea
 \label{eq:aion_saddle}
  &&a_{ion}(\p,t)=\left[\frac{2\pi}{iS''_{t_i,t_i}}\right]^{1/2}
  e^{-iS(\p,T,t_i)}\,\Upsilon(\p+\A(t_i)) ,
\eea
 where $t_i$ is the complex saddle point given by the condition
\begin{eqnarray}
&&\frac{[\p+\A(t_i)]^2}{2}+I_p=0.
 \label{eq:ionPPT}
\end{eqnarray}
The saddle point method selects specific moments of time $t_i$ when ionization occurs.
At these times the instantaneous electron energy $\frac{[\p+\A(t_i)]^2}{2}$ is equal
 to the energy of the ground state and therefore the instantaneous momentum of the electron
 hits the pole of the bound state wave-function in the momentum space.
In the vicinity of the pole, the wave-function in the momentum space is determined
by the asymptotic part of the wave-function in the coordinate space.
Thus, in contrast to one-photon ionization, which probes the bound wave function near the core,
 the strong field ionization probes the asymptotic part of the bound wave function:
 \begin{eqnarray}
 \label{eq:asymp_wf}
&&\langle\rmbf|g\rangle\simeq C_{\kappa l}\kappa^{3/2}\,
	\frac{e^{-\kappa r}}{\kappa r}Y_{lm}(\frac{\rmbf}{r}).
\end{eqnarray}
Equation (\ref{eq:asymp_wf}) restricts our analysis to short range potentials
(see \cite{Lisa12} for consistent analytical  treatment of strong-field ionization from
 a long range (Coulomb) potential), $\kappa=\sqrt{2I_p}$, $C_{\kappa l}$ is a constant,
$Y_{lm}(\frac{\rmbf}{r})$ reflects the angular structure of the bound state.
For Coulomb potential $-Q/r$, the asymptotic expression (\ref{eq:asymp_wf})
 must be multiplied by $(\kappa r)^{Q/\kappa}$:
 \begin{eqnarray}
 \label{eq:asymp_wf_l}
&&\langle\rmbf|g\rangle\simeq C_{\kappa l}\kappa^{3/2}\,
	\frac{e^{-\kappa r}}{\kappa r}(\kappa r)^{Q/\kappa}Y_{lm}(\frac{\rmbf}{r}).
\end{eqnarray}

Evaluating the Fourier transform of (\ref{eq:asymp_wf}) we obtain explicit
expression for $\Upsilon(\p)$ (see \cite{PPT1}):
  \begin{eqnarray}
 \label{eq:asymp_Y}
&&\Upsilon(\p)=\left(\frac{2\kappa}{\pi}\right) ^{1/2}C_{\kappa l}\,
	Y_{lm}(\frac{\p}{p}).
\end{eqnarray}
Evaluation of the spherical function $Y_{lm}(\frac{\p}{p})$ at the pole $p=\pm i \kappa$ yields
(see  \cite{PPT1} and also \cite{Barth11,Barth13} for circularly polarized fields) :
\begin{eqnarray}
 \label{eq:pole_Ylm}
&&Y_{lm}(\frac{\p}{p})|_{p=\pm i \kappa}=C_{lm}\left( \frac{\pm p_{\perp}}{\kappa}\right)^m \,e^{im\phi_p},\\
&& C_{lm}=\frac{1}{2^{|m|}|m|\!}\sqrt{\frac{(2l+1)(l+|m|)!}{4\pi(l-|m|)!}},
\end{eqnarray}
where $p_{\perp}$  is the electron transverse momentum at the detector and
 $\phi_p$ is the azimuthal angle of the electron momentum at the detector.
Taking into account that $S''_{t_i,t_i}=(\p+\A(t_i))\A'(t_i)$ can be written as
\begin{eqnarray}
 \label{eq:Stiti}
&&iS''_{t_i,t_i}=\mu\kappa F(t_i),\\\label{eq:mu}
&&\mu=-{\rm sign}(F(t_i)),
\end{eqnarray}
we obtain the final expression for the amplitude of single ionization burst at
complex time $t_i(\p)$, specified by the final momentum of the electron $\p$:
 \bea
 \label{eq:aion_Shsub}
  &&a_{ion}(\p)= 2 C\sqrt{\frac{1}{\mu F(t_i)}}
\left( \mu\frac{p_{\perp}}{\kappa}\right)^m
  e^{-iS(\p,T,t_i(\p))}\,e^{im\phi_p},\\
  \label{eq:C}
  && C=C_{lm}C_{\kappa l}.
\eea
Note that both the real and the imaginary component of $t_i(\p)$  depend on laser parameters. 
Therefore, the dependence of the pre-exponential factor (prefactor) in $a_{ion}(\p)$ on the laser 
field is not simply ${F}^{-1/2}$ as illustrated below for the ionization at the maximum of the laser cycle.
Indeed, the sub-cycle dynamics in the prefactor is much slower than in the exponent, thus
one can also use a simpler expression for $S''_{t_i,t_i}$ corresponding to its value at the maximum of the laser field (\cite{PPT1}):
 \begin{eqnarray}
 \label{eq:Stiti}
&&iS''_{t_i,t_i}=-\frac{\kappa^2}{\omega}\frac{\sqrt{1+\gamma^2}}{\gamma}.
\end{eqnarray}
Omitting the sub-cycle dynamics in the prefactor we obtain a simpler expression for the amplitude of single ionization burst at
 the time $t_i(\p)$, consistent with the one derived by \cite{PPT1}:
 \bea
 \label{eq:aion_Shs}
  &&a_{ion}(\p)= 2C\left[\frac{-\gamma \omega}{\kappa\sqrt{1+\gamma^2}}\right]^{1/2}
\left( \frac{\mu p_{\perp}}{\kappa}\right)^m
  e^{-iS(\p,T,t_i(\p))+im\phi_p}.
\eea
At the same level of approximation, i.e. neglecting the sub-cycle dynamics in the prefactor, the effects
 of the Coulomb potential are incorporated by simply adding the factor $\left( \frac{2\kappa^3}{F}\right)^{Q/\kappa}$:
 \bea
 \label{eq:aion_Cs}
  &&\hspace{-0.5cm}a_{ion}(\p)=\! 2C\left(\frac{2\kappa^3}{F}\right)^{\frac{Q}{\kappa}}\!\!\left[\frac{-\gamma \omega}{\kappa\sqrt{1+\gamma^2}}\right]^{\frac{1}{2}}
\left( \frac{\mu p_{\perp}}{\kappa}\right)^m
  e^{-iS(\p,T,t_i(\p))+im\phi_p}.
\eea
The sub-cycle Coulomb effects are derived in \cite{Lisa12}.
Note that in the rigorous analysis within the analytical R-matrix (ARM) approach, which consistently treats the Coulomb effects both in bound and continuum
states (\cite{Lisa12,jivesh13}), the pole in $\Upsilon(\p+\A(t_i))$ does not appear, because the radial integration is removed due to the use of the Bloch operator (\cite{Bloch}).
Therefore, it removes all technical aspects and additional terms associated with the presence and the strength of the pole.

Note that the expression for the
induced dipole, Eq.~(\ref{eq:anal_dip0}), contains terms that look
very much like the ionization amplitude  Eq.~(\ref{eq:aion_saddle}). This observation is
important, as it suggests the connection of the harmonic response
to ionization, as in the simple man model. However, the story is
more subtle: the stationary momenta $\p_s$ in the harmonic dipole
are complex-valued, while here they are real observable
quantities.


The integral Eq.(\ref{eq:aion_int}) has been extensively studied by
Keldysh, Popov, Perelomov, Terent'ev, and many others. The
semiclassical picture in \cite{PPT1,PPT2,PPT3,PPT4},
 enabled by the application of saddle point method, shows that
strong-field ionization can be understood as tunnelling through
the oscillating barrier created by the laser field.
The tunneling picture clarifies the sensitivity of strong field ionization
to the asymptotic 'tail' of the bound wave-function (see Eq.(\ref{eq:asymp_wf},\ref{eq:asymp_wf_l})),
 since it is this asymptotic part that 'leaks' through the barrier.
The modulus of the ionization amplitude is associated with the imaginary part
of the action $S$ in Eq.~(\ref{eq:aion_saddle}, \ref{eq:aion_Shsub}, \ref{eq:aion_Shs}, \ref{eq:aion_Cs}). This imaginary part is
only accumulated from $t_i$ to $t'_i$, since in the
photoionization problem the canonical momentum registered at the
detector is real and the integration over time also proceeds along
the real time axis between $t'_i$ and the observation time $t$.

\begin{figure}
\includegraphics[width=0.5\textwidth]{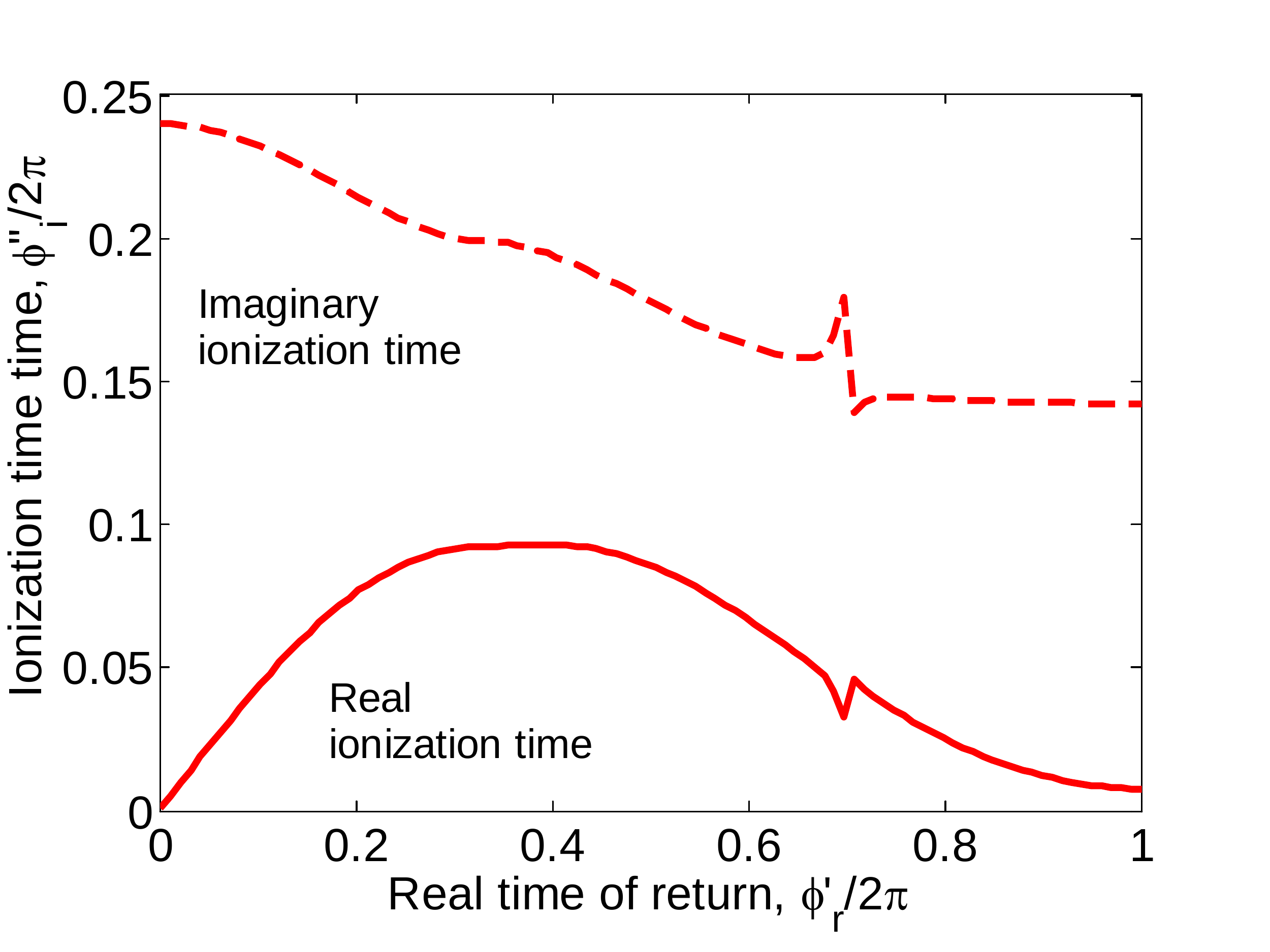}
\includegraphics[width=0.35\textwidth]{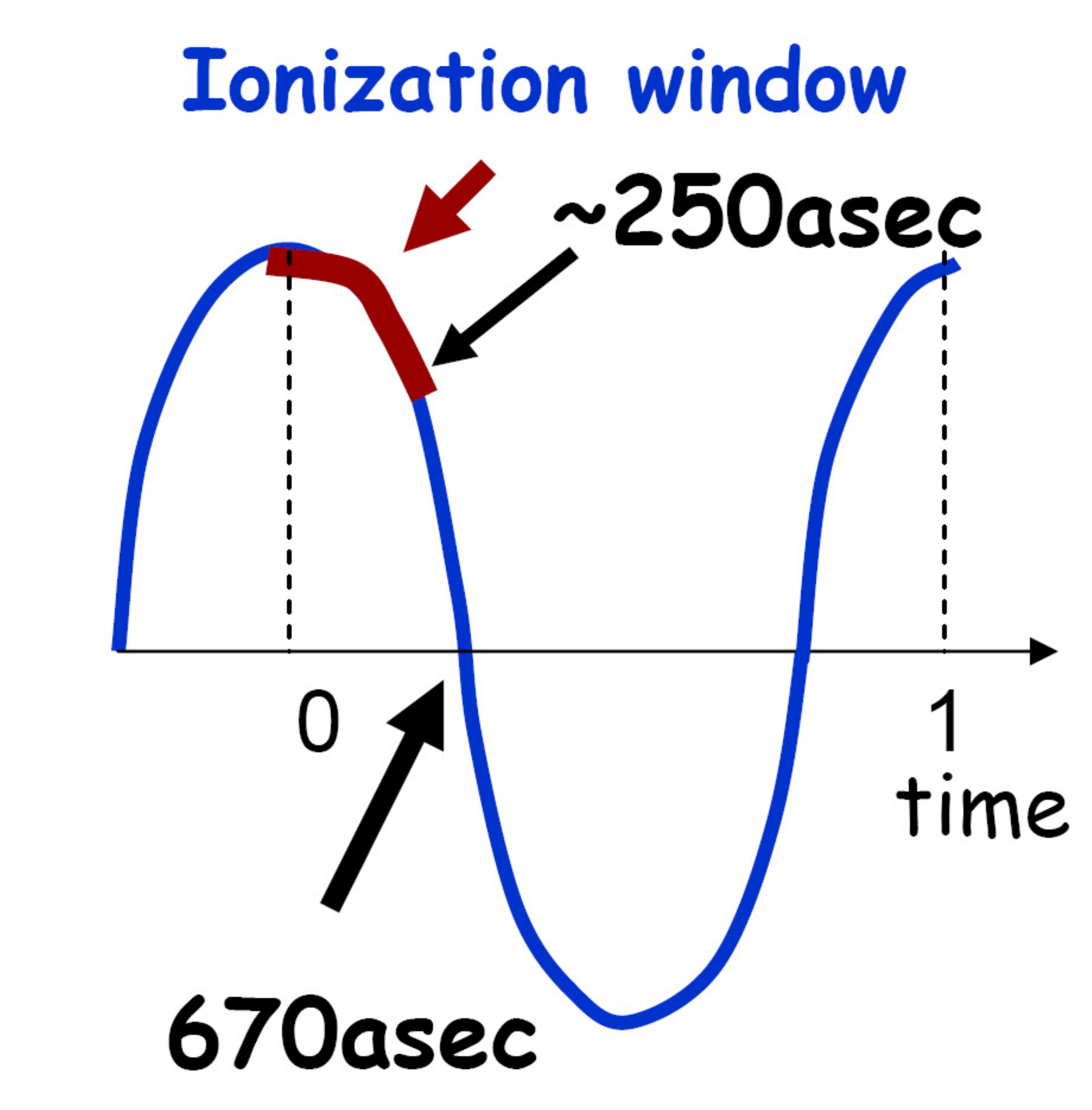}
\caption{Left panel: Real and imaginary ionization times vs real return time for $I_p=15.6$~eV, $I=1.3\cdot 10^{14}$~W/cm$^2$,
$\hbar\omega=1.5$~eV. Right panel: Cartoon illustrating the ionization
window. Ionization occurs around the field maximum within an
approximately 250~as time window (corresponding to the maximum value of
the real ionization time).} \label{fig:ion_time}
\end{figure}

This is why the complex saddle point $t_i$ is associated with the time at
which the electron enters the classically forbidden region -- the
tunnelling barrier -- while the real part of the complex saddle
point $t'_i$, after which  changes to the ionization amplitude
stop\footnote{Rigorously, this statement is only true for short range potentials.
Long-range electron-core interactions lead to additional modifications of the ionization amplitude after $t'_i$ (\cite{Lisa12}, \cite{Lisa12c}).},
  is associated with the time of exit from the classically
forbidden 'under-the-barrier' region.
 The same reasoning can be
extended to the ionization times  arising
within the semiclassical picture of harmonic generation, see
Fig.~(\ref{fig:ion_time}).
However, the ionization times in high harmonic generation are somewhat
different due to the fact that $\p_s$ are complex-valued.
In the next section we will consider the connection between these two times.

The imaginary ionization time defines the ionization probability.
Since the  imaginary component of the ionization time is larger
for short trajectories, these trajectories have a lower chance of
being launched compared to the long ones. The range over which the
real part of the ionization time changes within the quarter-cycle
defines the duration of the 'ionization window'. Typically, for a
$\lambda\simeq 800$ nm driving laser field and a laser intensity of $I\sim
10^{14}$~W/cm$^2$, the ionization times (their real
part) are spread within $\sim$250 attoseconds around the
instantaneous maximum of the laser field (see Fig.~\ref{fig:ion_time}). Thus, strong-field
ionization is an intrinsic attosecond process. Note that the
quantum 'ionization window' is shorter than the classical one (see
Fig.~\ref{fig:3step}), as according to the classical simple man
picture ionization happens at any phase of the laser field.

\begin{figure}
\includegraphics[width=0.484\textwidth]{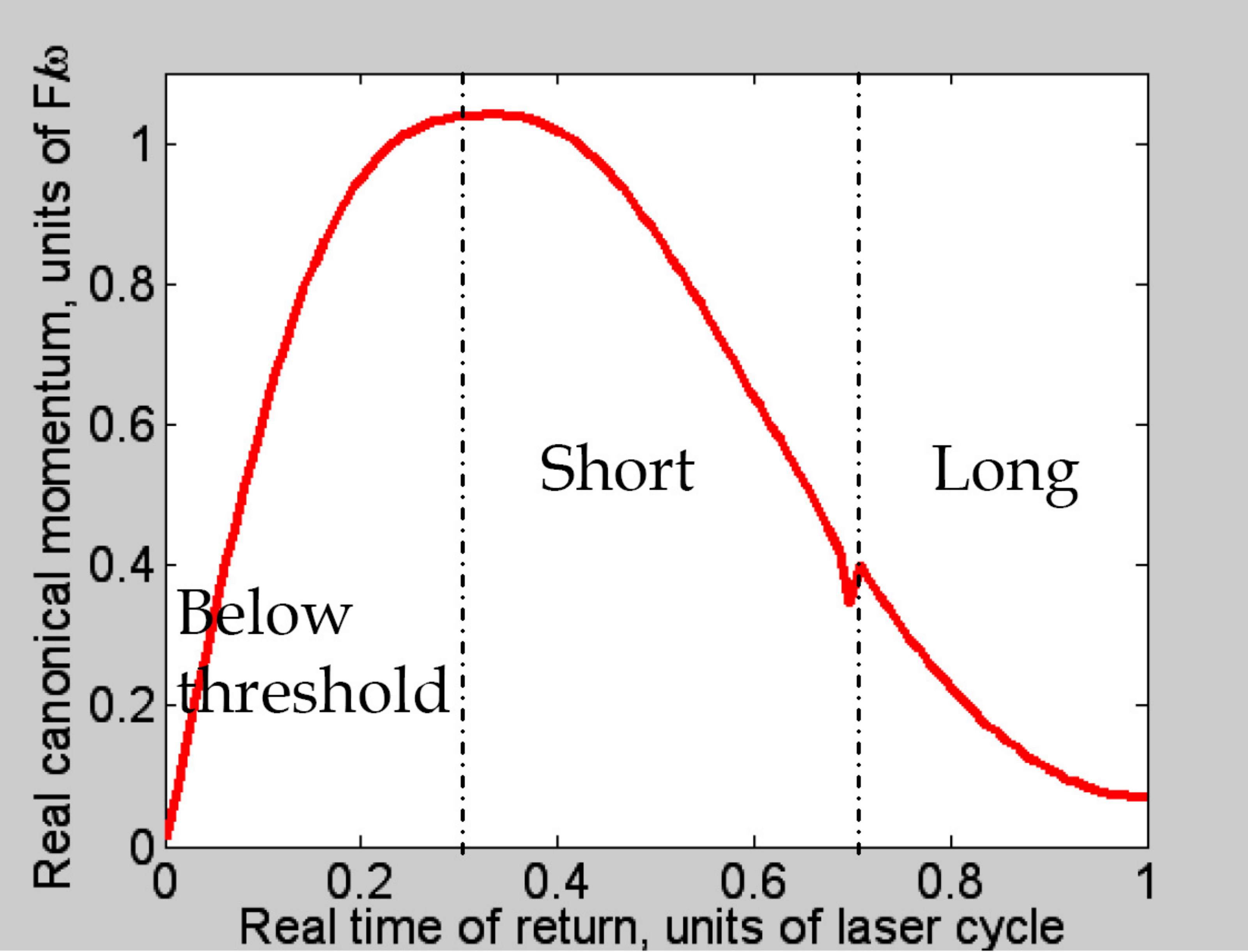}
\includegraphics[width=0.5\textwidth]{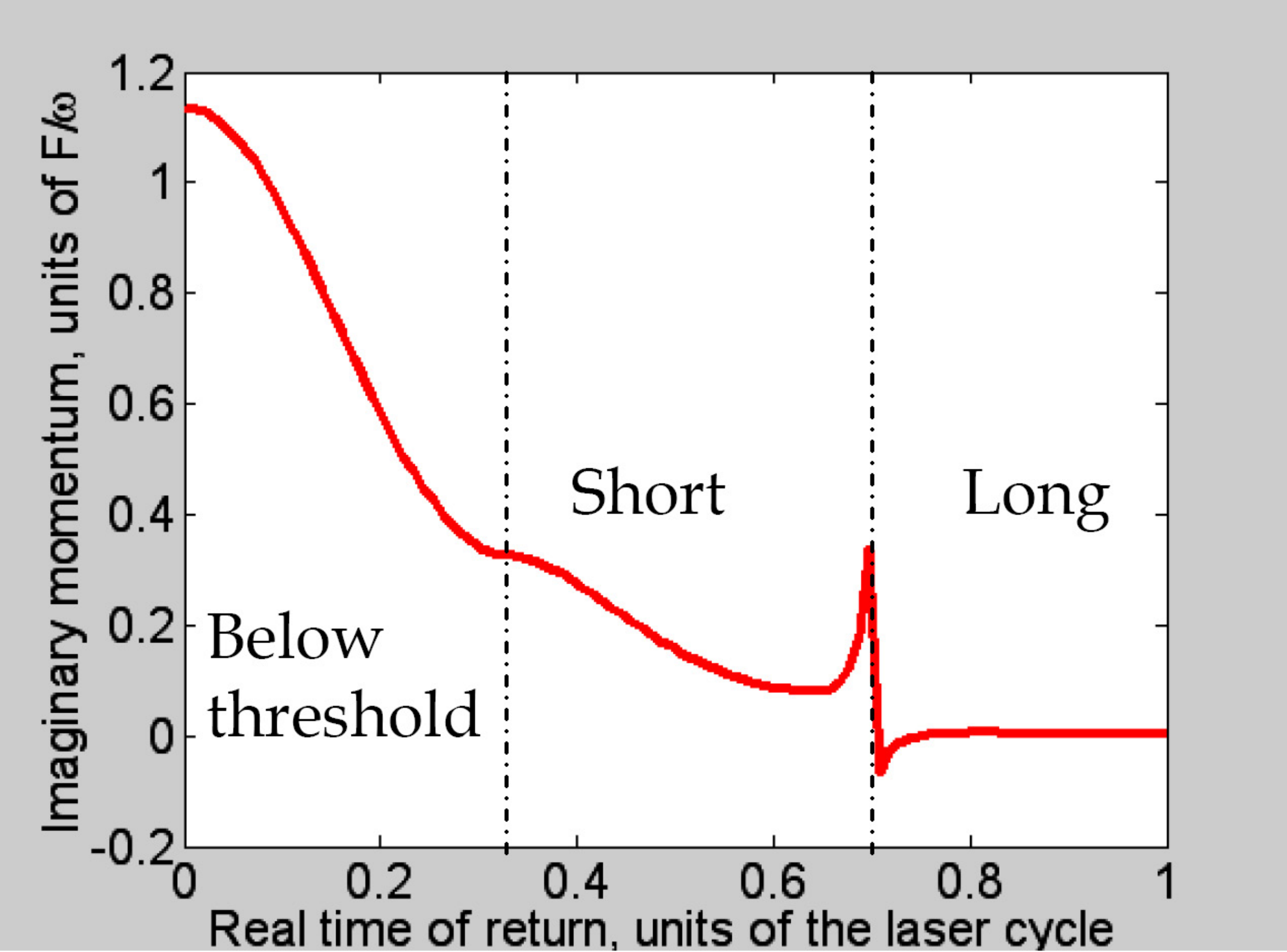}
\caption{Left panel: Real canonical momentum vs real return time
for $I_p=15.6$~eV, $I=1.3\cdot10^{14}$~W/cm$^2$, $\hbar\omega=1.5$~eV.
Right panel: Imaginary canonical momentum vs real return time.}
\label{fig:can_mom}
\end{figure}

Figure~\ref{fig:can_mom} shows the saddle point solutions for the
electron canonical momentum.

In photoionization, the electron canonical momentum is always real,
since it is the observable
registered at the detector. In contrast, in harmonic generation
the observable registered at the detector is the emitted photon,
and hence it is the photon energy that must be real. As a result, the
electron canonical momenta in HHG are complex. Electrons on long
trajectories have a very small imaginary canonical momentum.
Therefore, it is a very good approximation to associate long
trajectories with photoelectrons. Note that the maximum of the
real canonical momentum is about $p_{\rm max} \simeq A_0$. In the
photoelectron perspective $p_{\rm max}$ corresponds to an energy of
$2U_p$ at the detector - the cut-off energy for the so-called
direct photoelectrons, i.e. those that have not substantially
changed their momentum after ionization.

The imaginary part of the canonical momentum can be quite large
for short trajectories. The complex-valued solutions, not only for
the ionization times, but also for the recombination times and the
electron canonical momenta, challenge our understanding of the
underlying physical picture of harmonic generation. If the first
step of high harmonic generation is ionization, then why do these
liberated electrons have complex canonical momenta? Does  this mean
that these electrons have not been ionized? Can we factorize the
harmonic dipole into ionization, propagation and recombination?
The next section explores this opportunity.
 %
\section{Factorization of the HHG dipole: simple man on a complex plane}

Having derived the analytical expressions for the HHG dipole, can we
identify the simple man model in it, within the consistent quantum
approach? To do this, we need to factorize the harmonic dipole
into the three steps: ionization, propagation and recombination. That
is, we have to re-write the dipole as a product of the ionization
amplitude, the propagation amplitude and the recombination
amplitude.

Such factorization of the harmonic dipole is  not just curiosity
driven. It is important for extending the modelling of harmonic
emission to complex systems. Once the three steps are identified,
the respective amplitudes can be imported from different
approaches, tailored to calculate specifically ionization or
recombination in complex systems.

The factorization of the harmonic dipole runs into two types of
problems: technical and conceptual. The technical problems arise
from the fact that the original three-step (simple man) model is
formulated in the time domain. The three processes -- ionization,
propagation and recombination -- are the  sequence of subsequent
time-correlated events. The harmonic spectrum formally corresponds
to the harmonic dipole in the frequency domain, where  the three
processes become entangled: recall the contribution of different
quantum trajectories to the same photon energy. Thus, rigorous
factorization in the frequency domain is only possible in the
cut-off region, where short and long  solutions merge, see \cite{Frolov09}
\footnote{Note that the quantitative rescattering theory (see \cite{Le09})
postulates that one can factor out the recombination step in the frequency domain harmonic dipole.
This postulate is supported by the results of numerical simulations
demonstrating approximate factorization in the cut-off region, see \cite{Morishita08}.}.

The conceptual problem is due to the complex canonical momentum of
the electron responsible for HHG. Ionization in terms of
creating photoelectrons with real canonical momenta does not
appear to fit into the HHG picture. Can we build an alternative
model of HHG based entirely on photoelectrons, i.e. those electrons
which are indeed ionized at the first step?

Let us address these issues step by step, starting with the
factorization of the harmonic dipole in the frequency domain
(\cite{Frolov09,Kuchiev99,Morishita08}) and the time domain (\cite{Ivanov96}).
The former involves the factorization of Eq.~(\ref{eq:dipV4}),
the latter factorizes Eq.~(\ref{eq:dipV3_new}).


\subsection{Factorization of the HHG dipole in the frequency domain}

To re-write the harmonic dipole in the semi-factorized form, we
can take Eq.~(\ref{eq:anal_dip0}) and split the action integral $S$
that enters the phase of this expression into the following three time
intervals: from $t_i$ to $t'_i$, from $t'_i$ to $t'_r$, and from
$t'_r$ to $t_r$ (see Fig.~\ref{fig:action}). Then we can identify the group of terms that
looks like the ionization amplitude similar to that given by
Eq.~(\ref{eq:aion_saddle}),
    \bea
     \label{eq:aionPrim}
    &&a_{ion}(\p_s,t_i)=\left[\frac{2\pi}{i S''_{t_i,t_i}}\right]^{1/2}
    e^{-iS(\p_s,t'_i,t_i)}\,\Upsilon(\p_s+\A(t_i)).
    \eea
The ionization amplitude  is associated with the first time interval, from $t_i$ to
$t'_i$, and only the part of the action integral from $t_i$ to the
real time axis $t'_i={\rm Re}(t_i)$ enters this amplitude.
For a short-range potential (neglecting the sub-cycle effects in the prefactor):
\bea
 \label{eq:aion}
  &&a_{ion}(\p_s)= 2C\sqrt{\frac{- \tilde{\gamma} \omega}{\kappa\sqrt{1+\tilde{\gamma}^2}}}
\left( \frac{p_{s\perp}}{\kappa}\right)^m
  e^{-iS(\p_s,T,t_i)+im\phi_{p_s}}.
\eea
Constant $C$ is specified in Eq. (\ref{eq:C}).
The
momentum $\p_s$ is given by the full set of saddle point
conditions for $t_i, t_r$, and $\p_s$:
\begin{eqnarray}
&&\frac{[\p_{s}+\A(t_i)]^2}{2}+I_p=0,
 \nonumber \\
&&\int_{t_i}^{t_r}[\p_{s}+\A(t')]dt'=0,
 \nonumber \\
&&\frac{[\p_{s}+\A(t_r)]^2}{2}+I_p=N\omega.
 \label{eq:SaddleFull}
\end{eqnarray}
Note that $t_i$ in HHG and and $t_i$ in ionization are different, that is why in Eq. \ref{eq:aion_Shs} we use $\gamma$ and $\p$,
whereas in Eq. (\ref{eq:aion}) we use $\tilde{\gamma}$ (see Eq.\ref{eq:PandD}) and $\p_s$.
If imaginary part of $\p_s$ is equal to zero, then $\tilde{\gamma}=\gamma$.
For Coulomb potential (neglecting the sub-cycle effects in the prefactor):
\bea
 \label{eq:aion}
  &&a_{ion}(\p_s)=2 C\left(\frac{2\kappa^3}{F}\right)^{\frac{Q}{\kappa}}\!\sqrt{\frac{-\tilde{\gamma} \omega}{\kappa\sqrt{1+\tilde{\gamma}^2}}}
\left( \!\!\frac{p_{s\perp}}{\kappa}\!\!\right)^m
  e^{-iS(\p_s,T,t_i)+im\phi_{p_s}}.
\eea
Now consider the next time interval, from $t'_i$ to $t'_r$. The
prefactor arising from saddle point integration over the electron
momenta $\p$ leads to the term
\bea
    \label{eq:Spsps}
&&\frac{(2\pi)^{3/2}}{\sqrt{{\rm det}(i S''_{\p_s,\p_s})}}=\frac{(2\pi)^{3/2}}{(i(t_r-t_i))^{3/2}}.
 \eea
This term describes the free spreading of the electron wavepacket
between $t_i$ and $t_r$. Thus, we associate the following group of
terms with the propagation amplitude:
    \bea
    \label{eq:aprop}
  &&a_{prop}(\p_s,t_r,t_i)=\frac{(2\pi)^{3/2}}{(i(t_r-t_i))^{3/2}}e^{-iS(\p_s,t'_r,t'_i)}.
    \eea
Note that the denominator includes the complex-valued times $t_i$ and
$t_r$.

Finally, the recombination amplitude is represented by the
recombination matrix element $\d^{*}(\p_s+\A(t_r))$ and can be
associated with the following group of terms:
 \bea
 \label{eq:arec}
  &&\mathbf{a}_{rec}(\p_s,t_i)=\left[\frac{2\pi}{i S''_{t_r,t_r}}\right]^{1/2}e^{-iS(\p_s,t_r,t'_r)+iN\omega t_r}\,\d^{*}(\p_s+\A(t_r)),
\eea
where $S''_{t_r,t_r}=-\sqrt{2(N\omega-I_p)}{F}(t_r)$ for a linearly polarized field.
 As a result, the total dipole is formally written as
 \bea
 \label{eq:anal_dip}
  && \D (N\omega)=\sum_{j=1}^{4M}
\mathbf{a}_{rec}(\p_s,t^{(j)}_r)
a_{prop}(\p_s,t^{(j)}_r,t^{(j)}_i)
a_{ion}(\p_s,t^{(j)}_i),
\eea
where the index $j$ labels the saddle points.
However, in contrast to photoelectrons, the electrons involved in
HHG have complex canonical momenta $\p_s$. Therefore, the
imaginary part of the action is accumulated not only 'under the
barrier', from $t_i$ to $t'_i$, but also all the way between $t'_i$
and $t_r$. Thus, factoring out ionization as the first step of HHG
is not that convincing. Similarly, the recombination step involves
not only the recombination dipole, but also the possible change in the
amplitude due to the imaginary contribution to the action between
$t'_r$ and $t_r$. Thus, while we can formally associate several
groups of terms in the harmonic dipole~(\ref{eq:anal_dip}) with
amplitudes of ionization, propagation, and recombination, the
complex-valued electron momenta make such identification somewhat
stretched.

An additional point to note is that the three amplitudes are also
entangled due to the sum over the different saddle points in Eq.~(\ref{eq:anal_dip}).
Even if we only consider
contributions of the two most important trajectories, short and
long, the sum entangles their contributions and also mixes up the
contributions from different half-cycles. Importantly, a finite
pulse duration leads to a different mapping between the given
harmonic number and the ionization-recombination times for each
half-cycle.

A practical approach to factorization realized in the so-called quantitative rescattering theory (\cite{Le09}) is to assume that 
$\mathbf{a}_{rec}(\p_s(N),t^{(j)}_r(N))=\mathbf{a}_{rec}(N\omega)$ for all $j$ yielding:
 \bea
 \label{eq:anal_dip_lin}
  && \D (N\omega)=\mathbf{a}_{rec}(N\omega)\sum_{j=1}^{4M}
a_{prop}(\p_s,t^{(j)}_r,t^{(j)}_i)
a_{ion}(\p_s,t^{(j)}_i),
\eea
This approximation breaks down in the following cases:
\begin{enumerate}
  \item In two-color orthogonally polarized fields \cite{Morales12}. In this case more 
  than two trajectories returning at different angles can map into the same return energy \cite{Morales12}.
   Such trajectories must correspond to different recombination dipoles for different angles, violating \ref{eq:anal_dip_lin}.
  \item In the vicinity of the structural minimum of the recombination matrix element, 
  or when the phase of the matrix element changes rapidly (\cite{Smirnova09b}, \cite{serguei}). 
  \item When the sub-cycle dynamics associated with the electron interaction with the core potential can not be neglected.
\end{enumerate}
These technical problems can be remedied by looking at the dipole
in the time domain.
%
%

\subsection{Factorization of the HHG dipole in the time domain}

There are several advantages of using the time-domain dipole.  For
starters, if we do not perform the Fourier transform analytically,
the time $t_r$ no longer has to be complex. With the Fourier
integral performed using a standard FFT routine, we can keep $t_r$
on the real time axis. The number of saddle-point conditions is
also conveniently reduced to two (one of them, for the momentum
$\p_s$, is in general 3-dimensional)
\begin{eqnarray}
&&\frac{[\p_{s}+\A(t_i)]^2}{2}+I_p=0,
 \nonumber \\
&&\int_{t_i}^{t_r}[\p_{s}+\A(t')]\,dt'=0,
 \label{eq:SaddleTime}
\end{eqnarray}
with $t_r$ being the parameter, instead of the harmonic number $N$.

In the time domain, it is natural to sort the contributions to the induced
dipole according to the corresponding ionization bursts.  Then,
for each half-cycle $j$, there is a single ionization burst $j$ at time
$t^{(j)}_i$ that
contributes to the induced dipole as a function of the real return
time $t_r$, see left panel of Fig.~\ref{fig:ion_time}. After
saddle-point integration, this contribution is:
    \bea
    \label{eq:dipV3hc}
\D^{(j)}(t_r)&=& i \left[\frac{2\pi}{iS''_{t^{(j)}_i,t^{(j)}_i}}\right]^{1/2}
\frac{(2\pi)^{3/2}}{\sqrt{{\rm det}(iS''_{\p_s,\p_s})}} \,\times
\nonumber\\
&&\times \, \d^{*}(\p_s+\A(t_r))\,e^{-iS(\p,t_r,t^{(j)}_i)}\,\Upsilon(\p_s+\A(t^{(j)}_i)) , \nonumber
     \eea
  \bea
 S(\p_s,t_r,t_i)\equiv\frac{1}{2}\int_{t_i}^{t_r} [\p_s+\A(\tau)]^2 d\tau+I_p(t_r-t_i),
    \label{eq:Sactionhc}
    \eea
with ${\rm det}(iS''_{\p_s,\p_s})=\left[i(t_r-t^{(j)}_i)\right]^{3/2}$ (see also Eq. \ref{eq:Spsps}).
Just as in the frequency domain, up to a global phase factor the dipole can
be written as a product of three amplitudes:
    \bea
    \label{eq:hc_dip1}
    &&\D^{(j)}(t_r)= \mathbf{a}_{rec}(\p_s,t_r)a_{prop}(\p_s,t_r,t^{(j)}_i)a_{ion}(\p_s,t^{(j)}_i).
    \eea
The ionization and the propagation amplitudes entering this expression
are given by Eqs.~(\ref{eq:aionPrim},\ref{eq:aprop}). The
recombination amplitude is simply equal to the recombination
matrix element $\d^{*}(\p_s+\A(t_r))$, as we  have not performed
the Fourier transform yet. Equation (\ref{eq:hc_dip1}) is the
natural mathematical formulation of the three step model, which is
intrinsically sub-cycle.

If we ignore multiple returns and very long trajectories, then for
each $t_r$ there is only one ionization burst to deal with. As
opposed to the frequency domain, the contributions of the long and the short
trajectories from this ionization burst are not yet mixed -- they
are separated in time. This is very convenient if you need to look
at the contribution of only the short, or only the long trajectories: it
is straightforward to add a time-domain filter that would filter
out the unwanted contributions. Essentially, this would correspond
to making a window Fourier transform of the time-domain harmonic
dipole. The inclusion of the contribution of multiple returns is rarely
required for typical experimental conditions.

To model the full $\D(t_r)$ one needs to model ionization,
recombination and propagation separately for each half-cycle, and
then collect the contributions from each half-cycle (each ionization
burst):
 \bea
 \label{eq:sum_dip}
  &&\D(t_r)= \sum_j \D^{(j)}(t_r).
\eea

To obtain the harmonic spectrum, we have to perform the Fourier
transform, which is convenient to do numerically using a FFT
routine. There are two possible approaches to implement the
Fourier transform.

\textbf{Integration along Lewenstein's contour}. In this
approach, the Fourier transform is performed along the
time contour in the complex plane $t_r=t'_r+it''_r$. In this case the
argument of the recombination dipole $\p_s+\A(t_r)$ remains real
and so does the re-collision energy $E_{rec}(t_r)$. Since it is
difficult to {\it numerically} perform an integration along a complex
contour
    \bea
    \label{eq:FFT1comp}
      &&\D(N\omega)= \int dt_r \; e^{-N \omega t''_r}\, \D(t_r) \,e^{iN\omega t'_r},
    \eea
 one can use variable substitution and integrate over the real return times $t'_r$:
\bea
 \label{eq:FFT1}
  &&\D(N\omega)= \int dt'_r \left[ \frac{dt_r}{dt'_r} e^{- [E_{rec}(t_r)+I_p]t''_r}\right]\,  \D(t'_r)\, e^{iN\omega t'_r},\\
  && N\omega=E_{rec}(t_r)+I_p,\\
  &&E_{rec}(t_r)=[\p_s+\A(t_r)]^2/2.
\eea
The derivative in the square bracket is associated with the
variable substitution.

Note that Eq.~(\ref{eq:FFT1}) contains one approximation: the
term $e^{- (N\omega)t''_r} $ is modified according to the energy
conservation  $N\omega=E_{rec}(t_r)+I_p$. However, the integration
of Eq.~(\ref{eq:FFT1}) is not very convenient due to the
additional effort associated with the need to avoid the divergence
of $\frac{dt_r}{dt'_r}$ in the cut-off region (see
Fig.~\ref{fig:rec_en}).

\textbf{HHG dipole on the real time axis}. To keep things simple,
one can keep the half-cycle harmonic dipole on the
 real time axis:
 \bea
    \label{eq:hc_dip}
  &&\D^{(j)}(t)= \mathbf{a}_{rec}(\p_s,t) a_{prop}(\p_s,t,t^{(j)}_i)a_{ion}(\p_s,t^{(j)}_i),
    \eea
where the saddle points $\p_s$ and $t^{(j)}_i$ are given by the
Eqs.~(\ref{eq:SaddleTime}) and the index $j$ labels different solutions
corresponding to the same return time $t$.


In this 'real-time-axis' approach, the return time $t$ is a
parameter: we have to find $t_i$ and $\p_s$ for each $t$. This can
be done using a procedure similar to that described in the
previous section, only simpler. Specifically, we introduce the
dimensionless variables $\phi=\omega t$ and $p/(F/\w)=p_1+ip_2$.
For a linearly polarized field $p_{s,\perp}=0$.  For each real
$\phi$ we use the Eqs.~(\ref{eq:F11prim},\ref{eq:F21prim}) with
$\phi''_r=0$ and $\phi'_r\equiv\phi$:
    \bea
    \label{eq:F11pp}
    &&F_1(\phi)=p_{1}(\phi-\phi'_{i})+p_{2}\phi''_i-\cos(\phi'_{i})\cosh(\phi''_{i})+\cos(\phi)=0,\\
    \label{eq:F21pp}
    &&F_2(\phi)=-p_{1}\phi''_{i}+p_{2}(\phi-\phi'_{i})+\sin(\phi'_{i})\sinh(\phi''_{i})=0.
    \eea
We can now  use the
Eqs.~(\ref{eq:eion_r},\ref{eq:eion_i}) to express $p_1$ and $p_2$
in terms of $\phi'_i$ and $\phi''_i$. Then, we build the
surface $F(\phi)=F_1^2+F_2^2$ for each $\phi$. Next, we find the
minima on this surface. Alternatively, we can use $p_1$ and $p_2$ as our
variables, expressing $\phi'_i$ and
$\phi''_i$ via  $p_1$,  $p_2$, then the minima on the surface $F_1^2+F_2^2$ will yield the real, $p_1$, and the imaginary,
$p_2$, components of the canonical momentum, and then the Eqs.~(\ref{eq:ion_r},\ref{eq:ion_i})
yield the corresponding ionization times.

In this approach the divergence at the cut-off is
avoided, since the divergence occurs in the complex plane of the
return times when calculating the Fourier transform analytically using the saddle point
method.
The price to pay is that the recombination dipole has to be taken
at the complex arguments $\p_s+\textbf{A}(t)$ and the re-collision
energy $E_{rec}(t)$ has an imaginary part. In practice, one can use
the real part of the re-collision energy as the argument of the
recombination dipole. If one wants to avoid this approximation,
one has to extend the recombination dipoles into the complex plane of the
electron momenta.

Thus, one can formally factorize the harmonic dipole in the time
domain, overcoming the technical problems associated with the
factorization. However, one has to keep in mind that the ionization
 amplitude has to be modified to include complex canonical momenta and slightly different ionizaton times.
 Fortunately, it does not lead to changes in angular factors, because  $\Upsilon_n(\p_s+\A(t_i))$ remains the same.
 Indeed, both $\p_s$ and  $t_i$ are different in case of HHG and ionization, but the term $\p_s+\A(t_i)= i\mu\kappa$
 (see Eq. \ref{eq:mu}) is the same in both cases. The changes appear in the phase $S(\p_s,t_r,t_i)$ and the sub-cycle core effects,
  i.e. everywhere where $p_s$ and $t_i$ contribute separately.

 The conceptual problem associated with understanding the physical meaning of the
complex electron momenta, especially in the context of the  "ionization step", still remains.
The next section shows how, and to what extent, this problem
can be circumvented. It introduces the photoelectron model of HHG,
where the electron canonical momentum is restricted to the real
axis.

\section{The photoelectron model of HHG: the improved simple man}

In the standard simple man model, the electron motion between
ionization and recombination is modelled using classical
trajectories. Naturally, the electron velocity, the ionization time, and the
recombination time are all real-valued quantities. In the quantum
description, the rigorous approach based on the saddle point
method leads to trajectories with complex-valued momenta and complex-valued
ionization  and recombination times. The presence of complex
canonical momenta makes it difficult to identify the ionization
step.

The complex-valued canonical momenta and recombination times arise
from the requirement that the electron returns exactly to its
original position. Since the tunnelling electron accumulates an
imaginary displacement during its motion in the classically
forbidden region, the complex-valued momenta and return times must
compensate for this displacement.

This section shows that if we relax the return condition and
neglect the imaginary displacement between $t_i$ and ${\rm Re}(t_i)$, we
can obtain the same re-collision energy for real-valued canonical
momenta and for real-valued return times. We shall call this approach
the photoelectron model since it allows one to incorporate
standard strong-field ionization concepts in a natural manner. The
ionization amplitude would then correspond to creating an electron
with a real-valued canonical momentum, and the imaginary part of the
action integral would only be accumulated between $t_i$ and ${\rm
Re}(t_i)$.

\begin{figure}
\includegraphics[width=0.492\textwidth]{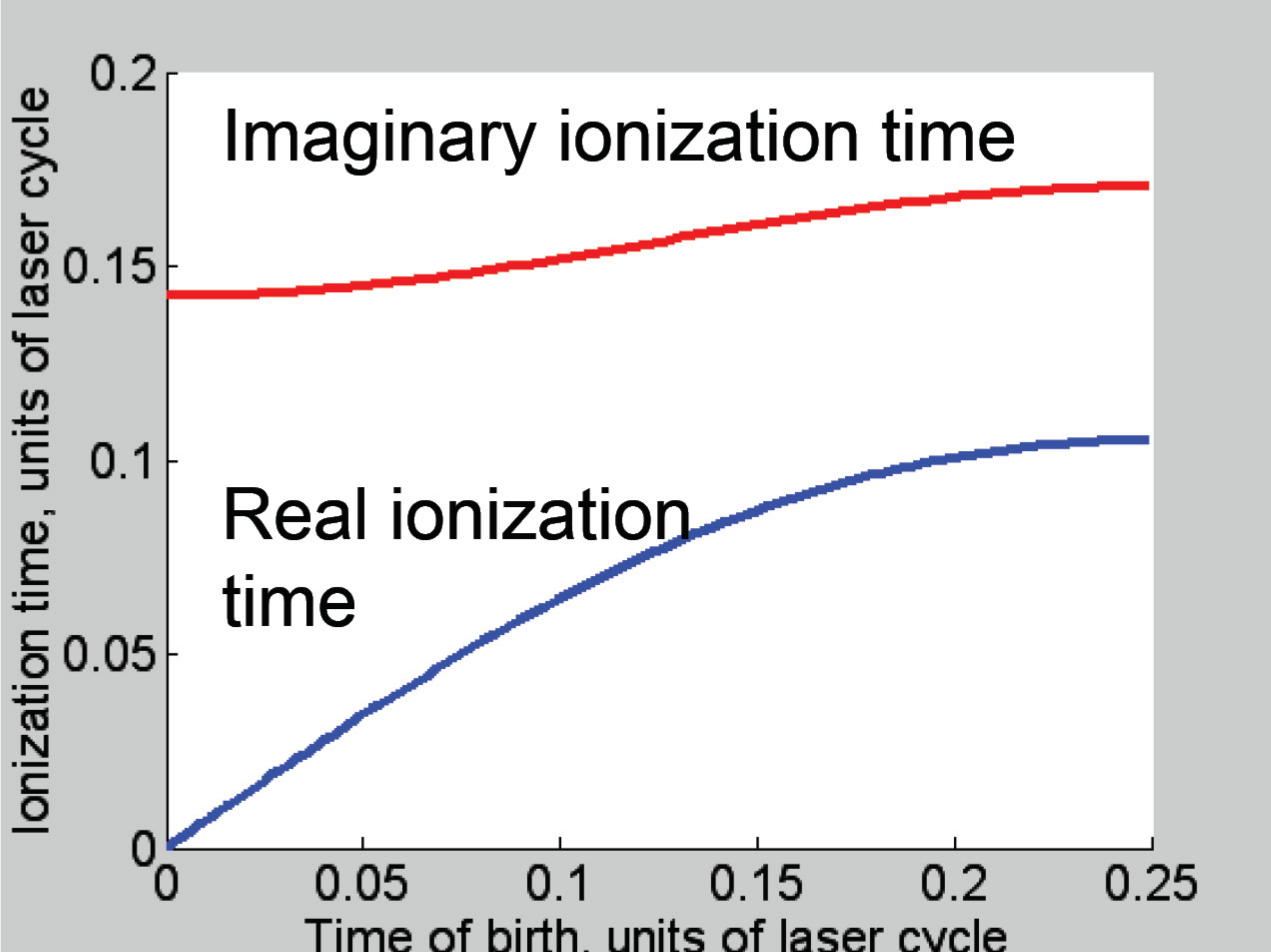}
\includegraphics[width=0.498\textwidth]{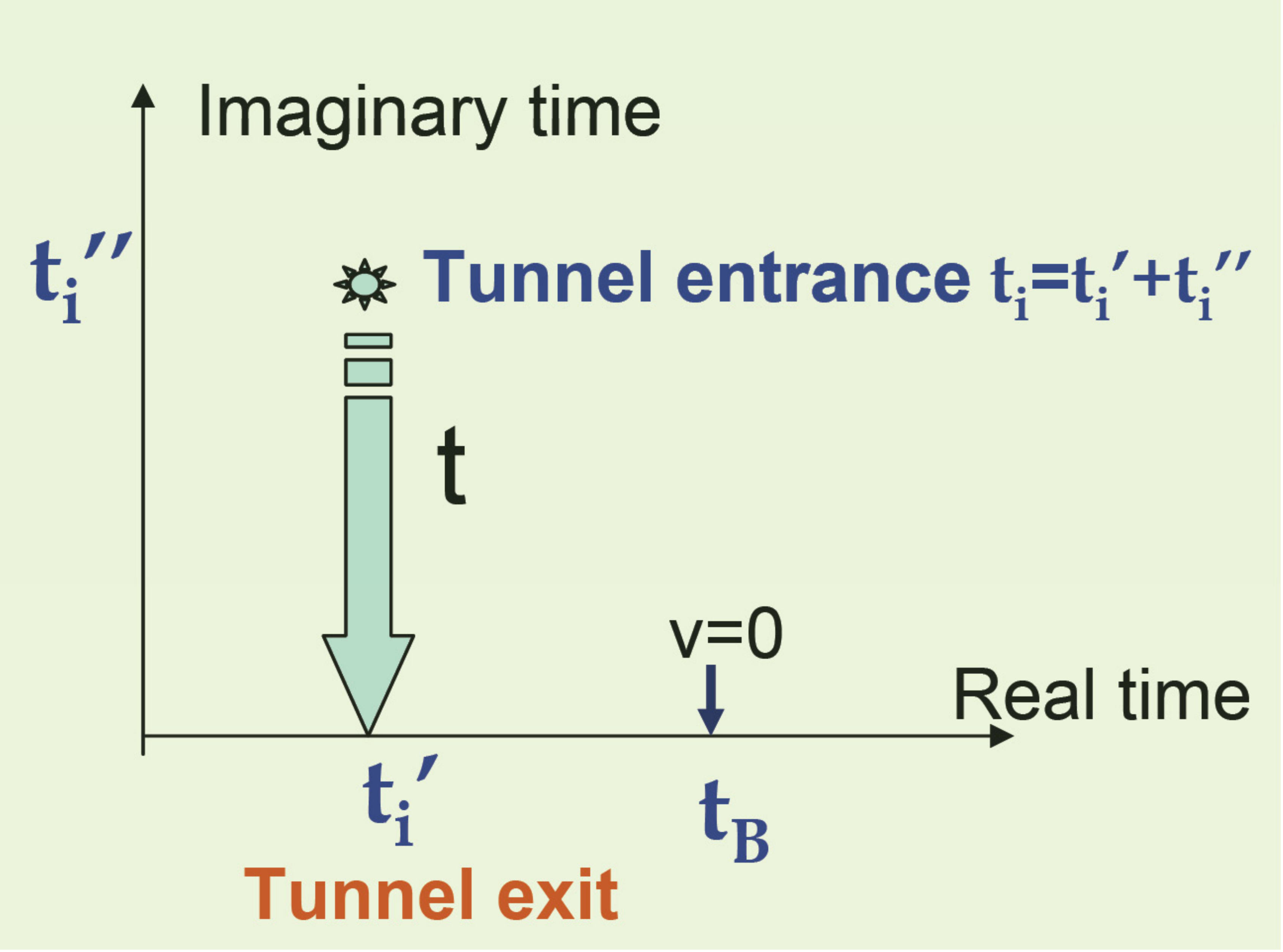}
\caption{Left panel: Complex ionization time for photoelectrons
$t_{i,ph}$ vs time of birth for $I_p=15.6$~eV, $I=1.3\cdot10^{14}$~W/cm$^2$, $\hbar\omega=1.5$~eV.
Right panel: Cartoon illustrating the
connection between $t_{i,ph}$ and $t_B$: the electron exits the
barrier with a negative velocity $\mathbf{v}(t_{i,ph})=\k (t_{i,ph})$  (directed towards the core).
Its velocity gradually decreases and becomes zero at the classical
ionization time ($\mathbf{v}(t_B)=\k (t_{B})=0$)  - the time of birth $t_B$.} \label{fig:clas_ion}
\end{figure}

In the classical model, one assumes that the electron trajectory
is launched at the real 'time of birth' $t_B$ with zero
instantaneous velocity. The electron instantaneous momentum at $t_B$ can be
written as $\k(t_B)=\textbf{p}+\textbf{A}(t_B)=\textbf{0}$, where the
canonical momentum $\textbf{p}$ is a constant of motion
(neglecting the core potential). The link between $t_B$ and
$\textbf{p}$, $\textbf{p}=-\textbf{A}(t_B)$, links $t_B$ via
$[\p+\A(t_i)]^2=-2I_p$ to the complex-valued ionization time $t_i$.
In particular, for a linearly polarized laser field we have
$[A(t_{i,ph})-A(t_B)]^2=-2I_p$. Note that this $t_{i,ph}$ is in
general different from the ionization time $t_i$ introduced in the
previous section, since now the electron canonical momentum is
forced to be real. The notation $t_{i,ph}$ stresses that this
ionization time corresponds to photoelectrons, i.e. to electrons with
real canonical momenta. Figure~\ref{fig:clas_ion} shows the mapping
between the time of birth and the complex time $t_{i,ph}$.

The photoelectron exits the tunnelling barrier at the real time,
${\rm Re}(t_{i,ph})$, and since ${\rm Re}(t_{i,ph})$ turns out to be
smaller than $t_B$, the electron velocity at ${\rm Re}(t_{i,ph})$
is directed towards the core.  It gradually decreases until
becoming equal to zero at $t_B$. The difference between ${\rm Re}(t_{i,ph})$
and $t_B$ is small near the peak of the oscillating
electric field, but increases as the field approaches zero. While
the times $t_B$ are always spread within one quarter-cycle, as in
the classical model, the times ${\rm Re}(t_{i,ph})$ are limited to
a shorter fraction of the quarter-cycle, see
Fig.~\ref{fig:clas_ion}.

\begin{figure}
\includegraphics[width=0.498\textwidth]{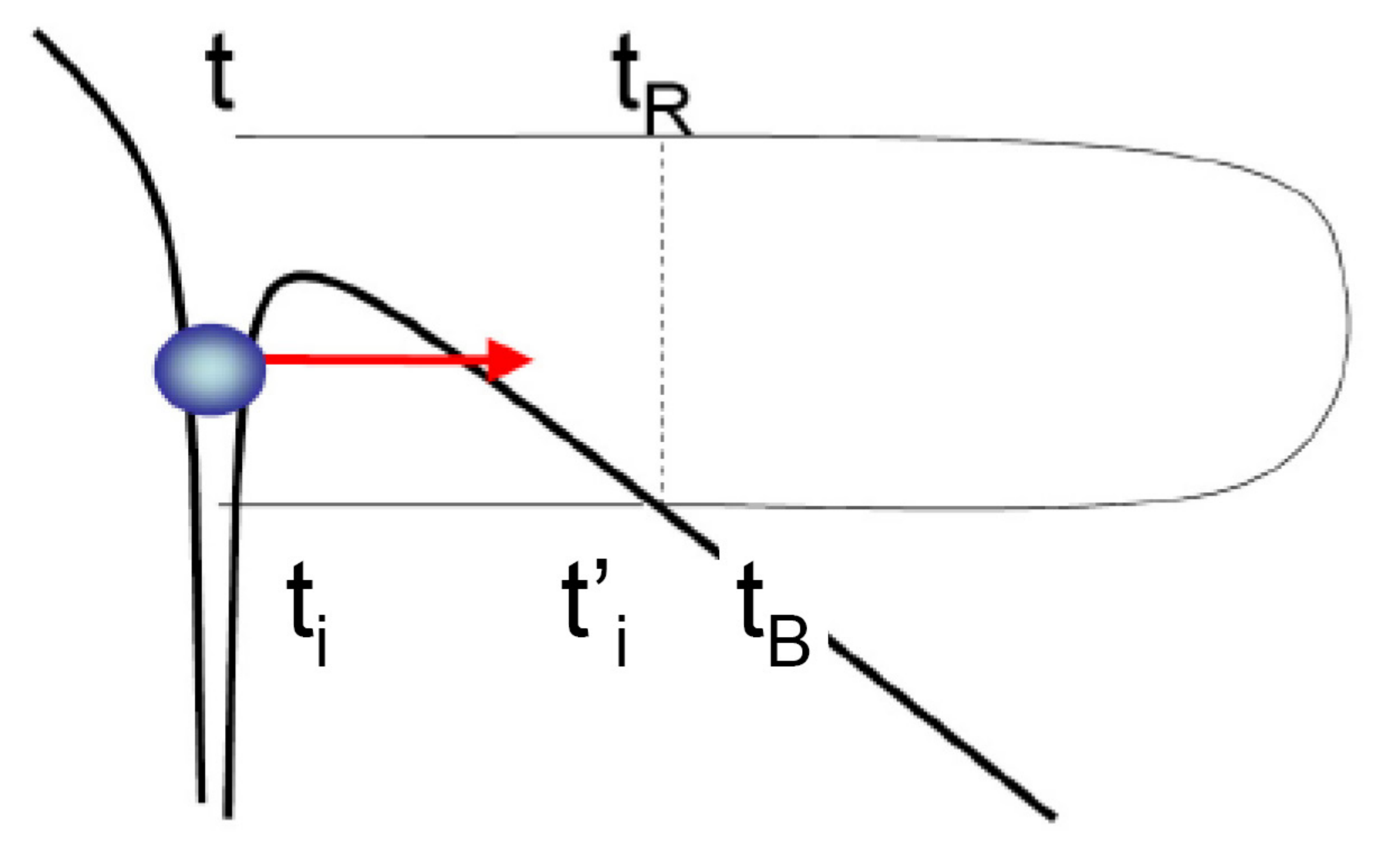}
\includegraphics[width=0.498\textwidth]{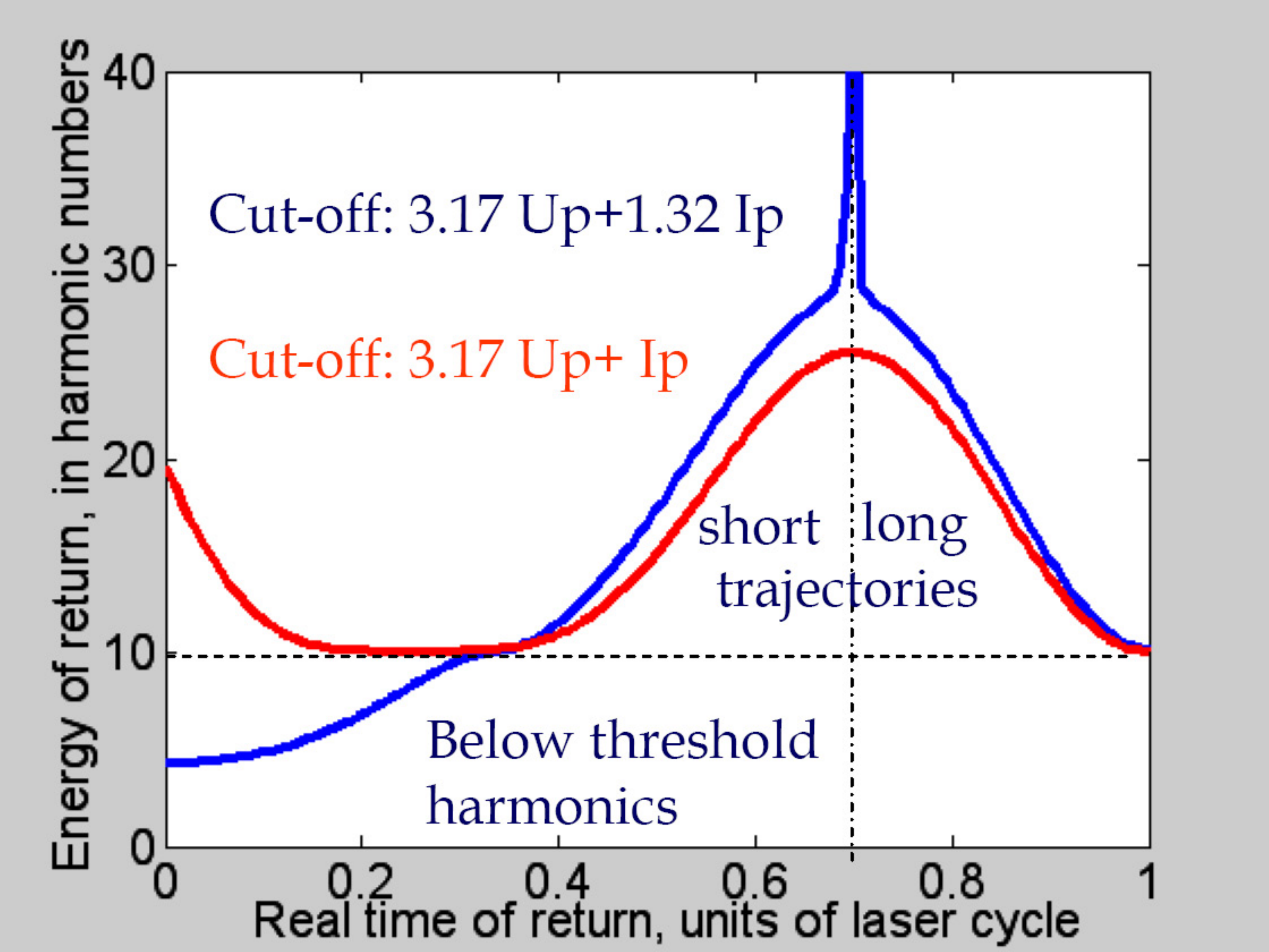}
\caption{Left panel: Physical picture of high harmonic generation
in coordinate space and the meaning of different times. The
electron enters the barrier at a complex time $t_i$ and exits the barrier at the
real time $t'_i={\rm Re}t_i$. Its velocity goes through zero at a (later) time
$t_B$. At the moment $t_R$ the electron returns to the
position it had at $t_B$, and at the moment $t$ it returns to the origin.
Right panel: The blue curve shows the electron return energy at the moment $t$
in the Lewenstein model, while the red curve shows the electron return energy in the
classical three-step model.} \label{fig:clas_rec}
\end{figure}

We now turn to the classical return time $t_R$. In the original
classical model, it is defined by the condition
\begin{eqnarray}\label{eq:return1}
    \int_{t_B}^{t_R}[p+A(\tau)] \, d \tau=\int_{t_B}^{t_R}[-A(t_B)+A(\tau)]\, d \tau=0.
\end{eqnarray}
However, since the electron is already offset from the origin at
$t_B$,
\begin{eqnarray}\label{eq:return2}
   \Delta z=\int_{t_{i,ph}}^{t_B} [A(\tau)-A(t_B)] \, d\tau,
\end{eqnarray}
it does not return to the origin at $t_R$, see
Fig.~\ref{fig:clas_rec}.

The energy $E(t_R)=[A(t_R)-A(t_B)]^2/2$ in the classical model is shown in the right panel of
Fig.~\ref{fig:clas_rec}, with
the cut-off at $3.17~U_p+I_p$. This cut-off is lower than in the
quantum treatment, precisely because the electron has not yet
returned to the core. The extra 0.32~$I_p$ in the quantum cut-off
law, $3.17~U_p+1.32~I_p$, is due to the extra energy accumulated by
the electron while covering the extra distance $\Delta z$
\footnote{Interestingly, if one defines the experimental cut-off
using the classical model, then the classical time-energy mapping
is very similar to the quantum: $t_R$ is very close to the real
part of $t_r$.  Since in the experiment the intensity is rarely
known exactly, it is very difficult to differentiate between the
classical (red) and the quantum ( blue) return energies in
Fig.~\ref{fig:clas_rec}.}.

Can we improve these results if we allow the photoelectrons to
travel a bit longer and allow them to return to the core? Why do not we
continue to monitor the electron trajectory at times $t>t_R$ and
register their energy at the time of return to the origin
$t_{r,ph}$, ignoring whatever imaginary displacement they might
have? There is just one problem with this plan: not all
trajectories return to the core since we have
limited the canonical momentum $p_{ph}=-A(t_B)$ to be no more than
$A_0$. With this in mind, we shall take the energy at the closest
approach to the origin as the return energy. We shall call this an
improved three-step model or \textbf{the photoelectron model}.

The model implies the neglect of the imaginary displacement and the
minimisation of the real displacement between $t_{i,ph}$ and
$t_{r,ph}$. The imaginary displacement has to be neglected since
we do not have imaginary canonical momenta and imaginary return
times to cover for it.

\begin{figure}
\includegraphics[width=0.498\textwidth]{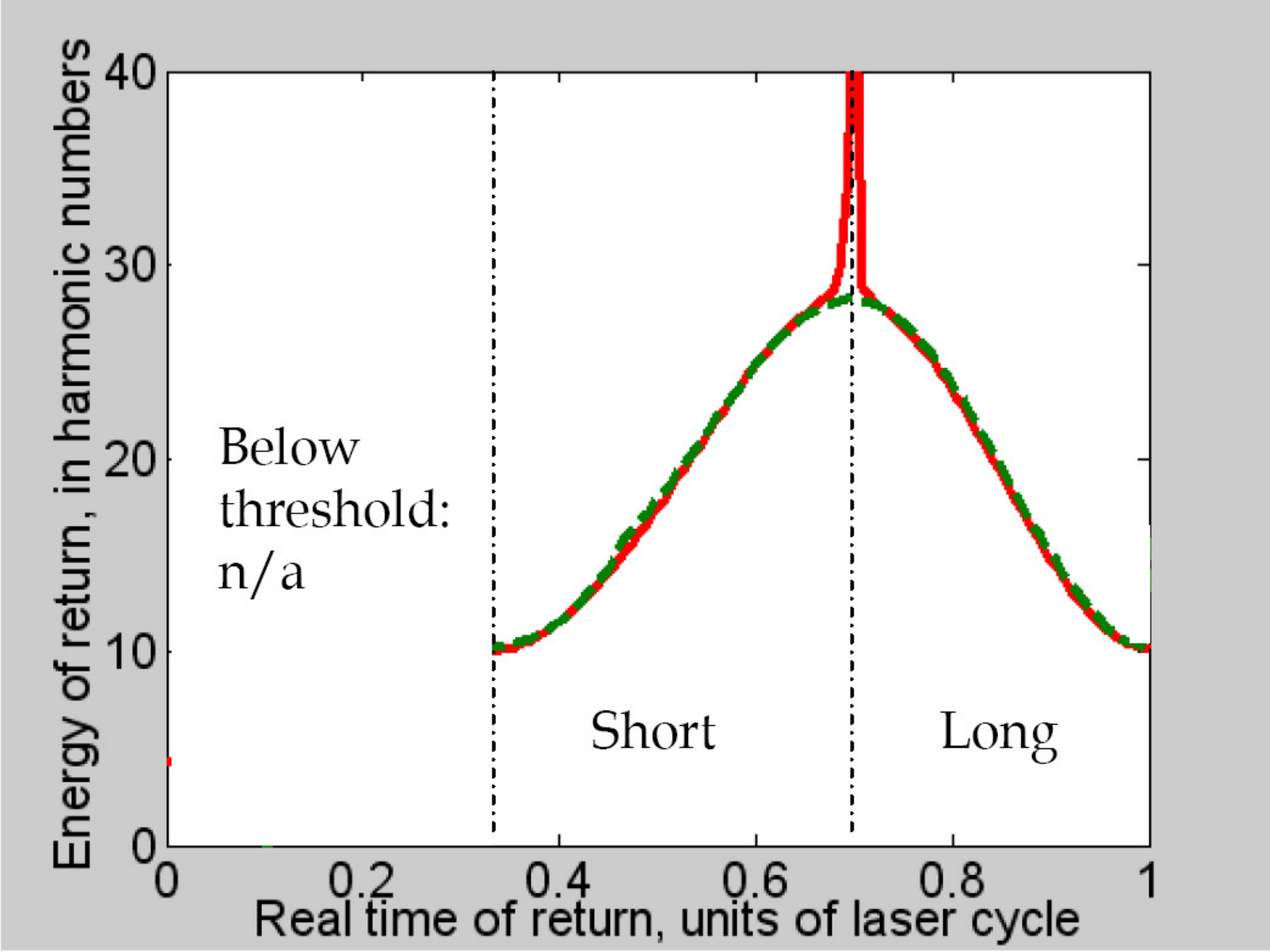}
\includegraphics[width=0.498\textwidth]{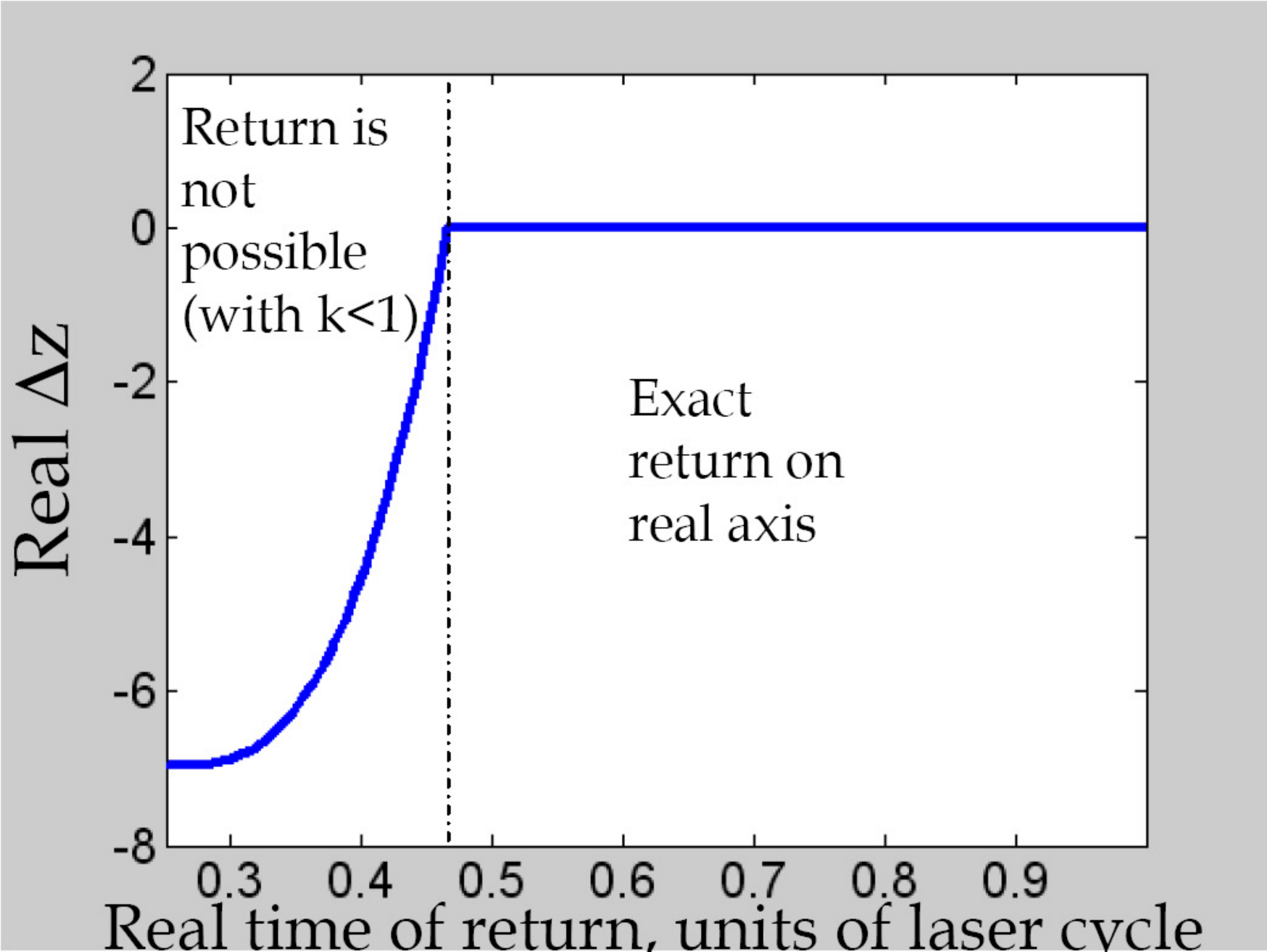}
\caption{Left panel: Energy of return for the Lewenstein model (red) and for the photoelectron model (green) vs real return time
for $I_p=15.6$~eV, $I=1.3 \cdot10^{14}$~W/cm$^2$, $\hbar\omega=1.5$~eV.
Right panel: Real part of the electron displacement in the photoelectron model.}
\label{fig:improved3st}
\end{figure}

The  photon energy resulting from the photoelectron model is
$E^{ph}_{rec}(t)+I_p=\frac{dS}{dt_{r,ph}}$, where
$S(p_{ph},t_{r,ph},t_{i,ph})$ is given by Eq.~(\ref{eq:Saction}).
It is in excellent agreement with the quantum
photon energy (see Fig.~\ref{fig:improved3st}, left panel) for
all those trajectories for which the real part of the electron
displacement from the origin passes through zero.
This is the case for the long trajectories
and for most of the short trajectories, except for the shortest
ones. These latter ones are 'born' at the end of the ionization window
and contribute to the lowest harmonics, just above the ionization
threshold.

For short trajectories, the electron is decelerated by the laser
field while returning to the core. Therefore, it needs a
sufficiently high drift momentum to reach the origin. Since we
have limited the canonical momentum $p=-A(t_B)$ below $A_0$, the
shortest trajectories cannot quite make it to the core. For them,
the time $t_{r,ph}$ corresponds to the closest approach to the
core. A non-zero real displacement yields a deviation of the
approximate action $S(p_{ph},t_{r,ph},t_{i,ph})$ from the real
part of the exact action defined in the previous section,
see the right panel of Fig.~\ref{fig:improved3st}.

The action in this model is reproduced very well, since it is the
time integral from the photon energy. Once the electron return
energy is well-reproduced, so is the action, even if the end
points $t'_i$, $t'_r$ are shifted.


From the mathematical perspective, the photoelectron model implies
that when we perform the integrals, we expand the action not at
the exact saddle point, but in its vicinity. In particular, we
shift the center of the expansion  for the canonical momentum from
the complex plane to the real axis. The error introduced in the
integral by this procedure is minimized if the new expansion point
lies within the saddle point region of the exact complex saddle
point of the multi-dimensional integral. Thus, the difference
$\Delta p=p_q-p_{ph}$ between the stationary point solution for
quantum orbits $p_q$ and the canonical momentum in the improved
three-step model $p_{ph}$ should be less than the size of the
stationary point region: $|\Delta p|<|\partial ^2 S/\partial
p^2|^{-1/2}=(t_r-t_i)^{1/2}$. We can estimate $|\Delta p|$ as
$|\Delta p|=|\Delta z/(t_r-t_i)|$, where $|\Delta z|$ includes the
neglected imaginary displacement. This estimate yields $|\Delta
z|<(t_r-t_i)^{1/2}$.

\begin{figure}
\includegraphics[width=0.498\textwidth]{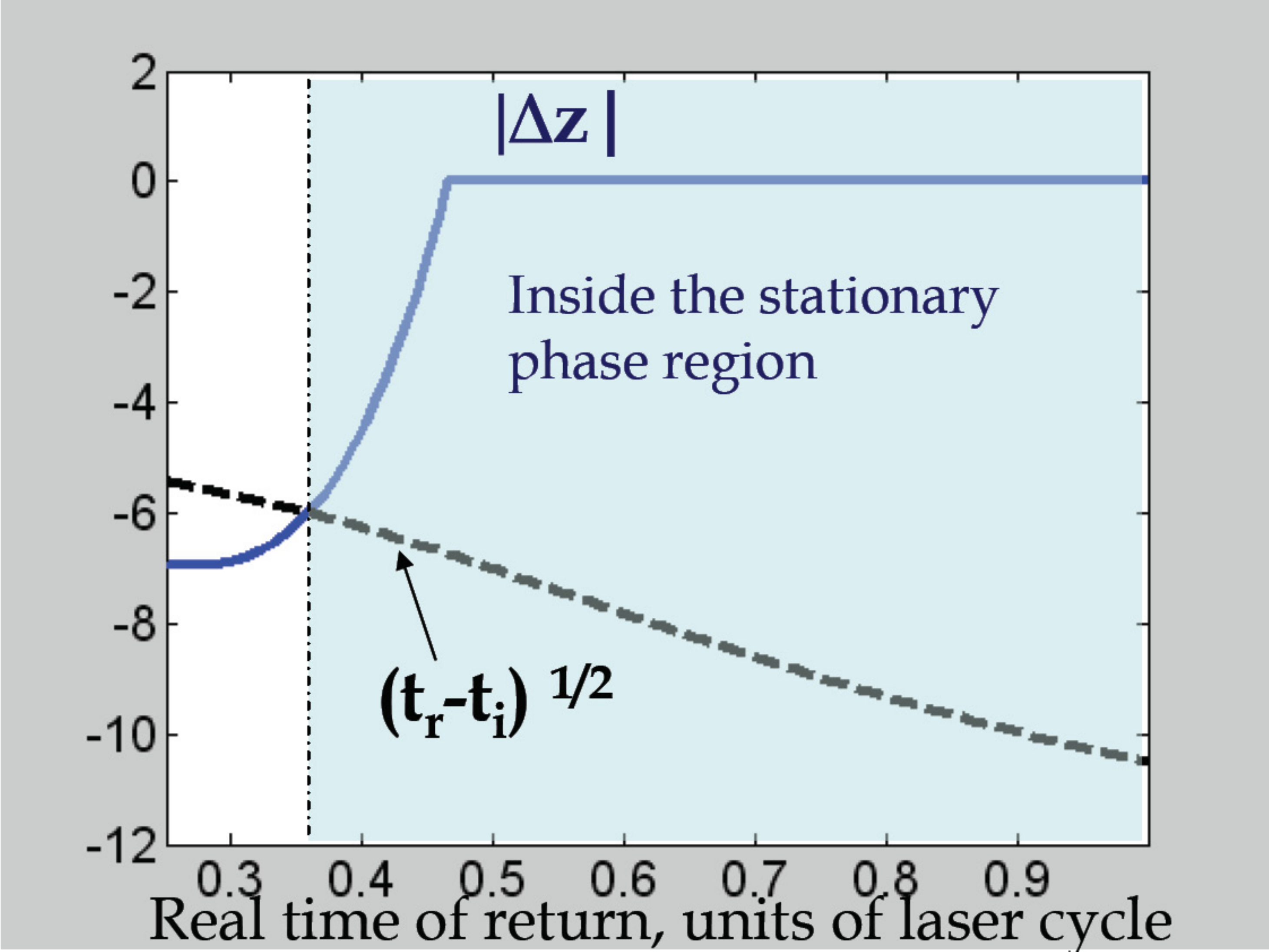}
\includegraphics[width=0.498\textwidth]{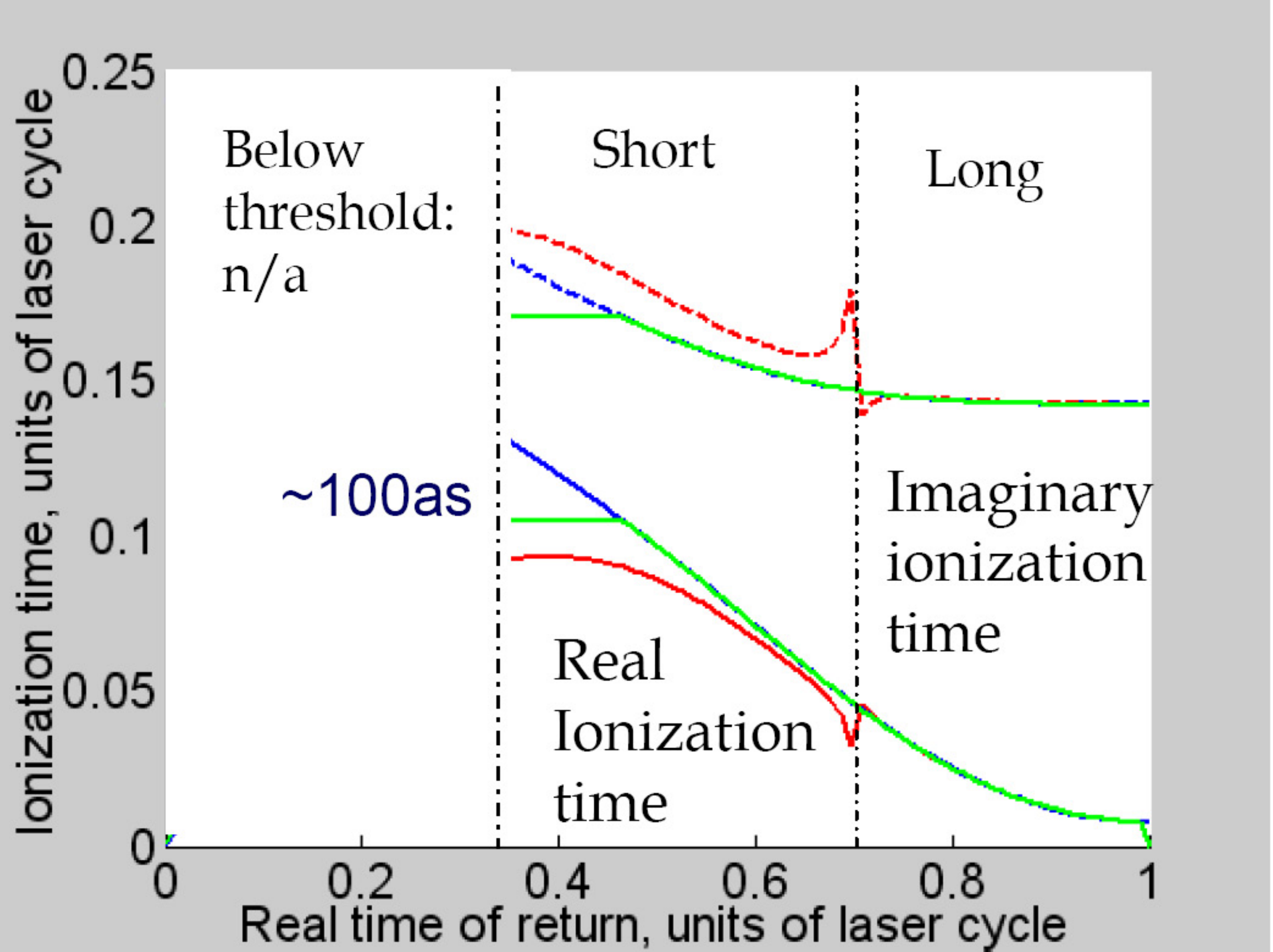}
\caption{Left panel: Applicability region of the photoelectron model. The condition $|\Delta
z|<(t_r-t_i)^{1/2}$ specifies the region of return times (filled) where the photoelectron model can be used.
The calculation is shown
for $I_p=15.6$~eV, $I=1.3\cdot10^{14}$~W/cm$^2$, $\hbar\omega=1.5$~eV.
Right panel: Real and imaginary ionization times for the Lewenstein model (red), the
photoelectron model with canonical momentum less than $A_0$ (green), and for
the canonical momentum not limited by this condition (blue).}
\label{fig:iontimesall}
\end{figure}

The left panel of Fig.~\ref{fig:iontimesall}
illustrates this condition for typical experimental parameters ($\omega=0.057$~a.u.,
$I_p=15.6$~eV, $I=1.3 \cdot 10^{14}$~W/cm$^2$): the improved three
step model cannot be applied for very short trajectories
returning earlier than $\omega t_{r,ph}=0.36$ or for harmonics
lower than $N=11$. Thus, for this particular set of parameters,
all above threshold harmonics are within the applicability
conditions of the improved three step model.

The right panel in Fig.~\ref{fig:iontimesall} compares the
ionization times resulting from the Lewenstein model and the
photoelectron model of HHG. The ionization times coincide for the long
trajectories. In this sense, the long trajectories indeed correspond
to photoelectrons. The difference between the ionization times for the
short trajectories is associated with the presence of imaginary
canonical momenta in the Lewenstein model. For the shortest
trajectories, the difference in the real ionization times is about 100
asec for the chosen laser parameters: the ionization window is
wider for the photoelectron model. As for the imaginary component
of the ionization times, they are smaller in the photoelectron
model. Therefore, short trajectories are less suppressed in this
model than in the full Lewenstein model.

Mathematically, implementing the photoelectron model requires only
one approximation - relaxing the return condition. Note that the
requirement of perfect return to the origin is an {\it artefact}
of neglecting the size of the ground state in the saddle point
analysis. If we take into account the size of the ground state,
then the return condition will naturally be relaxed: to be able to
recombine, the electron has to return to the core within the size
of the ground state. From this perspective, the extension of the
Lewenstein model to real systems, including molecules, should go
hand in hand with relaxing the return condition for its real part.

Measurement of ionization times might allow one to differentiate
between these two models and to pin down the nature of the electron
trajectories responsible for HHG. In particular, the interesting question is wether the complex momenta
 are the artefac of the $\delta$-like initial state, or are indeed relevant for realistic systems.

The accuracy of the first measurement \cite{Shafir12} was sufficient to distinguish
between $t_B$ and $t'_i$, but not high enough to distinguish delays between
 $t'_i$ (red curve, Fig.~\ref{fig:iontimesall}, left panel) and $t'_{i,ph}$ (green curve, Fig.~\ref{fig:iontimesall} , left panel).

\section{The multi-channel model of HHG: Tackling the multi-electron systems.}

In multielectron systems, there are multiple ways of energy
sharing between the liberated electron and the ion.
The ion can be left in its ground or in one of its excited electronic
states. These options are referred to as different ionization
channels. Multiple ionization channels lead to multiple HHG
channels: the returning electron can recombine with the ion in its
ground or in one of its excited states.

Multiple HHG channels present different pathways connecting the
same initial and final state - the ground state of the neutral
system - via different intermediate electronic states of the ion. Thus, high
harmonic emission in multielectron systems results from
multichannel interference, see \cite{Smirnova09}, i.e. the interference
of the harmonic light emitted in each channel. This interference
naturally records multielectron dynamics excited upon ionization
and probed by recombination, see \cite{Smirnova09}. How important are
these multiple channels? How hard is it to excite the ion during
strong-field ionization?

Strong-field ionization is exponentially sensitive to the
ionization potential $I_p$, suggesting that after ionization the
molecular ion is typically left in its ground electronic state. In
the Hartree-Fock picture, this corresponds to electron removal
from the highest occupied molecular orbital (HOMO). However,
multiple ionization channels can be very important in molecules
due to the geometry of the molecular orbitals and the proximity of
the excited electronic states in the ion to the ground state.

The formalism described above, in the sections 1.1-1.7, is essentially a
single-channel picture of HHG. It can be extended to multiple
channels.

First, we introduce the Hamiltonian of an N-electron neutral
molecule interacting with a laser field:
 \bea
 \label{eq:H^N}
 &&H^{N}=T^{N}_{e} +V^{N}_C+V^{N}_{ee}+V^{N}_L,
\nonumber
\eea
\bea
 &&V^{N}_C=-\sum_m\sum_{i=1}^{N}\frac{Q_m}{|\R_{m}-\rmbf_{i}|},
 \nonumber
 \eea
 \bea
 &&V^{N}_{ee}=\sum_{i \ne j}^{N}\frac{1}{|\rmbf_{i}-\rmbf_{j}|},
  \nonumber 
  \eea
  \bea
&&V^{N}_{L}=-\sum_{i} {\bf F}(t) \cdot \d_{i}= \sum_{i} {\bf F}(t)
\cdot \rmbf_{i}.
 \eea
Here, the nuclei are frozen at their equilibrium positions $\R_m$,
the index $m$ enumerates the nuclei with charges $Q_m$, the
superscript $N$ indicates the number of electrons involved,
$T^{N}_{e}$ is the electron kinetic energy operator, $V^{N}_C$
describes the Coulomb potential of the nuclei, $V^{N}_{ee}$
describes the electron-electron interaction, and $V^{N}_{L}$
describes the interaction with the laser field. Hats on top of
operators are omitted.

We will also use the Hamiltonian of the ion in the laser field,
$H^{(N-1)}$, and the Hamiltonian of an electron interacting
with the laser field, the nuclei, and the $(N-1)$ electrons of the
ion, $H^{e}=H^{N}-H^{(N-1)}$.

The Schr\"odinger equation for the N-electron wavefunction of the
molecule, initially in its ground electronic state
$\Psi^{N}_g(\rmbf)$, is
    \bea
  &&i \frac{\partial}{\partial t}\Psi^{N}(\rmbf,t)=H^{N}\Psi^{N}(\rmbf,t),
 \nonumber \\
 && \Psi^{N}(\rmbf,t=t_0)=\Psi^{N}_g(\rmbf).  
   \label{eq:Schroedinger}
    \eea
Similar to the single-electron case, its exact solution can be
written as
 \bea
\hspace*{-0.2cm}
| \Psi^{N}(t)\rangle= -i \int_{t_0}^{t} dt'\,\emph{U}^{N}(t,t')V^{N}_{L}(t')
 \emph{U}^{N}_0(t',t_0)\Psi^{N}_g(\rmbf)+\emph{U}^{N}_0(t,t_0)|\Psi^{N}_g\rangle.
      \label{eq:integralm}
\eea
Here the $U^{N}_0$ and $U^{N}$ are the N-electron propagators. The
former is determined by
 \bea
 &&i\partial U^{N}_0/\partial t=H^{N}_0U^{N}_0,\\
 &&U^{N}(t_0,t_0)=1,
\eea
where $H^{N}_0$ is the field-free Hamiltonian of the molecule:
$H^{N}_0=H^{N}-V^{N}_{L}$. The latter is the full
propagator determined by $i\partial U^{N}/\partial t=H^{N}U^{N}$.

The harmonic dipole reads
\bea
 \label{eq:dip_mult_0}
\nonumber
 \D(t)&=&-i \langle \emph{U}^{N}_0(t,t_0)\Psi^{N}_g(\rmbf)|\d|
 \int_{t_0}^{t} dt'\emph{U}^{N}(t,t')\times \\ &&\times V^{N}_{L}(t')\emph{U}^{N}_0(t',t_0)\Psi^{N}_g(\rmbf)\rangle +c.c.
\eea

Just as in the one-electron case (Eq.~\ref{eq:laser_free_prop}), propagation without the laser
field is simple as long as the energy $E_g$ and the wavefunction
of the initial state of the neutral molecule or atom are known:
 \bea
  &&\emph{U}^{N}_0(t',t_0)\Psi^{N}_g(\rmbf)=e^{-iE_g(t'-t_0)}\Psi^{N}_g(\rmbf).
  \eea
Finding the full propagator $\emph{U}^{N}(t,t')$ is just as hard
as solving the multi-electron TDSE.

To simplify the analysis, we will make the following two
approximations. First, we shall neglect the correlations between the
electrons in the ion and the liberated electron after ionization.
In this case, the full propagator factorizes into two independent
parts describing the evolution of the continuum electron and the
evolution of the ion in the laser field between ionization and
recombination: $\emph{U}^{N}(t,t')\simeq
\emph{U}^{(N-1)}(t,t')\emph{U}^{e}(t,t')$. Second,  we will  derive the results for short range potentials, just
like we did in the single electron case considered above:
$\emph{U}^{e}(t,t')\simeq \emph{U}^{e}_V(t,t')$, and supply the corrections due to Coulomb effects.

One can improve upon these two approximations  by including the
electron-electron correlations during ionization perturbatively,~\cite{Walters10},
and by using the eikonal-Volkov states,~\cite{Smirnova08},
for the continuum electron, instead of the plane wave Volkov states.
The eikonal-Volkov states include the laser
field fully, the interaction of the continuum electron with the
core in the eikonal approximation, and also take into account the
interplay between these two interactions (the so-called
Coulomb-laser coupling, \cite{Smirnova07}).

A consistent approach, which includes both
electron-electron correlations and long-range effects in strong field ionization can be developed within the time-dependent analytical
R-matrix (ARM) method (\cite{lisa12b}). This method (i) splits the configuration space into the
inner and outer region, uses quantum chemistry in the inner region, (ii) the eikonal-Volkov propagation in the outer region,(~\cite{Smirnova08}), and (iii) the Bloch operator (\cite{Bloch}) to match the solutions in two regions.

Moreover, if we can
factorize the dipole response into the usual steps -- ionization,
propagation, recombination, we can think of improving each of the
three steps separately, e.g. by using improved ionization and
recombination amplitudes that include the electron-electron
correlation beyond the perturbation theory.

Just like in the one-electron formalism considered above, we will
introduce the identity resolved on the momentum states of the continuum
electron, but now we also have to include the electronic states of
the ion \footnote{Here we use the field-free states of the ion.
If the limited amount of basis states is used, then one should try to find the optimal "laser-dressed" basis.}:
\be
I=\int d\p\, \sum_{n} \mathbb{A}|n^{(N-1)}\otimes \p^{n}_t \ra
\la n^{(N-1)}\otimes \p^{n}_t|\mathbb{A},
\label{eq:identity}
\ee
where $\mathbb{A}$ denotes the antisymmetrizing operator.

The harmonic dipole becomes
\bea
\label{eq:dip_mult}
\D(t)&=& -i \langle \Psi^{N}_g|\d| \int_{t_0}^{t} dt'\, e^{iE_g (t-t')}\int d\p\,\emph{U}^{(N-1)}(t,t')|{n}^{(N-1)}\ra \times \nonumber\\
&& \times \,
\emph{U}^{e}_V(t,t')|\p^{n}_{t'}\ra
 \la \p^{n}_{t'}{n}^{(N-1)}| V^{N}_{L}(t')|\Psi^{N}_g\rangle +c.c.
\eea

Note a crucial change compared to the single-channel case (Eq.~\ref{eq:dipV2}): the
appearance of the laser-induced dynamics between the bound states
of the ion, described by the propagator
$\emph{U}^{(N-1)}(t,t')|{n}^{(N-1)}\rangle$. This dynamics
can be calculated if the dipole couplings, $d_{mn}$, between all
essential states, as well as  their eigenenergies, $E_n$, are
known.

\begin{figure}
\includegraphics[width=1\textwidth]{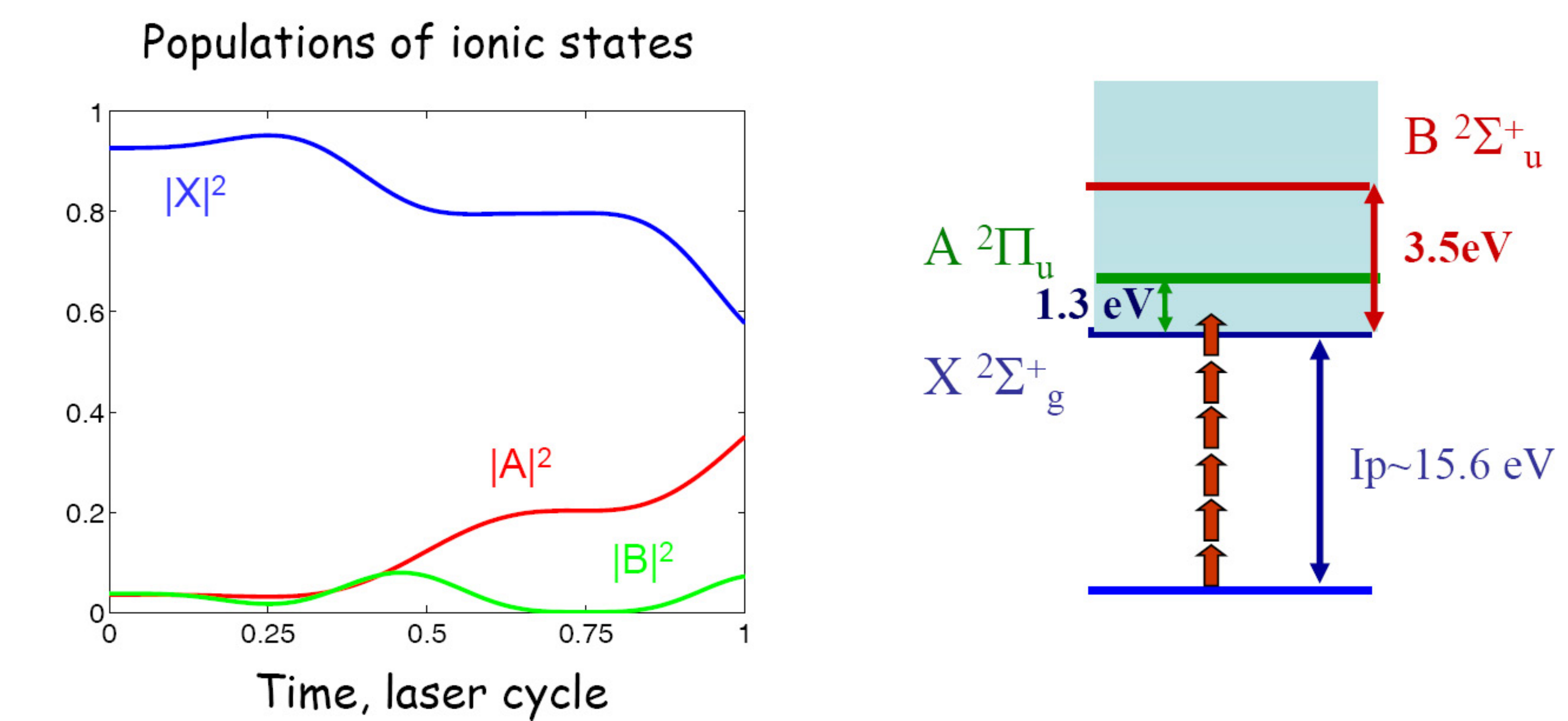}
\caption{Left panel: Sub-cycle dynamics in the N$_2^+$ ion aligned
at $\theta=50^o$ to the laser field polarization: populations of
the field-free ionic states X (blue), A (red), and B (green) in a
$I=0.8 \cdot10^{14}$~W/cm$^2$, 800 nm laser field.
 Right panel: Electronic states of the N$_2^+$ ion.} \label{fig:nadyn}
\end{figure}

Consider, for example, the case of an N$_2$ molecule with three
essential states in the N$_2^+$ ion, denoted as $X$, $A$ and $B$,
see Fig.\ref{fig:nadyn}.
The time-dependent transition amplitudes
$a_{mn}(t,t')$ between the state $n^{(N-1)}$ populated at the moment $t'$
and the state $m^{(N-1)}$ at the moment $t$ are given by $a_{mn}(t,t')=\la
{m}^{(N-1)}|\emph{U}^{(N-1)}(t,t')|{n}^{(N-1)}\rangle$.
It is  a solution of the following system of differential equations:
    \bea
    \frac{d(\mathbf{A}^{(n)})}{dt}=\left[\mathbf{H}+\mathbf{V}(t)\right]\mathbf{A}^{(n)},
    \eea
where, for our three ionic states, the Hamiltonian of the ion is
$\mathbf{H}= \begin{pmatrix}
                    E_1 & 0 & 0 \\
                    0 & E_2 & 0 \\
                    0 & 0 & E_3
                   \end{pmatrix}
$, with the energies  $E_n$ of the three states.  The interaction
between these three states is described by the matrix of the
laser-induced couplings, $V_{mn}(t)=-\d_{mn}\cdot{\bf F}(t)$, that is
$\mathbf{V}(t) =
 \begin{pmatrix}
0 & V_{12}(t) & V_{13}(t) \\
V_{21}(t) & 0 & V_{23}(t) \\
V_{31}(t) & V_{32}(t) & 0
 \end{pmatrix}$.
Finally, $\mathbf{A}^{(n)}=\begin{pmatrix}
a_{1n}(t,t') \\
a_{2n}(t,t') \\
a_{3n}(t,t')
\end{pmatrix}$
is the vector describing the population amplitudes of all
essential ionic states, starting from the state  $n^{(N-1)}$ at time $t'$.

Let us introduce channel specific Dyson orbitals
$\Psi^{D}_n(\rmbf)\equiv\la
{n}^{(N-1)}|\Psi^{N}_g(\rmbf)\rangle$.
 These are the
overlaps between the $N$-electron wavefunction of the ground state
of the neutral and the $(N-1)$-electron wavefunction of the ionic
state $|n^{(N-1)}\ra$. Let us assume that the dipole operator that starts
ionization at the moment $t'$ in Eq.~(\ref{eq:dip_mult}) acts only
on the  electron, that will be liberated (i.e. we neglect the exchange-like effects in ionization).
 In this case, the multielectron dipole
$\D_{mn}$, which corresponds to leaving the ion in the state $n^{(N-1)}$
after ionization and then recombination with the ion in the state
$m^{(N-1)}$, can be re-written in a form very similar to the
one-electron case (Eq.~(\ref{eq:dipV3_new})):
    \bea
&&\D^{(mn)}(t)= i \int_{t_0}^{t} dt'\int d\p \, \d_m^{*}(\p+\A(t))\,a_{mn}(t,t')\,e^{-iS_n(\p,t,t')}\times \nonumber \\
&&\hspace*{1.5cm}\times {\bf F}(t')\d_n(\p+\A(t')),
   \eea
   \bea
    \label{eq:dionm}
  &&  \d_n(\p+\A(t))=\langle \p+\A(t)|\d|\Psi^{D}_n\rangle,
    \nonumber 
    \eea
    \bea
    \label{eq:drecm}
  &&  \d_m(\p+\A(t))=\langle \p+\A(t)|\langle n^{(N-1)}|\d|\Psi^{N}_g\rangle,
    \nonumber \\
  &&  S_n(\p,t,t')=\frac{1}{2}\int_{t'}^{t} [\p+\A(\tau)]^2 d\tau+I_{p,n}(t-t'). \nonumber
    \label{eq:Sactionm}
    \eea
This expression is remarkably similar to one-electron dipole (\ref{eq:dipV3_new}).
The transformation similar to (\ref{eq:grib}) is also valid in this case, yielding
 \bea
    \label{eq:dipmV31}
&&\D^{(mn)}(t)= i \int_{0}^{t} dt'\int d\p \, \d_m^{*}(\p+\A(t))\,a_{mn}(t,t')\,e^{-iS(\p,t,t')}\times \nonumber \\
&&\hspace*{1.5cm}\times \Upsilon_n(\p+\A(t')),\\
    \nonumber
  &&  \Upsilon_n(\p+\A(t))=\left[\frac{\left[\p+\A(t) \right]^2}{2} +I_{p,n}\right] \langle \p+\A(t)|\Psi^{D}_n\rangle,
    \label{eq:Y_n}
    \eea
where $I_{p,n}$ is the ionization potential to the state $n$ of the ion and the matrix $a_{mn}(t,t')$
 is calculated while setting $E_n$ to zero.

\begin{figure}
\includegraphics[width=1\textwidth]{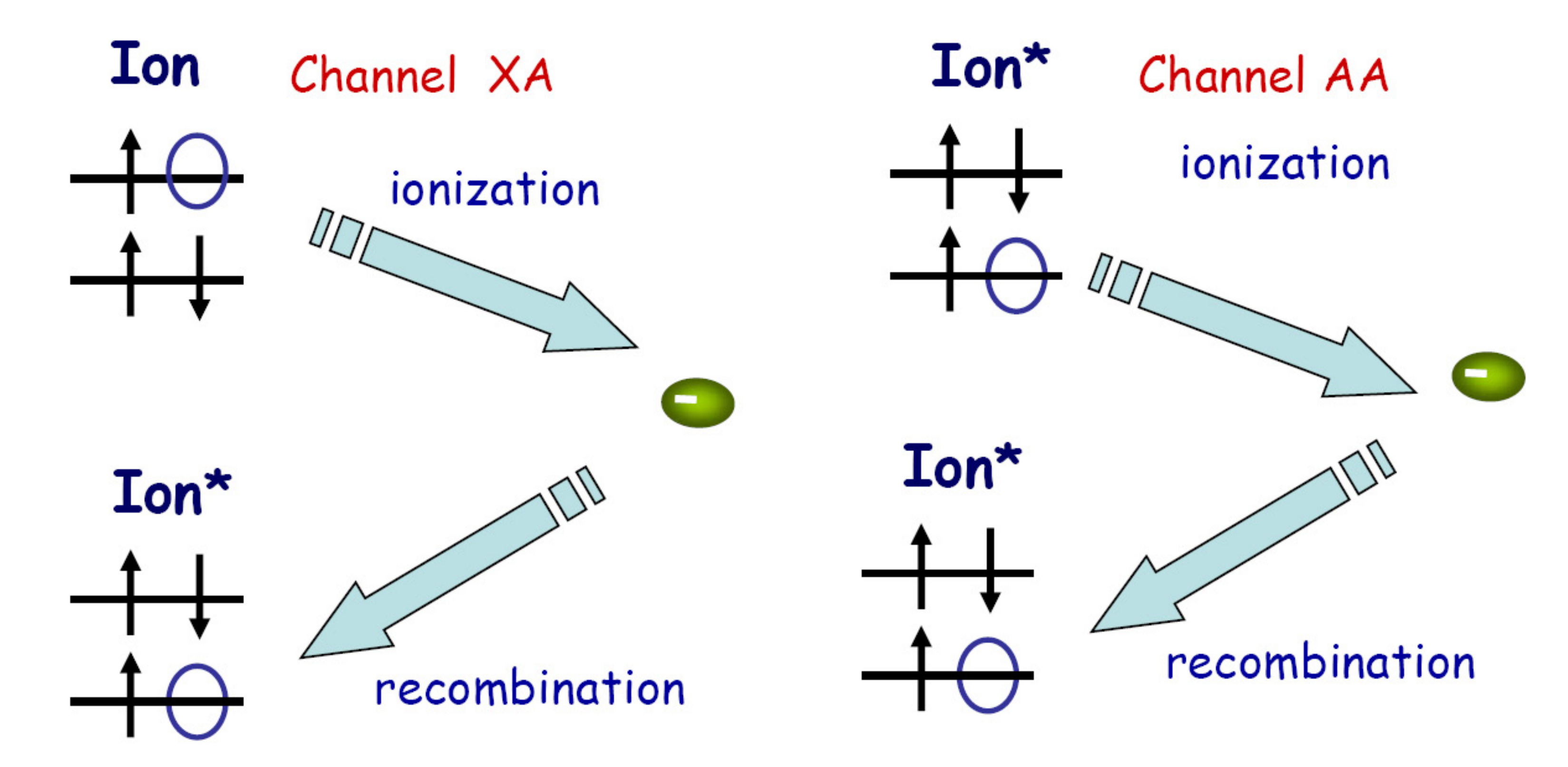}
\caption{Left panel: Cross-channel in HHG associated with ionization from and
recombination to different orbitals. This channel is due to real
excitations induced by the laser field between ionization and
recombination. Right panel: Diagonal channel in HHG associated with
ionization from and recombination to the same orbital.} \label{fig:hole}
\end{figure}

The total harmonic signal results from the coherent superposition of
the dipoles $\D_{mn}$ associated with each ionization-recombination
channel:
    \bea
    \label{eq:dipmV3t}
    &&\D(t)= \sum_{m,n} \D^{(mn)}(t).
    \eea
Substantial sub-cycle transitions, such as those shown in
Fig.~\ref{fig:nadyn} for the N$_2^+$ ion in typical experimental
conditions, have a crucial impact on the harmonic radiation. They lead
to the appearance of the cross-channels in HHG (the off-diagonal
elements $\D^{(mn)}$ for $m \neq n$ in Eq.~(\ref{eq:dipmV3t})) since the state
of the ion changes between the ionization and the recombination, see
Fig.~\ref{fig:hole}.
These channels are indeed substantial in high
harmonic generation from the N$_2$ molecules, see \cite{Mairesse10}, as
illustrated in Fig.~\ref{fig:nadyn}.

In the recent literature on high harmonic generation one can often
come across a rather loose language, which refers to different
ionization and recombination channels as associated with different
Hartree-Fock molecular orbitals. This language  should not be
taken literally as a statement on the applicability of the
Hartree-Fock picture and on the physical reality of the
Hartree-Fock orbitals as observable physical quantities.  Loosely
speaking, removing an electron from the highest occupied molecular
orbital (HOMO) creates the ion in the ground state. Removing an
electron from one of the lower lying orbitals (e.g. HOMO-1,
HOMO-2) creates the ion in one of its excited states. Thus, the reference
to the orbitals should only be understood as a language for describing
ionization and recombination channels associated with different
multielectron states of the ion -- and those are physically
relevant and observable. In the orbital language, electron removal
from an orbital creates a hole in this orbital. The laser induced
dynamics in the ion, moves the hole between the orbitals in the
time window between ionization and recombination, see
Fig.~\ref{fig:hole}.

Application of the saddle point method in each channel leads to the following half-cycle dipole
for the given ionization -- recombination channel:
 \bea
 \label{eq:hc_dipmult}
  &&\D^{(j,mn)}(t)= \mathbf{a}^m_{rec}(\p_s,t)a^{mn}_{prop}(\p_s,t,t_i^{(j)})a^{n}_{ion}(\p_s,t_i^{(j)}),\\
  \label{eq:hc_ionmult}
  &&a^{n}_{ion}(\p_s,t_i)=\left[\frac{2\pi}{S''_{t_i,t_i}}\right]^{1/2}e^{-iS(\p_s,t_i^{'(j)},t_i^{(j)})}\Upsilon_n(\p_s+\A(t_i^{(j)})),\\
  &&a^{mn}_{prop}(\p_s,t,t_i)=\frac{(2\pi)^{3/2}}{(t-t_i)^{3/2}}e^{-iS(\p_s,t,t_i^{'(j)})} a_{mn}(t,t_i^{(j)}),\\
  &&\mathbf{a}^m_{rec}=\d_m^{*}(\p_s+\A(t)).
\eea
Neglecting the sub-cycle dynamics in the prefactor of $a^{n}_{ion}(\p_s,t_i)$ we can substitute Eq. (\ref{eq:hc_ionmult}) by the following expression, which includes the Coulomb effects in ionization (the sub-cycle Coulomb effects see \cite{Lisa12}):
\bea
 \label{eq:hc_ionmult1}
  &&a^{n}_{ion}(\p_s,t_i)=2\!\left(\!\frac{2\kappa^3}{F}\!\right)^{\frac{Q}{\kappa}}\!\! \sqrt{\frac{-\tilde{\gamma} \omega}{\kappa\sqrt{1+\tilde{\gamma}^2}}}e^{-iS(\p_s,t'_i,t_i)}\Upsilon_n(\p_s\!+\!\A(t_i)).
\eea
A term similar to $\Upsilon_n$ also arises within the time-dependent analytical
 R-matrix approach applied to multi-channel strong field ionization \cite{lisa12b}.
However in \cite{lisa12b}, the radial integration is removed due to the use of the Bloch operator.
Thus in \cite{lisa12b}, the pole in $\Upsilon_n$ does not arise even when the long-range potential is taken into account.
Function $\Upsilon_n(\p_s+\A(t_i))$ encodes the angular structure of the Dyson orbital in the asymptotic region,
which is more complex than the one arising in the asymptotic of the atomic wave-function ( \ref{eq:asymp_wf_l}) leading to Eq. (\ref{eq:pole_Ylm}).
The simple expressions for the asymptotic of the Dyson orbital for small molecules can be found in \cite{Murray11}.

Here, we have considered the harmonic dipole on the real time axis. Note
that the propagation amplitude is modified
compared to the one in the one-electron case (Eq.~\ref{eq:aprop})
to include the laser
-- induced dynamics in the ion $a_{mn}(t)$. The full dipole for
each ionization-recombination channel is the sum over the
different half-cycles and the harmonic spectrum results from the
FFT of the full dipole $\D_{mn}(N\omega)$:
\bea
    \label{eq:sum_dip_multi}
    &&\D^{(mn)}(t)= \sum_j \D^{(j,mn)}(t),\\
    &&\D^{(mn)}(N\omega)= \int dt \,\D^{(mn)}(t) e^{iN\omega t}.
    \eea
The complete harmonic response is obtained by adding coherently the
contributions of all ionization -- recombination channels.

\section{Outlook}

Having factorized the dipole, we can use improved amplitudes for
each step. These are the key components of the current theoretical
work in high harmonic spectroscopy of molecules.

\textbf{Improving ionization}. Improved ionization amplitudes can
be taken from semi-analytical and/or numerical approaches. The task is to define
the function $\Upsilon(\p+\A(t))$ for a realistic system and include long-range and polarization (\cite{Suren}) effects.
For
example, one can use the results of \cite{Murray11}, where the
ionization amplitude is represented as:
    \bea
    \label{eq:aion_imp}
   \widetilde{a}^{n}_{ion}(\p_s,t_i)=\mathbb{R}_{\mathfrak{lm}}(I_p,F) e^{-iS(\p_s,t'_i,t_i)}.
    \eea
The exponent describes the sub-cycle dynamics of strong-field
ionization, i.e. is the same as for the atomic case and a
short-range potential. The prefactor
$\mathbb{R}_{\mathfrak{lm}}(I_p,F)$ accounts for the influence of
the core potential and the shape of the initial state on the
ionization rate. For atoms, this prefactor has been derived in
the seminal papers of Perelomov, Popov and Terent'ev (see
\cite{PPT1,PPT2,PPT3,PPT4}) and improved in \cite{Popruzhenko}.
A simple recipe for incorporating
their results into the sub-cycle ionization amplitudes can be
found in \cite{YudinIvanov01}.
Fully consistent treatment of long-range, polarization  and orbital effects can be
developed within the time-dependent analytical R-matrix (ARM) approach,
see \cite{Lisa12,lisa12b}.

\textbf{Improving propagation}. In addition to the dynamics in the
molecular ion, including the laser-induced transitions between
different ionic states, the second most important modification of
the propagation amplitudes is the incorporation of possible
transverse nodal structure in the continuum wavepackets. The nodal
planes in the continuum wavepacket arise during tunnelling from
bound states. For example, in the CO$_2$ molecule, the HOMO and the
corresponding Dyson orbital have nodal planes parallel and
perpendicular to the molecular axis. Consequently, in the case of
tunnel ionization with the molecular axis aligned parallel to the
polarization of the ionizing field, the nodal plane will not only
reduce the ionization rate, but will also be imprinted on the
shape of the electronic wavepacket that emerges after ionization.
Propagation between ionization and recombination will lead to the
spreading of the wavepacket, but it will not remove the presence
of the node as the wavepacket returns to the core
\cite{Smirnova09,Smirnova09a,Smirnova09b}. Clearly, this
aspect of propagation is important for the recombination
amplitude.

Consider, for example, ionization from a state with angular
momentum $L=1$. Its projection on the laser polarization is either
$L_z=0$ (no nodal plane along the electric field) or $L_z=1$
(nodal plane along the electric field). After tunnelling, in the
plane orthogonal to the laser polarization, in the momentum space the
continuum wavepackets are proportional to $\left( \frac{p_{\perp}}{\kappa}\right)^m$, (see Eq.\ref{eq:aion_Shsub} or \ref{eq:aion_Cs}):
    \bea
    \label{eq:wavepackets1}
    &&\Psi_{L_z=0}(p_{\perp})\propto e^{-\frac{p_{\perp}^2}{2}\tau},
    \nonumber \\
    &&\Psi_{L_z=1}(p_{\perp})\propto
    \frac{p_{\perp}}{\kappa} e^{i\phi_p} e^{-\frac{p_{\perp}^2}{2}\tau},
    \eea
where $\tau={\rm Im}(t_i)$, $\kappa=\sqrt{2I_p}$, and $\phi_p$ is
the angle between $p_{\perp}$ and the $x$-axis. As we can see, the
presence of the nodal plane for $L_z=1$ leads to the additional term
$p_{\perp}/\kappa$. We now propagate these wavepackets until the
recombination time $t_r$. Fourier transforming back into the
coordinate space, in the plane orthogonal to the laser polarization, we
get
    \bea
    \label{eq:wavepackets2}
    &&\Psi_{L_z=0}(\rho)\propto e^{-\frac{\rho^2}{2(t_r-t_i)}}
    \nonumber \\
    &&\Psi_{L_z=1}(\rho)\propto
    \frac{\rho}{\kappa (t_r-t_i)}
    e^{i\phi}e^{-\frac{\rho^2}{2(t_r-t_i)}}
    \eea
where $\rho$ is the transverse radial coordinate and $\phi$ is
the angle between the radial vector and the x-axis. Recalling that
$x=\rho\cos\phi$, we see that if we combine the $L_z=\pm 1$ states
to form the real-valued spherical harmonic $p_x$, the presence of
the nodal plane effectively changes the dipole operator ${\bf d}$
to ${\bf d}\cdot x/(\kappa (t_r-t_i))$. In \cite{Smirnova09,Smirnova09a,Smirnova09b}
 such modifications of recombination operators has been used to account for the appearance of nodal planes.

In most experiments with molecular HHG to-date, the alignment
distribution is rather broad. Even if the molecular ensemble is,
on average, aligned parallel to the laser polarization, for most
molecules the characteristic alignment angle would be sufficiently
different from that associated with the nodal plane. In this case,
the relative importance of the nodal planes in recombination is
reduced. However, for well-aligned molecular ensembles this would
become a significant factor.

\textbf{Improving recombination}.   The recombination step can be
significantly improved beyond the SFA, if one uses the
recombination dipoles $\d_m^{*}(\p_s+\A(t))$ calculated using
ab-initio approaches. For example, the quantitative rescattering
theory (see \cite{Lin10}) relies on using the Schwinger variational
method  to calculate the field-free recombination matrix elements.
Alternatively, one can use the R-matrix approach (see \cite{Harvey09,Alex}).
Both allow one to incorporate the full complexity of the recombination process,
including the channel coupling due to the electron-electron
correlation and automatically include the exchange effects in recombination \cite{santra, patchkovskii,suren}. The drawback of these methods, at the moment, is the
absence of the laser field in the calculations of the
recombination amplitudes. This approximation  breaks down in case of
a structured continuum \cite{serguei}, common for many molecules.
The impact of the IR field on such continuum states has been recently demonstrated
experimentally \cite{ott}, substantiating the prediction of \cite{serguei}.
In the approach described by
\cite {Smirnova09},  the eikonal-Volkov
approximation for the continuum states was used to obtain improved dipoles
in the single-channel approximation with exchange. The eikonal-Volkov approximation
fully includes the interaction of the continuum electron with the laser field,
but the interaction with the core potential is only
included in the eikonal approximation, and the correlation-induced
channel coupling is neglected. Improving the recombination amplitudes
to account for all these effects -- the channel coupling due to the
electron-electron correlation, the core potential, and the laser
field, is one of the key theoretical challenges today.

With each of the three steps in the harmonic response improved,
the original SFA-based theory turns from purely qualitative into a little more realistic.
The separation of the three steps, crucial for our ability
to improve each of them separately, benefits from the high intensity
of the driving field and the large oscillation amplitude of the active
electron. The high field intensity also lies at the heart of the main
difficulties in building an adequate theoretical description.
Nevertheless, the effort is worth the investment: the combination
of attosecond temporal and Angstrom spatial resolution is
extremely valuable. High harmonic spectroscopy appears to be well
suited for tracking the multielectron dynamics induced by the
ionization process.

It is very attractive to replace the ionization step induced by
the IR field with the one-photon ionization induced by a
controlled attosecond XUV pulse, phase-locked to the strong IR field (see \cite{SchaferXUV}).
The latter would drive the continuum electron. Such an
arrangement should allow one to move from dealing with outer
valence electrons to dealing with inner valence and deeper lying
electrons. This appears to be an exciting regime for tracking the
hole dynamics (\cite{Kulef}) initiated by inner-valence or deeper ionization.
Importantly, for deeply bound orbitals, the effect of the IR
driving field on the core-rearrangement and the hole dynamics
should be substantially less than for the outer-valence electrons.

High harmonic spectroscopy helps to record the relative phases between different
ionization channels, which provide information about electron rearrangement during
ionization and define the initial conditions for the hole migration both in the inner valence
(\cite{Kulef}) and outer valence (\cite{Smirnova09}) shells.
These initial conditions are sensitive to the frequency, intensity and duration of
the ionizing pulse, opening opportunities for controlling hole migration and, possibly,
related chemical reactions \cite{Weinkauf}.

\section{Acknowledgements}
We are grateful to Maria Richter for reading the manuscript and suggesting many important corrections,
Szczepan Chelkowski and Thomas Schultz for useful comments, and Felipe Morales for his help in preparing the manuscript.
We thank Pascal Sali$\grave{e}$res and Alfred Maquet for encouraging us to undertake this project.
Finally, we acknowledge the stimulating atmosphere of the KITPC workshop
"Attosecond Science - Exploring and Controlling Matter on Its
Natural Time Scale" in Beijing.

\section{Appendix A: Supplementary derivations}
In this Section we prove that the transformation (\ref{eq:grib}):
\bea
    \label{eq:gribA}
    \hspace*{-1cm}
    && \int_{t_0}^{t}dt' \,e^{-iS(\p,t,t')}\,{\bf F}(t')\,\d(\p+\A(t'))=\\\nonumber &&=\int_{t_0}^{t}dt' \,e^{-iS(\p,t,t')}\Upsilon(\p+\A(t')) ,\\
    &&\Upsilon(\p+\A(t'))=\left[ \frac{(\p+\A(t'))^2}{2}+I_p \right]\langle \p+\A(t')|g\rangle,
     \label{eq:YA}
 \eea 
 is applicable in case of high harmonic generation. By definition
 \bea
    \label{eq:gribA1}
    \hspace*{-1cm}
    && \int_{t_0}^{t}dt' \,e^{-iS(\p,t,t')}\,{\bf F}(t')\,\d(\p+\A(t'))\equiv\\\nonumber &&\equiv e^{-iI_pt}\int_{t_0}^{t}dt' \langle \Psi^V_{\p}(t';t)|-V_L|g(t')\rangle.
 \eea 
 Adding and subtracting the kinetic energy operator $\frac{\widehat{\p^2}}{2}$ (\cite{Wilhelm}, \cite{Gribakin97}) we obtain:
\bea
    \label{eq:gribA2}
    \hspace*{-1cm}
    && \int_{t_0}^{t}dt' \langle \Psi^V_{\p}(t';t)|-\frac{\widehat{\p^2}}{2}-V_L+\frac{\widehat{\p^2}}{2}|g(t')\rangle\equiv
    \\\label{eq:gribA3}     
   \hspace*{-1cm}  && \equiv
   \int_{t_0}^{t}dt'\left\{ i\frac{\langle\partial \Psi^V_{\p}(t';t)|}{\partial t'}|g(t')\rangle+\left[ \frac{(\p+\A(t'))^2}{2} \right]\langle\Psi^V_{\p}(t';t)|g(t')\rangle\right\}.
 \eea 
 Here we have used that $-i\frac{\langle\partial \Psi^V_{\p}(t';t)|}{\partial t'}=\left[\frac{\widehat{\p^2}}{2}+V_L\right]\langle \Psi^V_{\p}(t';t)|$.
 Integrating by parts the first term in Eq.(\ref{eq:gribA3}) we obtain:
 \bea    
 \label{eq:gribA4}     
   \hspace*{-1cm}  &&
   \int_{t_0}^{t}dt' i\frac{\langle\partial \Psi^V_{\p}(t';t)|}{\partial t'}|g(t')\rangle=
   \\\label{eq:gribA5}     
   \hspace*{-1cm}  && =
   -i\int_{t_0}^{t}dt'\langle\Psi^V_{\p}(t';t)|}{\frac{\partial|g(t')\rangle}{\partial t'}+\langle\Psi^V_{\p}(t';t)|g(t')\rangle |_{t_0}^{t}.
   \eea 
   The transformation (\ref{eq:gribA}) can be recovered using $i\frac{\partial|g(t')\rangle}{\partial t'}=-I_p|g(t')\rangle$
    and taking into account that the boundary term $\langle\Psi^V_{\p}(t';t)|g(t')\rangle |_{t_0}^{t}$ does not contribute to the high harmonic dipole.    
    Indeed the contribution of the boundary term to high harmonic dipole is:
 \bea    
 \label{eq:gribA6} 
 \hspace*{-1cm}  &&  
    \int d\p\, \d ^{*}(\p+\A(t))\left[\langle\Psi^V_{\p}(t;t)|g(t)\rangle- \langle\Psi^V_{\p}(t_0;t)|g(t_0)\rangle  \right]=
   \\\label{eq:gribA7}     
   \hspace*{-1cm}  && =\int d\p\, \langle g(t)|\hat \d|\Psi^V_{\p}(t;t)\rangle \langle\Psi^V_{\p}(t;t)|g(t)\rangle-
   \\\label{eq:gribA8}     
   \hspace*{-1cm}  && -\int d\p\, \langle g(t)|\hat \d|\Psi^V_{\p}(t;t)\rangle \langle\Psi^V_{\p}(t_0;t)|g(t_0)\rangle.
   \eea 
  The term (\ref{eq:gribA7} ) is equal to zero, the term (\ref{eq:gribA8} ) tends to zero when $t_0\rightarrow-\infty$.
  Indeed, 
   \bea
      \hspace*{-1cm}  && \int d\p\, \langle g(t)|\hat \d|\Psi^V_{\p}(t;t)\rangle \langle\Psi^V_{\p}(t;t)|g(t)\rangle=\langle g(t)|\hat \d|g(t)\rangle=0,   
   \eea 
   while the second term (\ref{eq:gribA8} ) is:
   \bea    
   \label{eq:gribA10}     
   \hspace*{-1cm}  && \int d\p\, \langle g(t)|\hat \d|\Psi^V_{\p}(t;t)\rangle \langle\Psi^V_{\p}(0;t)|g(0)\rangle=
   \\\hspace*{-1cm}  && =
   \int d\p\, e^{-i\frac{1}{2}\int_{t_0}^t d\tau \left[\p+\A(\tau)\right]^2} \langle g(t)|\hat \d|\p+\A(t)\rangle \langle\p|g\rangle.
   \eea    
 This term corresponds to the projection of the ground state onto the basis of plane waves at $t_0\rightarrow-\infty$ followed by recombination of the resulting 
 oscillating wave-packet back to the ground state at time $t$. Spreading of the free electron wave-packet over infinite time $(t-t_0)\rightarrow+\infty$ makes this projection negligible.
\section{Appendix B: The saddle point method}
The saddle point method is one of the key techniques in the analytical
strong-field theory.
It is an asymptotic method, which allows one to analytically evaluate the
integrals from highly oscillating functions, such as the integral in Eq.~(\ref{eq:dipV3}).

 \subsection{Integrals on the real axis}

How would one calculate the following integral,
\begin{eqnarray}
    I=\int_a^b f(x)e^{\lambda h(x)} dx
    \label{eq:Lect1Eq1}
\end{eqnarray}
for some smooth functions $f(x)$ and $h(x)$, without knowing much
about them, or if they look ugly and complicated? All we know is
that they are real-valued functions on the real axis $x$.

In general, one could think that there is not much one can do.
Fortunately, this is not the case if the positive and real $\lambda$ is
large, $\lambda\gg 1$ -- then, the integral can be calculated.

\subsubsection{Contribution of the end points}

The first idea that comes to my mind when looking at such an integral
is to try integration by parts. This approach works just fine under
certain circumstances, see below. The first stumbling block meets
you right at the gate: how does one integrate by parts if both
$h(x)$ and $f(x)$ are unknown?

The trick is simple:
\begin{eqnarray}
I&=&\int_a^b f(x)e^{\lambda h(x)} dx
\nonumber \\
&=&\int_a^b dx \frac{f(x)}{\lambda h'(x)} \lambda h'(x)
e^{\lambda h(x)}
\nonumber \\
&=&\frac{1}{\lambda}\frac{f(x)}{h'(x)}  e^{\lambda
h(x)}|_a^b -\frac{1}{\lambda}\int_a^b dx e^{\lambda h(x)} \left[\frac{d}{dx}\left(\frac{f(x)}{h'(x)}\right)\right].
\label{eq:Lect1Eq2}
\end{eqnarray}
We  started with an integral that did not have a small
parameter $1/\lambda$ in front. Now we have two terms: the first
comes from the contributions at the end points. The second term is
another integral, now with a small parameter in front. Dealing
with it in the same way as with the original integral, we will get
terms proportional to $1/\lambda^2$, and so on.

Thus, we conclude that the main contribution to the integral comes
from the end points, and is given by the first term:
\begin{eqnarray}
    I&=&\int_a^b f(x)e^{\lambda h(x)} dx
       \nonumber \\
    &=& \frac{1}{\lambda}\left[\frac{f(b)}{h'(b)}  e^{\lambda
    h(b)}-\frac{f(a)}{h'(a)}  e^{\lambda
    h(a)}\right] +O(\lambda^{-2}).
    \label{eq:Lect1Eq3}
\end{eqnarray}
This result is applicable unless there is a  problem with the
second term in Eq.~(\ref{eq:Lect1Eq2}) -- the integral
\begin{eqnarray}
    -\frac{1}{\lambda}\int_a^b dx e^{\lambda h(x)} \left[\frac{d}{dx}\left(\frac{f(x)}{h'(x)}\right)\right].
    \label{eq:Lect1Eq3b}
\end{eqnarray}
The problem arises if $h'(x)=0$ somewhere between the two end
points of the integral. What do we do then? Obviously, the
points where $[f(x)/h'(x)]$ diverges can bring major
contributions to the integral.

Given that $\lambda\gg 1$,  the way the exponential function
changes between $a$ and $b$ is most important. The first
possibility is $h'\neq 0$ in the integration interval. Then, the
integral is accumulated at the end points, and the end point where
$h(x)$ is larger dominates. In general, for an exponential
function $e^{\lambda h(x)}$ the main contribution to the integral
will come from the region where it reaches its maximum value --
and hence where $h'(x)=0$.

Suppose that, somewhere between $a$ and $b$, the derivative $h'=0$.
If the function $h(x)$ has a minimum, the contribution of this
minimum won't be competitive with the contributions from the end
points (remember that $\lambda$ is large and positive). But if it
has a maximum, then the main contribution to the integral comes
from the region near the maximum. The way to handle this situation
is described in the next section.

\subsubsection{The Laplace Method}

Let us consider an integral from a function $f(x)$ shown in Fig.~\ref{Fig1-1}.
\begin{figure}
\includegraphics[width=3.5in,angle=0]{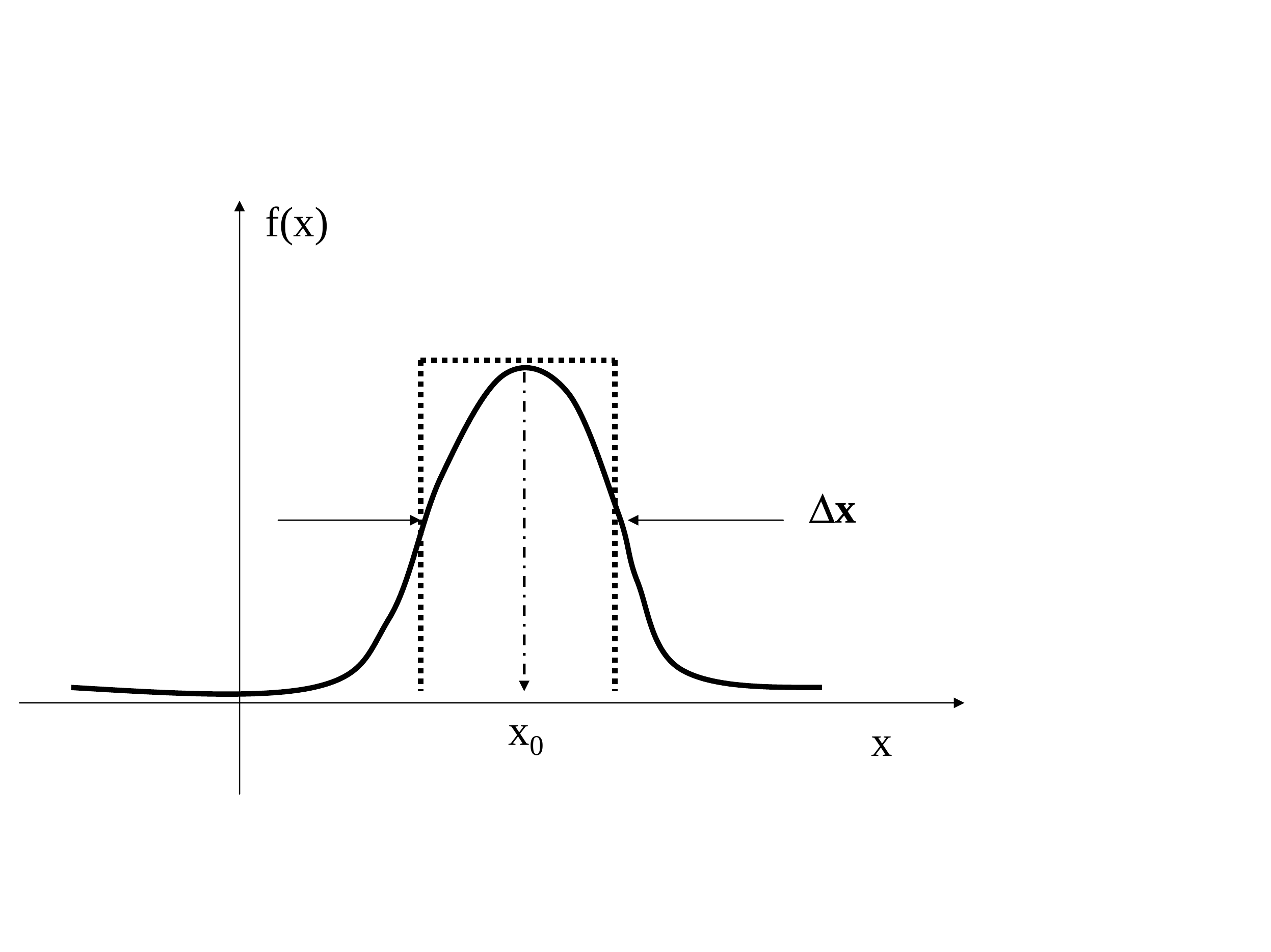}
\caption{Integral from a bell-shaped curve with a maximum at $x_0$ and a width of $\Delta x$.} \label{Fig1-1}
\end{figure}
The function is
bell-shaped, has a maximum at the point $x_0$, where its first
derivative is, of course, equal to zero, and quickly falls off to
each side of $x_0$.

Calculation of this integral is very simple - all we need
is to find the effective width $\Delta x$ of the bell-shaped
curve, and then the integral is
\begin{equation}
I=\int^{\infty}_{-\infty} f(x) dx=f(x_0) \Delta x.
\end{equation}
Let us first try some simple estimates of the width $\Delta x$. In
order to do it, we expand $f(x)$ around $x_0$ in a Taylor series,
remembering that the first derivative is zero at this point:
\begin{eqnarray}
f(x)\approx f(x_0)+\frac{1}{2} f''(x_0) (x-x_0)^2 = f(x_0)-\frac{1}{2} |f''(x_0)| (x-x_0)^2.
\end{eqnarray}
Notice that I have explicitly used the fact that the second
derivative at the local maximum is negative.

A potential candidate for the width $\Delta x$ is the full width
at half maximum (FWHM). The half-width $\Delta x/2$ at each side is
given by
\begin{equation}
f(x_0)-\frac{1}{2} |f''(x_0)| (\Delta x/2)^2=f(x_0)/2.
\end{equation}
This gives us $\Delta x=\sqrt{4 f_0/|f''|}$.

A more accurate calculation of the required width comes from the
following trick, which will also smoothly bring us into the saddle
point method
\begin{equation}
I=\int f(x) dx=\int e^{\ln f(x)} dx.
\end{equation}
This transformation allows us to reduce
the  integral to a familiar Gaussian form. We proceed by expanding
$\ln f(x)$ in a Taylor series, remembering that $f'(x_0)=0$ and
$f''(x_0)=-|f''(x_0)|$:
\begin{equation}
I=\int e^{\ln f(x)} dx\simeq \int e^{\ln
f(x_0)-\frac{|f''(x_0)|}{f(x_0)} \frac{(x-x_0)^2}{2}} dx.
\end{equation}
Recalling that the integral from a Gaussian is
\begin{equation}
\int^{\infty}_{-\infty} e^{-a x^2} dx= \sqrt{\pi /a},
\end{equation}
and setting the limits of our integral to $\pm \infty $, we get
the final answer
\begin{equation}
I=f(x_0)\sqrt{2\pi f(x_0)/|f''(x_0)|}.
\end{equation}

As you can see, the width $\Delta x$ turned out to be pretty close
to the FWHM.


\subsubsection{Saddle point method: the steepest descent in a complex plane}

We now move to the {\it saddle point method} which is used for
integrals of complex-valued functions:
\begin{equation}
I=\int _C e^{\lambda f(z)} dz.
\end{equation}
where $\lambda$ is large and positive, and the rest is hidden in
$f(z)$. The integral is to be taken over a contour C, and the only
good thing about this contour is that its ends, somewhere far away
from the place of action, do not contribute to the value of the
integral.

There are assumed to be no poles, so that we are allowed to deform the
contour C as we wish. The key of the steepest descent is a clever
modification of the integration contour.

First, note that a complex function $f(z)$ has a real part and an imaginary
part, $f(z)=u(z)+i v(z)\equiv u(x,y)+i v(x,y)$, where $x$ and $y$
are the real and the imaginary parts of $z$, $z=x+iy$.

Let us now look at the integral more closely and recall the
previous section, where the integration was based on expanding the
function around a maximum and reducing the integrand to a
Gaussian. In our case we have a function $\exp(\lambda u +
i\lambda v)$ that changes its absolute value very rapidly due
to the $\lambda u$ part. It also oscillates rapidly due to the
$\lambda v$ part. The trick of the steepest descent is to modify
the contour of integration in such a way that it will go through a
place where the real part $u$ climbs to a maximum along the
contour and then quickly falls, while the imaginary part $v$ stays
constant along the same contour, freezing any fast oscillations.

It may not be obvious at first glance that such a modification of
the contour is possible. But it is.

We start in a manner entirely analogous to the previous section.
Let us assume that the function $f(z)$ has a zero derivative at the
point $z_0={x_0,y_0}$, where $x_0$ and $y_0$ are coordinates in the
complex plane; the point $z_0$ lies somewhere between the left and
the right ends of the contour $C$.

If $f_z(z_0)=0$, then both the real and the imaginary parts of $f$
must have zero derivatives there:
\begin{equation}
u_{x}=v_{x}=0,  \ \ \ \ v_{y}=u_{y}=0.
\end{equation}
Thus, not only at $z_0$ the absolute magnitude of our function
goes through an extremum, but also the oscillating part is
stationary. Another important observation is that the gradients of
the two functions, $\nabla u$ and $\nabla v$, are always
orthogonal to each other:
\begin{equation}
{\nabla v} \cdot \nabla u=u_x v_x+ u_y v_y=0.
\end{equation}
This is the consequence of the Cauchy-Riemann conditions:
\begin{equation}
u_{x}=v_{y}, \ \ \ \ v_{x}=-u_{y}.
\end{equation}


The gradient points into the direction for which the function changes. If we
move along the gradient of $u$, following the path of its steepest
rise and fall through the point $z_0$, we are also moving
orthogonal to the gradient of $v$. Thus, $v$ will stay constant,
and fast oscillations are frozen. We see that the desired
modification of the contour is indeed possible.

How should the landscape of $u(x,y)$ look like? Due to the same
Cauchy-Riemann conditions the functions $u$ and $v$ can only have
{\it saddle points} at $z_0$:
\begin{equation}
u_{xx}+u_{yy}=0, \ \ \ \ v_{xx}+v_{yy}=0.
\end{equation}
Real mountain peaks, which go down in all directions, only happen
at singularities, and we decided that there would be no
singularities in $f(z)$.


 Therefore, the landscape of the function $u$ around the point $z_0$ must
look something like shown in Fig.~\ref{Fig1-2}.
\begin{figure}
\includegraphics[width=4.0in,angle=0]{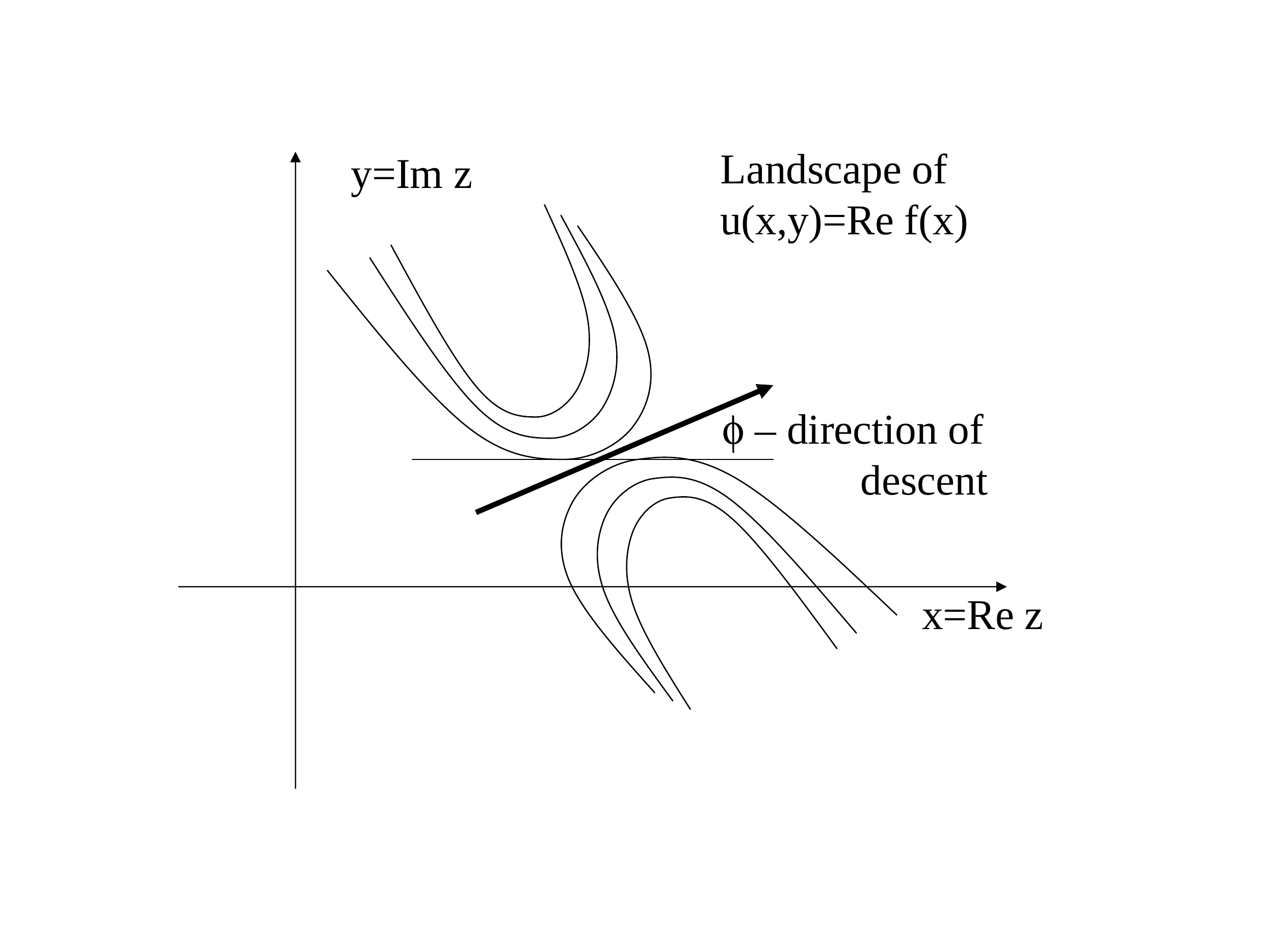}
\caption{Saddle point method. The landscape of $u(x,y)={\rm Re}
f(z)$ around the point $z_0$ where $f'=0$.}
\label{Fig1-2}
\end{figure}

All we have to do now is to find the correct path of the {\it
steepest descent through the saddle point}, such that $u$ will
rise as quickly as possible as we approach the saddle point and
then decrease as quickly as possible as we leave the saddle point.
The Cauchy-Riemann conditions promise us that, while we are at it,
$v$ will stay constant.

Let us expand $f(z)$ in a Taylor series around $z_0$, remembering
that $f'(z_0)=0$:
\begin{equation}
f(z)\approx  f(z_0)+\frac{1}{2}f''(z_0)(z-z_0)^2.
\end{equation}
Of course, $f''(z_0)$ is a complex number, which we will denote as
$f''(z_0)=\alpha \exp(i\theta)$. If our path traverses the saddle
point  at some angle $\phi$, then $z-z_0=\rho \exp(i\phi)$ and
\begin{equation}
\frac{1}{2}f''(z_0)(z-z_0)^2=\frac{1}{2}\alpha \rho^2
e^{i(\theta+2\phi)}.
\end{equation}
Now, the simple trick is to choose the angle $\phi$ properly -- we
set
\begin{equation}
e^{i(\theta+2\phi)}=-1
\end{equation}
and keep the angle $\phi$ given by the above condition fixed,
changing only $\rho$, so that $dz=d(z-z_0)=\exp(i\phi) d\rho$.

If we do this, the integral along such a path will look as
\begin{equation}
I=e^{\lambda f(z_0)} e^{i\phi} \int_{C'} e^{-\lambda \alpha
\frac{\rho^2}{2}} d\rho,
\end{equation}
where the deformed contour $C'$ is going through the saddle point
as a straight line at an angle $\phi$. Note that, indeed, there are
no oscillations along such path, and the real integrand decays as a
Gaussian.

The integration limits are now extended to plus and minus infinity
and the integral is done:
\begin{equation}
I=e^{\lambda f(z_0)} e^{i\phi} \sqrt{\frac{2\pi}{\lambda \alpha}}.
\end{equation}
Recall that $\alpha \equiv |f''(z_0)|$.

At this point we are almost done, but three important remarks are
still in order.

First, there is ambiguity in the definition of the direction
$\phi$ from  $\exp(i(\theta+2\phi))=-1 $. Indeed, the total angle
$\theta+2\phi$ could be both plus and minus $\pi$. Thus, formally,
without looking at the landscape shown in Fig.~\ref{Fig1-2}, we have a choice
of two $\phi$:
\begin{equation}
\phi_1 =-\theta /2 +\pi/2, \ \ \ \ \phi _2=-\theta /2 -\pi/2.
\end{equation}
The whole idea of the method is to choose such a direction that you
never have to cross the 'mountains' in the landscapes of $u(x,y)$
and $v(x,y)$. You should choose the direction (deforming the
contour $C$) that takes you from the valley, through the saddle
point, into another valley. Otherwise, you will also have to
include the contributions of the 'mountains' into the integral.
Usually, it is the first choice that works, but one should take a
look at the landscape and check. The wrong option will go in an
obviously wrong way, crossing into the tops of the mountains rather
than staying all the way in the valley and smoothly climbing to the
saddle. We shall see an example of it in the next section.

Second, if there are several saddle points, i.e. $f(z)$ has many
points where its derivative is zero, the integral will be the sum
of the contributions from all these points. Then the individual phases
$\phi$ for each saddle point become very important.

Third, there is a modification of the prefactor when dealing with multi-dimensional integrals:
\bea
&&I=\int_C e^{\lambda f(z)}dz,\\
&&I\simeq \left(\frac{2\pi}{\lambda}\right)^{n/2}e^{\lambda f(z_0)}\frac{1}{\sqrt{-f_{zz}(z_0)}},
\eea
where $f_{zz}$ is the Hessian matrix (the matrix of the second derivatives of the function f).

\subsection{Stationary phase method}

The stationary phase method is a simple application of the saddle
point method to a function with a purely imaginary phase:
\begin{equation}
I=\int g(x) e^{i\lambda f(x)} dx,
\end{equation}
where $g(x)$ is a benign, very slowly changing function which does
not do much - just makes sure that the integrand goes to zero at
infinity. The constant $\lambda$ is again real and positive, the
integral is supposed to be performed along the real axis and the
function $f(x)$ is purely real on the real axis. Intuitively, it
is clear that if the function $\exp(i\lambda f(x))$ is oscillating
very quickly, its integral averages to zero unless there are some
points where the oscillations freeze. These areas are the regions
where the phase of the oscillation, $f(x)$, stays nearly constant,
i.e. areas around the point where the derivative turns to zero,
$f'=0$, see Fig.~\ref{Fig1-3}.
\begin{figure}
\includegraphics[width=3.7in,angle=0]{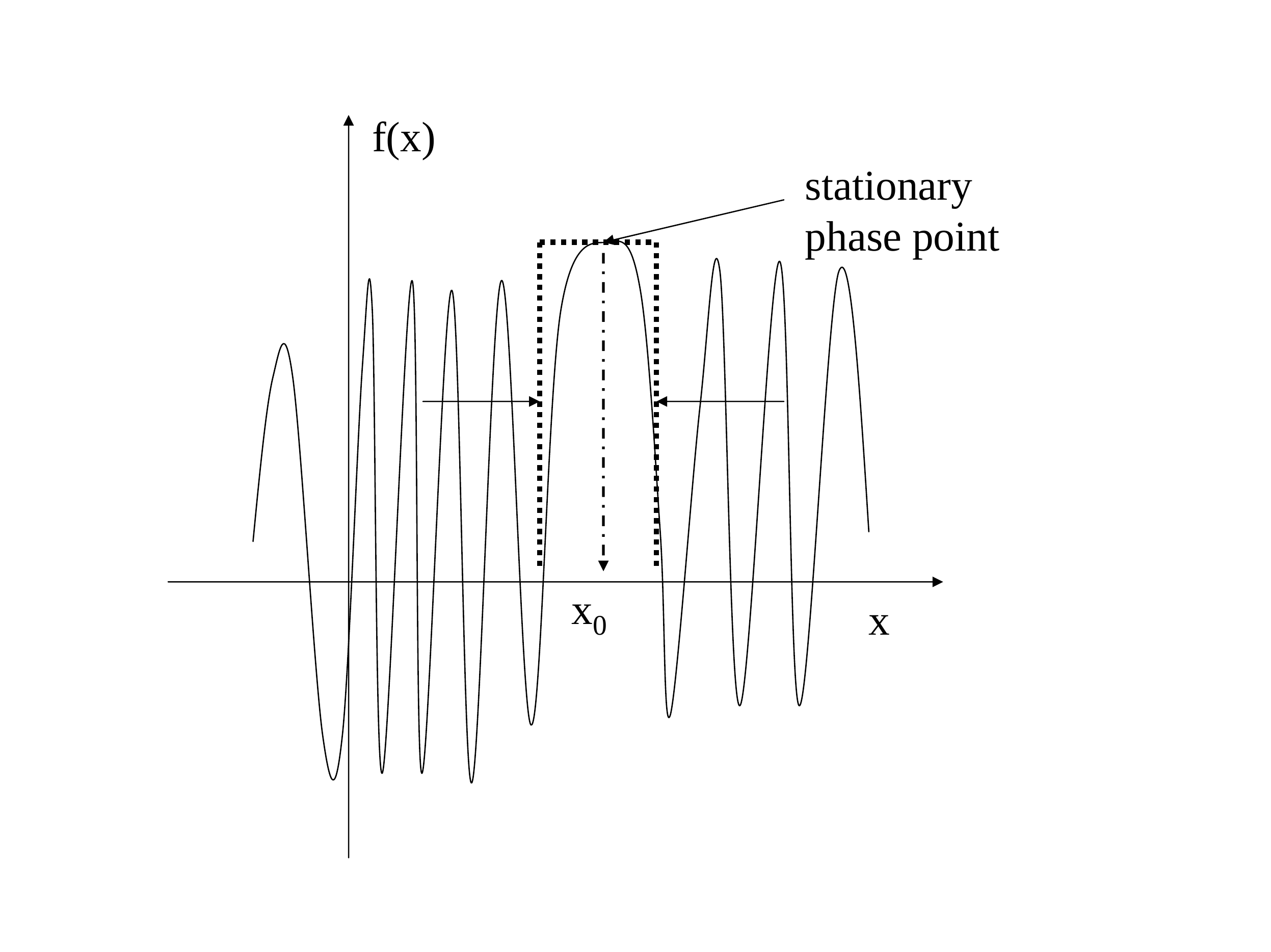}
\caption{Stationary phase method. The integral comes from the area
where the integrand does not oscillate as much.} \label{Fig1-3}
\end{figure}

The problem can be turned into that studied in
the previous section. Again, suppose that the derivative $f'=0$ at
some point $x_0$. We use the same Taylor expansion around this
point and denote $x-x_0$ as, say, $\xi$. The integral is
approximated as
\begin{equation}
I=g(x_0) e^{i\lambda f(x_0)} \int e^{i\lambda
f''(x_0)\frac{\xi^2}{2}} d\xi
\end{equation}
and we will assume that $f''=\alpha >0$ ($f$ is a real-valued
function and $x_0$ is on the real axis, hence $f''$ is real). The
case $f''<0$ is handled in an identical manner.

Calculation of the integral
\begin{equation}
\int e^{i\lambda \alpha \frac{\xi^2}{2}} d\xi
\end{equation}
follows the exact prescription of the saddle point method.
Obviously, the phase $\theta$ of the second derivative (see
previous section) is $\theta=\pi/2$ (i.e. $i\alpha=\alpha
\exp(i\pi/2)$) and the contour of integration has to be turned at
an angle $\phi$ to the real axis, such that $\theta +2\phi =\pm
\pi$. This yields the two possible choices of $\phi$:
$\phi=+\pi/4$ and $\phi=-3\pi/4$; the answer for the integral is:
\begin{equation}
I= g(x_0) e^{i\lambda f(x_0)}\sqrt{\frac{2\pi}{\alpha
\lambda}}e^{i\phi}.
\end{equation}

\begin{figure}
\includegraphics[width=3.7in,angle=-0]{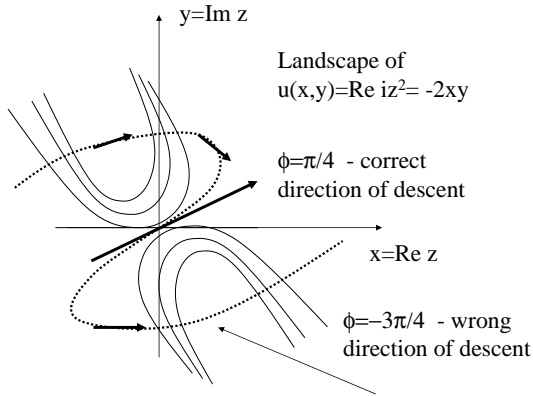}
\caption{Stationary phase method. Correct and incorrect paths of
the steepest descent for $f(z)=iz^2$.} \label{Fig1-4}
\end{figure}
To find the correct choice of $\phi$, one has to look at the
landscape of $u(x,y)={\rm Re}(iz^2)=-2xy$. The landscape is shown in
Fig.~\ref{Fig1-4}. The correct choice is obviously the first one,
$\phi=+\pi/4$, the second would mean that the contour has to be
deformed as shown in Fig.~\ref{Fig1-4} with the dashed line, going through high
mountains on the way to the saddle to cross it in the opposite
direction of $\phi=-3\pi/4$.

So, the final answer is
\begin{equation}
I=g(x_0) e^{i\lambda f(x_0)}\sqrt{\frac{2\pi}{\alpha
\lambda}}e^{i\pi/4}.
\end{equation}


\section{Appendix C: Treating the cut-off region: regularization of the divergent stationary phase solutions}

In this subsection we briefly outline the idea of the so-called uniform approximation
- one of the approaches for handling the merging stationary points.
The regularization involves two steps. First, we need to find a specific real
return time $t_r=t_{r0}$ and the associated $t_i=t_{i0}$, $p_s=p_{s0}$ ,
such that $\partial^2 S(t_{r0},t_{i0},p_{s0})/\partial t^2=0$. In practice,
one can simply pick the real return time corresponding to the cut-off energy.
The  next step  requires the expansion of the total action in Eq.~(\ref{eq:Saction})
around $t=t_{r0}$ in a Taylor series up to the
third order: \bea
 \label{eq:uni}
S(t,t_{i0},p_{s0})=S(t_{r0})+(t-t_{r0})
S'_{tt}+\frac{(t-t_{r0})^2}{2} S''_{tt} +\frac{(t-t_{r0})^3}{6}
S'''_{ttt}, \eea where all the derivatives  of
$S(t,t_{i0},p_{s0})$ are taken at $t_{r0}$. Finally, one substitutes the
expansion (\ref{eq:uni}) into the harmonic dipole,
\bea
&&D(N\omega)\propto  \int_{-\infty}^{\infty} dt e^{iN\omega t}
e^{-iS(t,t_{i0},p_{s0})+iN\omega t} +c.c., \eea and uses the Airy
function \bea
&&\int_{-\infty}^{\infty} dt \cos(at^3\pm x t)\equiv \frac{\pi}{(3a)^{1/3}}Ai\left[ \frac{\pm x}{(3a)^{1/3}}\right].
\eea
Now we introduce the 'cut-off harmonic number' $N_0$ and the distance from the cut-off $N=N- N_0$:
\bea
&& N_0\omega=E_{rec}(t_{r0})+I_p,
\eea
here $E_{rec}(t_{r0})=p_{s0}+A(t_{r0})$ is the re-collision energy at time $t_{r0}$, and $N_0$ does not have to be integer.
The dipole near the cut-off is expressed via the Airy function:
\bea
\int_{-\infty}^{\infty} dt e^{iN\omega t} e^{-iS(t,t_{i0},p_{s0})+iN\omega t} +c.c.=\int_{-\infty}^{\infty} d\xi \cos(\frac{\chi}{6}\xi^3\pm \Delta N\omega \xi), \eea
so that
\bea
D(N\omega)\propto \frac{2\pi}{(\chi/2)^{1/3}}Ai\left[ \frac{\Delta N \omega}{(\chi/2)^{1/3}}\right],
\eea
where $\chi \equiv -S'''_{ttt}(t_{r0})$ and
can be estimated by $\chi \cong v(t_{r0})F_0 \omega$, given that $F'_t(\chi)\cong F_0 \omega$ and $F(\chi)\cong 0$.
Using the asymptotic expansion of the Airy functions,
we obtain simple expressions for the dipole just before and after the cut-off.
Before the cut-off of the harmonic spectra (for $\Delta N<0$), the dipole oscillates,
$Ai \sim \cos(-(\Delta N \omega)^{3/2}(8/9\chi)^{1/2})$, after the cut-off,
the harmonic dipole exponentially decreases, $Ai \sim \textrm{exp}(-(\Delta N \omega)^{3/2}(8/9\chi)^{1/2})$.
The oscillations of the harmonic dipole before the cut-off are due to the interference of the short and the long trajectories.

\section{Appendix D: Finding saddle points for the Lewenstein model}
In section 1.5 we have described how one can find all saddle point solutions
in the Lewenstein model for a fixed harmonic number $N$.
Here we present an alternative and equivalent  approach of finding
the saddle point solutions, i.e. solving the Eqs.~(\ref{eq:ion0},\ref{eq:ret0},\ref{eq:rec0}),
which can be used in all cases, but is particularly convenient if the
Fourier transform is performed numerically.
The idea is to solve the Eqs.~(\ref{eq:ion0},\ref{eq:ret0},\ref{eq:rec0})
'forward', i.e. to fix the grid of the real recombination times and find all the other
saddle point solutions, and the corresponding harmonic number $N$.
The recombination condition  Eq. (\ref{eq:rec0}) $(p_{s,\parallel}=p'+ip'')$
can be re-written as follows:
\bea
\left(\Delta p'+i\Delta p''\right)^2&=&2(N\omega-I_p),\\
\Delta p'&\equiv& p'-A_0\sin(\phi_r')\cosh(\phi_r''),\\
\Delta p''&\equiv& p''-A_0\sinh(\phi_r'')\cos(\phi_r'),
\eea
yielding
\bea
&&\left(\Delta p'\right)^2-\left(\Delta p''\right)^2+2i\Delta p'\Delta p''=2(N\omega-I_p).
\eea
Since the right-hand side of this equation is real, we obtain that $\Delta p'\Delta p''=0$.
For above threshold harmonics $(N\omega-I_p)>0$ and $\Delta p'\ne0$, $\Delta p''=0$.
For below threshold harmonics $(N\omega-I_p)<0$ and $\Delta p'=0$, $\Delta p''\ne0$.
Separating the imaginary and the real parts in the Eqs.~(\ref{eq:ion0},\ref{eq:ret0}),
we obtain the four equations quoted in the main text, see Eqs.~(\ref{eq:F11prim},\ref{eq:F21prim},\ref{eq:eion_r},\ref{eq:eion_i}).
Supplementing these equations for above threshold harmonics with $\Delta p''=0$ yields
\bea
\label{eq:above}
 &&p_2=\sinh(\phi_r'')\cos(\phi_r'),
 \eea
and for below threshold  harmonics with $\Delta p'=0$, yielding
\bea
\label{eq:below}
 &&p_1=\sin(\phi_r')\cosh(\phi_r''),
 \eea
  we obtain five equations.

\textbf{For above threshold harmonics}: for each fixed $\phi_r'$
we use the Eqs.~(\ref{eq:eion_r},\ref{eq:eion_i}) to express $\phi_i'$, $\phi_i''$ via
$p_2$ and $p_1$ and then we use Eq.~(\ref{eq:above}) to exclude $p_2$ and substitute
$\phi_i'(p_1,\phi_r'')$, $\phi_i''(p_1,\phi_r'')$ and $p_2(p_1,\phi_r'')$ into the Eqs.~(\ref{eq:F11prim},\ref{eq:F21prim}).
Using the gradient method, we can now find the minima of the function $F=F_1^2+F_2^2$
in the plane of $p_1$ and $\phi_r''$ for each fixed $\phi_r'$.
The minima define the saddle point solutions for $p_1$ and $\phi_r''$.
Knowing $p_1$ and $\phi_r''$, we find $\phi_i'$, $\phi_i''$, $p_2$ from the Eqs.~(\ref{eq:eion_r},\ref{eq:eion_i},\ref{eq:above}).
Finally, the corresponding harmonic number can be calculated from $\left(\Delta p'\right)^2=2(N\omega-I_p)$,
yielding $N\omega=A_0^2\left(p_1-\sin\phi_r'\cosh\phi_r'')^2\right/2+I_p$.
Naturally, the harmonic number defined this way does not have to be integer.

\textbf{For below threshold harmonics}, the procedure is essentially the same.
For each fixed $\phi_r'$, we use the Eqs.~(\ref{eq:eion_r},\ref{eq:eion_i})
to express $\phi_i'$, $\phi_i''$ via $p_2$ and $p_1$ and then we use Eq.~(\ref{eq:below})
to exclude $p_1$ and substitute $\phi_i'(p_2,\phi_r'')$, $\phi_i''(p_2,\phi_r'')$ and
$p_1(p_2,\phi_r'')$ into the Eqs.~(\ref{eq:F11prim} and \ref{eq:F21prim}).
Using the gradient method, we can now find the minima of the function $F=F_1^2+F_2^2$
in the plane of $p_2$ and $\phi_r''$ for each fixed $\phi_r'$.
The minima define the saddle point solutions for $p_2$ and $\phi_r''$.
Knowing $p_2$ and $\phi_r''$ we find $\phi_i'$, $\phi_i''$, $p_1$
from the Eqs.~(\ref{eq:eion_r},\ref{eq:eion_i},\ref{eq:below}).
Finally, the corresponding harmonic number can be calculated from $\left(\Delta p''\right)^2=2(I_p-N\omega)$,
yielding $N\omega=I_p-A_0^2\left(p_2-\sinh\phi_r''\cos\phi_r')^2\right/2$.

In this method, it is convenient to determine the return time $\phi_{tr}'$
corresponding to the threshold harmonic number $N_{t}=I_p/\omega$.
This can be easily done, since, at the threshold, $p_2=\sinh(\phi_r'')\cos(\phi_r')$ and $p_1=\sin(\phi_r')\cosh(\phi_r'')$.
Thus, we can use these equations together with the Eqs.~(\ref{eq:eion_r},\ref{eq:eion_i} )
to express $\phi_i'$, $\phi_i''$ via $\phi_r'$ and $\phi_r''$ and then use the Eqs.~(\ref{eq:F11prim},\ref{eq:F21prim})
to find a minimum of the function $F=F_1^2+F_2^2$ in the plane of $\phi_r'$ and $\phi_r''$, representing the
threshold values of $\phi_r'$ and $\phi_r''$.
Once the threshold value $\phi_{rt}'$ of $\phi_r'$ is known,
one can separately implement the procedures described above
for below, $\phi_r'<\phi_{rt}'$, and above, $\phi_r'>\phi_{tr}'$, threshold harmonics.

\bibliographystyle{wivchauy}
\bibliography{Smirnova}

\let\be\undefined
\let\ee\undefined
\let\bea\undefined
\let\eea\undefined
\let\ra\undefined
\let\la\undefined
\let\bef\undefined
\let\enf\undefined
\let\p\undefined
\let\rmbf\undefined
\let\pc\undefined
\let\x\undefined
\let\y\undefined
\let\q\undefined
\let\R\undefined
\let\w\undefined
\let\k\undefined
\let\F\undefined
\let\A\undefined
\let\D\undefined
\let\d\undefined

%% file: main.bbl
\begin{thebibliography}{67}
\providecommand{\natexlab}[1]{#1}
\providecommand{\url}[1]{\texttt{#1}}
\providecommand{\urlprefix}{URL }
\expandafter\ifx\csname urlstyle\endcsname\relax
  \providecommand{\doi}[1]{doi:\discretionary{}{}{}#1}\else
  \providecommand{\doi}{doi:\discretionary{}{}{}\begingroup
  \urlstyle{rm}\Url}\fi
\input{babelbst.tex}
\newcommand{\Capitalize}[1]{\uppercase{#1}}
\newcommand{\capitalize}[1]{\expandafter\Capitalize#1}

\bibitem[{Baker \emph{\bbletal{}}(2006)Baker, Robinson, Haworth, Teng, Smith,
  Chirila, Lein, Tisch, \bbland{} Marangos}]{Baker06}
Baker, S., Robinson, J.S., Haworth, C.A., Teng, H., Smith, R.A., Chirila, C.C.,
  Lein, M., Tisch, J.W.G., \bbland{} Marangos, J.P. (2006) Probing proton
  dynamics in molecules on an attosecond time scale. \emph{Science},
  \textbf{312}, 424.

\bibitem[{Barth \bbland{} Smirnova(2011)}]{Barth11}
Barth, I. \bbland{} Smirnova, O. (2011) Nonadiabatic tunneling in circularly
  polarized laser fields: Physical picture and calculations. \emph{Phys. Rev.
  A}, \textbf{84}, 063\,415.

\bibitem[{Barth \bbland{} Smirnova(2013)}]{Barth13}
Barth, I. \bbland{} Smirnova, O. (2013) Nonadiabatic tunneling in circularly
  polarized laser fields ii: Derivation of formulas. \emph{Phys. Rev. A},
  \textbf{87}, 013\,433.

\bibitem[{Becker \emph{\bbletal{}}(2002{\natexlab{a}})Becker, Grasbon, Kopold,
  B., Paulus, \bbland{} Walther}]{Becker02}
Becker, W., Grasbon, F., Kopold, R., B., M.D., Paulus, G.G., \bbland{} Walther,
  H. (2002{\natexlab{a}}) Above-threshold ionization: from classical features
  to quantum effects. \emph{Advances In Atomic, Molecular, and Optical
  Physics}, \textbf{48}, 35.

\bibitem[{Becker \emph{\bbletal{}}(2002{\natexlab{b}})Becker, Grasbon, Kopold,
  Milosevic, Paulus, \bbland{} Walther}]{Wilhelm}
Becker, W., Grasbon, F., Kopold, R., Milosevic, D.B., Paulus, G.G., \bbland{}
  Walther, H. (2002{\natexlab{b}}) Above-threshold ionization: From classical
  features to quantum effects. \emph{Adv. At. Mol. Opt. Phys.}, \textbf{48},
  36.

\bibitem[{Bloch(1957)}]{Bloch}
Bloch, C. (1957) Une formulation unifiee de la theorie des reactions
  nucleaires. \emph{Nuclear Physics}, \textbf{4}, 503.

\bibitem[{Brunel(1987)}]{Brunel87}
Brunel, F. (1987) Not-so-resonant, resonant absorption. \emph{Phys. Rev.
  Lett.}, \textbf{59}, 52.

\bibitem[{Brunel(1990)}]{Brunel90}
Brunel, F. (1990) Harmonic generation due to plasma effects in a gas undergoing
  multiphoton ionization in the high-intensity limit. \emph{J. Opt. Soc. Am.
  B}, \textbf{7}, 521.

\bibitem[{Corkum(1993)}]{Corkum93}
Corkum, P.B. (1993) Plasma perspective on strong field multiphoton ionization.
  \emph{Phys. Rev. Lett}, \textbf{71}, 1994.

\bibitem[{Corkum \emph{\bbletal{}}(1989)Corkum, Burnett, \bbland{}
  Brunel}]{Brunel89}
Corkum, P.B., Burnett, N.H., \bbland{} Brunel, F. (1989) Above-threshold
  ionization in the long-wavelength limit. \emph{Phys. Rev. Lett.},
  \textbf{62}, 1259.

\bibitem[{Frolov \emph{\bbletal{}}(2003)Frolov, Manakov, Pronin, \bbland{}
  Starace}]{Frolov_prl}
Frolov, M.V., Manakov, N.L., Pronin, E.A., \bbland{} Starace, A.F. (2003)
  Model-independent quantum approach for intense laser detachment of a weakly
  bound electron. \emph{Phys. Rev. Lett.}, \textbf{91}, 053\,003.

\bibitem[{Frolov \emph{\bbletal{}}(2009)Frolov, Manakov, Sarantseva, Emelin,
  Ryabikin, \bbland{} Starace}]{Frolov09}
Frolov, M.V., Manakov, N.L., Sarantseva, T.S., Emelin, M.Y., Ryabikin, M.Y.,
  \bbland{} Starace, A.F. (2009) Analytic description of the high-energy
  plateau in harmonic generation by atoms: Can the harmonic power increase with
  increasing laser wavelengths? \emph{Phys. Rev. Lett.}, \textbf{102},
  243\,901.

\bibitem[{Gaarde \emph{\bbletal{}}(2008)Gaarde, Tate, \bbland{}
  Schafer}]{Gaarde2008}
Gaarde, M.B., Tate, J.L., \bbland{} Schafer, K.J. (2008) Macroscopic aspects of
  attosecond pulse generation. \emph{J. Phys. B: At. Mol. Opt. Phys.},
  \textbf{41}, 132\,001.

\bibitem[{Gallagher(1988)}]{Gallagher88}
Gallagher, T. (1988) Above-threshold ionization in low-frequency limit.
  \emph{Phys. Rev. Lett.,}, \textbf{61}, 2304.

\bibitem[{Gordon \bbland{} Santra(2006)}]{santra}
Gordon, A. \bbland{} Santra, R. (2006) Three-step model for high-harmonic
  generation in many-electron systems. \emph{Phys. Rev. Lett}, \textbf{96},
  073\,906.

\bibitem[{Gribakin \bbland{} Kuchiev(1997)}]{Gribakin97}
Gribakin, G.F. \bbland{} Kuchiev, M.Y. (1997) Multiphoton detachment of
  electrons from negative ions. \emph{Phys. Rev. A}, \textbf{55}, 3760.

\bibitem[{Grynberg \emph{\bbletal{}}(2010)Grynberg, Aspect, \bbland{}
  Fabre}]{Grynberg}
Grynberg, G., Aspect, A., \bbland{} Fabre, C. (2010) \emph{Introduction to
  Quantum Optics: From the Semi-classical Approach to Quantized Light},
  Cambridge University Press.

\bibitem[{Haessler \emph{\bbletal{}}(2010)Haessler, Boutu, Giovanetti-Teixeira,
  Ruchon, Auguste, Diveki, Breger, Maquet, Carre, Taieb, \bbland{}
  Salieres}]{Stefan}
Haessler, S.and~Caillat, J., Boutu, W., Giovanetti-Teixeira, C., Ruchon, T.,
  Auguste, T., Diveki, Z., Breger, P., Maquet, A., Carre, B., Taieb, R.,
  \bbland{} Salieres, P. (2010) Attosecond imaging of molecular electronic
  wavepackets. \emph{Nature Phys.}, \textbf{6}, 200.

\bibitem[{Harvey \emph{\bbletal{}}(2012)Harvey, Morales, \bbland{}
  Smirnova}]{Alex}
Harvey, A.G., Morales, F., \bbland{} Smirnova, O. (2012) The r-matrix method
  for attosecond spectroscopy. \emph{Theory of electron-molecule collisions for
  astrophysics, biophysics and low temperature plasmas: opportunities and
  challenges}, \textbf{106}, 16.

\bibitem[{Harvey \bbland{} J.(2009)}]{Harvey09}
Harvey, A. \bbland{} J., T. (2009) The r-matrix method for attosecond
  spectroscopy. \emph{J. Phys. B}, \textbf{42}, 095\,101.

\bibitem[{Huillier \bbland{} Balcou(1993)}]{Huillier93}
Huillier, A. \bbland{} Balcou, P. (1993) High-order harmonic generation in rare
  gases with a 1-ps 1053-nm laser. \emph{Phys Rev Lett}, \textbf{70}, 774.

\bibitem[{Ivanov \emph{\bbletal{}}(1996)Ivanov, Brabec, \bbland{}
  Burnett}]{Ivanov96}
Ivanov, M.Y., Brabec, T., \bbland{} Burnett, N. (1996) Coulomb corrections and
  polarization effects in high-intensity high-harmonic emission. \emph{Phys.
  Rev. A.}, \textbf{54}, 742--745.

\bibitem[{Kaushal \bbland{} Smirnova(2013)}]{jivesh13}
Kaushal, J. \bbland{} Smirnova, O. (2013) Non-adiabatic coulomb effects in
  strong field ionisation in circularly polarised laser fields i: Ionisation
  rates. \emph{arXiv preprint arXiv:1302.2609}.

\bibitem[{Keldysh(1965)}]{Keldysh64}
Keldysh, L.V. (1965) Ionization in the field of a strong electromagnetic wave.
  \emph{Sov. Phys. JETP}, \textbf{20}, 1307.

\bibitem[{Kopold \emph{\bbletal{}}(2002)Kopold, Becker, \bbland{}
  Milosevic}]{Kopold02}
Kopold, R., Becker, W., \bbland{} Milosevic, D.B. (2002) Quantum orbits: a
  space-time picture of intense-laser-induced processes in atoms. \emph{Journal
  of Modern Optics}, \textbf{49}, 1362.

\bibitem[{Krause \emph{\bbletal{}}(1992{\natexlab{a}})Krause, Schafer,
  \bbland{} Kulander}]{Kulander92}
Krause, J.L., Schafer, K.J., \bbland{} Kulander, K.C. (1992{\natexlab{a}})
  High-order harmonic generation from atoms and ions in the high intensity
  regime. \emph{Phys. Rev. Lett.}, \textbf{68}, 3535.

\bibitem[{Krause \emph{\bbletal{}}(1992{\natexlab{b}})Krause, Schafer,
  \bbland{} Kulander}]{Krause92}
Krause, J.L., Schafer, K.J., \bbland{} Kulander, K.C. (1992{\natexlab{b}})
  High-order harmonic generation from atoms and ions in the high intensity
  regime. \emph{Phys. Rev. Lett}, \textbf{68}, 3535.

\bibitem[{Krausz \bbland{} Ivanov(2009)}]{Krausz09}
Krausz, F. \bbland{} Ivanov, F. (2009) Attosecond physics. \emph{Rev. Mod.
  Phys.}, \textbf{81}, 163.

\bibitem[{Kuchiev(1987)}]{Kuchiev87}
Kuchiev, M.Y. (1987) Atomic antenna. \emph{Sov. Phys. JETP Letters},
  \textbf{45}, 319.

\bibitem[{Kuchiev \bbland{} Ostrovsky(1999)}]{Kuchiev99}
Kuchiev, M.Y. \bbland{} Ostrovsky, V.N. (1999) Quantum theory of high harmonic
  generation as a three-step process. \emph{Phys Rev A}, \textbf{60}, 3111.

\bibitem[{Kulander \emph{\bbletal{}}(1993)Kulander, Krause, \bbland{}
  Schafer}]{Kulander93}
Kulander, K., Krause, J.L., \bbland{} Schafer, K.J. (1993) \emph{NATO ASI
  Series}, Plenum Press, New York and London.

\bibitem[{Le \emph{\bbletal{}}(2009)Le, Lucchese, Tonzani, Morishita, \bbland{}
  Lin}]{Le09}
Le, A.T., Lucchese, R.R., Tonzani, S., Morishita, T., \bbland{} Lin, C.D.
  (2009) Quantitative rescattering theory for high-order harmonic generation
  from molecules. \emph{Phys. Rev. A}, \textbf{80}, 013\,401.

\bibitem[{Lein(2005)}]{Lein05}
Lein, M. (2005) Attosecond probing of vibrational dynamics with high-harmonic
  generation. \emph{Phys Rev Lett}, \textbf{94}, 053\,004.

\bibitem[{Lewenstein \emph{\bbletal{}}(1994)Lewenstein, Balcou, Ivanov,
  L'Huillier, \bbland{} Corkum}]{Lewenstein94}
Lewenstein, M., Balcou, P., Ivanov, M.Y., L'Huillier, A., \bbland{} Corkum,
  P.B. (1994) Theory of high-harmonic generation by low-frequency laser fields.
  \emph{Phys. Rev. A.}, \textbf{49}, 2117.

\bibitem[{Lin \emph{\bbletal{}}(2010)Lin, Le, Chen, Morishita, \bbland{}
  Lucchese}]{Lin10}
Lin, C.D., Le, A.T., Chen, Z., Morishita, T., \bbland{} Lucchese, R. (2010)
  Strong-field rescattering physics—self-imaging of a molecule by its own
  electrons. \emph{J. Phys. B: At. Mol. Opt. Phys.}, \textbf{43}, 122\,001.

\bibitem[{Lunnemann \emph{\bbletal{}}(2008)Lunnemann, Kuleff, \bbland{}
  Cederbaum}]{Kulef}
Lunnemann, S., Kuleff, A., \bbland{} Cederbaum, L. (2008) Ultrafast charge
  migration in 2-phenylethyl-n,n-dimethylamine. \emph{Chem. Phys. Lett},
  \textbf{450}, 232.

\bibitem[{Macklin \emph{\bbletal{}}(1993)Macklin, Kmetec, \bbland{}
  Gordon}]{Macklin93}
Macklin, J.J., Kmetec, J.D., \bbland{} Gordon, C.L. (1993) High-order harmonic
  generation using intense femtosecond pulses. \emph{Phys Rev Lett},
  \textbf{70}, 766.

\bibitem[{Mairesse \emph{\bbletal{}}(2010)Mairesse, Higuet, Dudovich, Shafir,
  Fabre, M\'evel, Constant, Patchkovskii, Walters, Ivanov, \bbland{}
  Smirnova}]{Mairesse10}
Mairesse, Y., Higuet, J., Dudovich, N., Shafir, D., Fabre, B., M\'evel, E.,
  Constant, E., Patchkovskii, S., Walters, Z., Ivanov, M.Y., \bbland{}
  Smirnova, O. (2010) High harmonic spectroscopy of multichannel dynamics in
  strong-field ionization. \emph{Phys. Rev. Lett.}, \textbf{104}, 213\,601.

\bibitem[{Morales \emph{\bbletal{}}(2012)Morales, Barth, Serbinenko,
  Patchkovskii, \bbland{} O.}]{Morales12}
Morales, F., Barth, I., Serbinenko, V., Patchkovskii, S., \bbland{} O., S.
  (2012) Shaping polarization of attosecond pulses via laser control of
  electron and hole dynamics. \emph{J. Mod. Opt.}, \textbf{59}, 1303.

\bibitem[{Morishita \emph{\bbletal{}}(2008)Morishita, Le, Chen, \bbland{}
  Lin}]{Morishita08}
Morishita, T., Le, A.T., Chen, Z., \bbland{} Lin, C.D. (2008) Accurate
  retrieval of structural information from laser-induced photoelectron and
  high-order harmonic spectra by few-cycle laser pulses. \emph{Phys. Rev.
  Lett}, \textbf{100}, 013\,903.

\bibitem[{Murray \emph{\bbletal{}}(2011)Murray, Spanner, Patchkovskii,
  \bbland{} Ivanov}]{Murray11}
Murray, R., Spanner, M., Patchkovskii, S., \bbland{} Ivanov, M.Y. (2011) Tunnel
  ionization of molecules and orbital imaging. \emph{Phys. Rev. Lett.},
  \textbf{106}, 173\,001.

\bibitem[{Ott \emph{\bbletal{}}(2013)Ott, Kaldun, Raith, Meyer, Laux, Evers,
  Keitel, Greene, \bbland{} T.}]{ott}
Ott, C., Kaldun, A., Raith, P., Meyer, K., Laux, M., Evers, J., Keitel, C.H.,
  Greene, C.H., \bbland{} T., P. (2013) Lorentz meets $\textsc{F}$ano spectral
  line shapes: A universal phase and its laser control. \emph{arXiv:1301.1454}.

\bibitem[{Patchkovskii \emph{\bbletal{}}(2012)Patchkovskii, Smirnova, \bbland{}
  Spanner}]{serguei}
Patchkovskii, S., Smirnova, O., \bbland{} Spanner, M. (2012) Attosecond control
  of electron correlations in one-photon ionization and recombination. \emph{J.
  Phys. B}, \textbf{45}, 131\,002.

\bibitem[{Patchkovskii \emph{\bbletal{}}(2006)Patchkovskii, Zhao, Brabec,
  \bbland{} Villeneuve}]{patchkovskii}
Patchkovskii, S., Zhao, Z., Brabec, T., \bbland{} Villeneuve, D.M. (2006) High
  harmonic generation and molecular orbital tomography in multielectron
  systems: beyond the single active electron approximation. \emph{Phys. Rev.
  Lett}, \textbf{97}, 123\,003.

\bibitem[{Perelomov \bbland{} Popov(1967)}]{PPT3}
Perelomov, A.M. \bbland{} Popov, V.S. (1967) \emph{Sov. Phys. JETP,},
  \textbf{25}, 336.

\bibitem[{Perelomov \emph{\bbletal{}}(1966)Perelomov, Popov, \bbland{}
  Terent'ev}]{PPT1}
Perelomov, A.M., Popov, V.S., \bbland{} Terent'ev, M.V. (1966) \emph{Sov. Phys.
  JETP,}, \textbf{23}, 924.

\bibitem[{Perelomov \emph{\bbletal{}}(1967)Perelomov, Popov, \bbland{}
  Terent'ev}]{PPT2}
Perelomov, A.M., Popov, V.S., \bbland{} Terent'ev, M.V. (1967) \emph{Sov. Phys.
  JETP,}, \textbf{24}, 207.

\bibitem[{Popov \emph{\bbletal{}}(1968)Popov, Kuznetsov, \bbland{}
  Perelomov}]{PPT4}
Popov, V.S., Kuznetsov, V.P., \bbland{} Perelomov, A.M. (1968) \emph{Sov. Phys.
  JETP,}, \textbf{26}, 222.

\bibitem[{Popruzhenko \emph{\bbletal{}}(2008)Popruzhenko, Mur, Popov, \bbland{}
  Bauer}]{Popruzhenko}
Popruzhenko, S.V., Mur, V.D., Popov, V.S., \bbland{} Bauer, D. (2008) Strong
  field ionization rate for arbitrary laser frequencies. \emph{Phys. Rev.
  Lett}, \textbf{101}, 193\,003.

\bibitem[{Salieres \emph{\bbletal{}}(2001)Salieres, Carré, Le~Deroff, Grasbon,
  Paulus, Walther, Kopold, Becker, Milosevic, Sanpera, \bbland{}
  Lewenstein}]{Salieres01}
Salieres, P., Carré, B., Le~Deroff, L., Grasbon, F., Paulus, G.G., Walther,
  H., Kopold, R., Becker, W., Milosevic, D.B., Sanpera, A., \bbland{}
  Lewenstein, M. (2001) Feynman's path-integral approach for intense-laser-atom
  interactions. \emph{Science}, \textbf{292}, 902.

\bibitem[{Schafer \emph{\bbletal{}}(1993)Schafer, Yang, DiMauro, \bbland{}
  Kulander}]{Schafer93}
Schafer, K.J., Yang, B., DiMauro, L.F., \bbland{} Kulander, K.C. (1993) Above
  threshold ionization beyond the high harmonic cutoff. \emph{Phys. Rev.
  Lett.,}, \textbf{70}, 1599.

\bibitem[{Schafer \emph{\bbletal{}}(2004)Schafer, Gaarde, A., J., \bbland{}
  Keller}]{SchaferXUV}
Schafer, K., Gaarde, M.B., A., H., J., B., \bbland{} Keller, U. (2004) Strong
  field quantum path control using attosecond pulse trains. \emph{Phys. Rev.
  Lett}, \textbf{92}, 023\,003.

\bibitem[{Shafir \emph{\bbletal{}}(2012)Shafir, Soifer, Bruner, Dagan,
  Mairesse, Patchkovskii, Ivanov, Smirnova, \bbland{} Dudovich}]{Shafir12}
Shafir, D., Soifer, H., Bruner, B.D., Dagan, M., Mairesse, Y., Patchkovskii,
  S., Ivanov, M.Y., Smirnova, O., \bbland{} Dudovich, N. (2012) Resolving the
  time when an electron exits a tunnelling barrier. \emph{Nature},
  \textbf{485}, 343.

\bibitem[{Smirnova \emph{\bbletal{}}(2009{\natexlab{a}})Smirnova, Mairesse,
  Patchkovskii, Dudovich, Villeneuve, Corkum, \bbland{} Ivanov}]{Smirnova09}
Smirnova, O., Mairesse, Y., Patchkovskii, S., Dudovich, N., Villeneuve, D.,
  Corkum, P., \bbland{} Ivanov, M.Y. (2009{\natexlab{a}}) High harmonic
  interferometry of multi-electron dynamics in molecules. \emph{Nature},
  \textbf{460}, 972.

\bibitem[{Smirnova \emph{\bbletal{}}(2007{\natexlab{a}})Smirnova, Mouritzen,
  Patchkovskii, \bbland{} Ivanov}]{Smirnova07}
Smirnova, O., Mouritzen, A.S., Patchkovskii, S., \bbland{} Ivanov, M.Y.
  (2007{\natexlab{a}}) Coulomb–laser coupling in laser-assisted
  photoionization and molecular tomography. \emph{J. Phys. B: At. Mol. Opt.
  Phys.}, \textbf{40}, F197.

\bibitem[{Smirnova \emph{\bbletal{}}(2009{\natexlab{b}})Smirnova, Patchkovskii,
  Mairesse, Dudovich, \bbland{} Ivanov}]{Smirnova09b}
Smirnova, O., Patchkovskii, S., Mairesse, Y., Dudovich, N., \bbland{} Ivanov,
  M.Y. (2009{\natexlab{b}}) Strong-field control and spectroscopy of attosecond
  electron-hole dynamics in molecules. \emph{Proceedings of the National
  Academy of Sciences (PNAS)}, \textbf{106}, 16\,556.

\bibitem[{Smirnova \emph{\bbletal{}}(2009{\natexlab{c}})Smirnova, Patchkovskii,
  Mairesse, Dudovich, Villeneuve, Corkum, \bbland{} Ivanov}]{Smirnova09a}
Smirnova, O., Patchkovskii, S., Mairesse, Y., Dudovich, N., Villeneuve, D.,
  Corkum, P., \bbland{} Ivanov, M.Y. (2009{\natexlab{c}}) Attosecond circular
  dichroism spectroscopy of polyatomic molecules. \emph{Phys. Rev. Lett.},
  \textbf{102}, 063\,601.

\bibitem[{Smirnova \emph{\bbletal{}}(2007{\natexlab{b}})Smirnova, Spanner,
  \bbland{} Ivanov}]{SmirnovaJMO07}
Smirnova, O., Spanner, M., \bbland{} Ivanov, M. (2007{\natexlab{b}}) Anatomy of
  strong field ionization ii: to dress or not to dress? \emph{J. Mod. Optics},
  \textbf{54}, 1019.

\bibitem[{Smirnova \emph{\bbletal{}}(2008)Smirnova, Spanner, \bbland{}
  Ivanov}]{Smirnova08}
Smirnova, O., Spanner, M., \bbland{} Ivanov, M. (2008) Analytical solutions for
  strong field-driven atomic and molecular one- and two-electron continua and
  applications to strong-field problems. \emph{Phys. Rev. A}, \textbf{77},
  033\,407.

\bibitem[{Sukiasyan \emph{\bbletal{}}(2010)Sukiasyan, Smirnova, Brabec,
  \bbland{} Ivanov}]{Suren}
Sukiasyan, S.and~Patchkovskii, S., Smirnova, O., Brabec, T., \bbland{} Ivanov,
  M.Y. (2010) Exchange and polarization effect in high-order harmonic imaging
  of molecular structures. \emph{Phys. Rev. A}, \textbf{82}, 043\,414.

\bibitem[{Torlina \emph{\bbletal{}}(2012)Torlina, Ivanov, Walters, \bbland{}
  Smirnova}]{lisa12b}
Torlina, L., Ivanov, M., Walters, Z.B., \bbland{} Smirnova, O. (2012)
  Time-dependent analytical r-matrix approach for strong-field dynamics. ii.
  many-electron systems. \emph{Phys. Rev. A}, \textbf{86}, 043\,409.

\bibitem[{Torlina \emph{\bbletal{}}(2013)Torlina, Kaushal, \bbland{}
  Smirnova}]{Lisa12c}
Torlina, L., Kaushal, J., \bbland{} Smirnova, O. (2013) Nonadiabatic coulomb
  effects in strong field ionization in circularly polarized laser fields ii:
  Photoelectron spectra. \emph{in preparation}.

\bibitem[{Torlina \bbland{} Smirnova(2012)}]{Lisa12}
Torlina, L. \bbland{} Smirnova, O. (2012) Time-dependent analytical r-matrix
  approach for strong-field dynamics. i. one-electron systems. \emph{Phys. Rev.
  A}, \textbf{86}, 043\,408.

\bibitem[{Walters \bbland{} Smirnova(2010)}]{Walters10}
Walters, Z.B. \bbland{} Smirnova, O. (2010) Attosecond correlation dynamics
  during electron tunnelling from molecules. \emph{J. Phys. B: At. Mol. Opt.
  Phys.}, \textbf{43}, 161\,002.

\bibitem[{Weinkauf \emph{\bbletal{}}(1997)Weinkauf, Schlag, Martinez, \bbland{}
  .}]{Weinkauf}
Weinkauf, R., Schlag, E., Martinez, T., \bbland{} ., L.R. (1997) Nonstationary
  electronic states and site-selective reactivity. \emph{Journal of Physical
  Chemistry A}, \textbf{101}, 7702.

\bibitem[{Yudin \bbland{} Ivanov(2001)}]{YudinIvanov01}
Yudin, G.L. \bbland{} Ivanov, M.Y. (2001) Nonadiabatic tunnel ionization:
  Looking inside a laser cycle. \emph{Phys. Rev. A}, \textbf{64}, 013\,409.

\bibitem[{Zair \emph{\bbletal{}}(2008)Zair, Holler, Guandalini, Schapper,
  Biegert, Gallmann, Keller, Wyatt, Monmayrant, Walmsley, Cormier, Auguste,
  Caumes, \bbland{} Sali\`eres}]{Zair08}
Zair, A., Holler, M., Guandalini, A., Schapper, F., Biegert, J., Gallmann, L.,
  Keller, U., Wyatt, A.S., Monmayrant, A., Walmsley, I.A., Cormier, E.,
  Auguste, T., Caumes, J.P., \bbland{} Sali\`eres, P. (2008) Quantum path
  interferences in high-order harmonic generation. \emph{Phys. Rev. Lett.},
  \textbf{100}, 143\,902.

\end{thebibliography}
